\documentclass[%
twocolumn,
superscriptaddress,
longbibliography,
showpacs,
preprintnumbers,
nofootinbib,
amsmath,amssymb,
aps,
prb,
]{revtex4-1}

\usepackage{graphicx}
\usepackage{dcolumn} % Align table columns on decimal point
\usepackage{bm} % bold math
\usepackage{siunitx}
\usepackage[english]{babel}

\usepackage{epstopdf}

\usepackage{hyperref}% add hypertext capabilities

\usepackage[final]{changes}
\definechangesauthor[name=C. Pfleiderer, color=blue]{CP}

\newcommand*{\plotfont}{\fontfamily{phv}\selectfont}
\DeclareTextFontCommand{\textplotfont}{\plotfont}

\begin{document}

\newcommand{\cpr}{CePd$_{1-x}$Rh$_{x}$}

\title{Neutron Depolarization due to Ferromagnetism and Spin Freezing 
in CePd$_{1-x}$Rh$_x$}

\author{M. Seifert}
\affiliation{Physik-Department, Technical University Munich, D-85747 Garching, Germany}

\author{P. Schmakat}
\affiliation{Physik-Department, Technical University Munich, D-85747 Garching, Germany}
\affiliation{Heinz Maier-Leibnitz Zentrum (MLZ), Technical University Munich, D-85748 Garching, Germany}

\author{M. Schulz}
\affiliation{Heinz Maier-Leibnitz Zentrum (MLZ), Technical University Munich, D-85748 Garching, Germany}

\author{P. Jorba}
\affiliation{Physik-Department, Technical University Munich, D-85747 Garching, Germany}

\author{V. Hutanu}
\affiliation{Heinz Maier-Leibnitz Zentrum (MLZ), Technical University Munich, D-85748 Garching, Germany}

\author{C. Geibel}
\affiliation{Max-Planck-Institute for Chemical Physics of Solids, D-01187 Dresden, Germany}

\author{M. Deppe}
\affiliation{Max-Planck-Institute for Chemical Physics of Solids, D-01187 Dresden, Germany}

\author{C. Pfleiderer}
\affiliation{Physik-Department, Technical University Munich, D-85747 Garching, Germany}
\affiliation{Centre for QuantumEngineering (ZQE), Technical University of Munich, D-85748 Garching, Germany}
\affiliation{Munich Center for Quantum Science and Technology (MCQST), Technical University Munich, D-85748 Garching, Germany}

\date{\today}

%%%%%%%%%%%%%%%%%%%%%%%%%%%%%%%%%%%
\begin{abstract}
We report neutron depolarization measurements of the suppression of long-range ferromagnetism and the emergence of magnetic irreversibilities and spin freezing in {\cpr} around $x^*\approx0.6$. Tracking the temperature versus field history of the neutron depolarization, we find clear signatures of long-range Ising ferromagnetism below a Curie temperature $T_{\rm C}$ for $x=0.4$ and a spin freezing of tiny ferromagnetic clusters below a freezing temperature $T_{\rm F1}$ for $x>x^*$. Under zero-field-cooling/field-heating and for $x>x^*$ a reentrant temperature dependence of the neutron depolarization between $T_{\rm F2}<T_{\rm F1}$ and $T_{\rm F1}$ is microscopically consistent with a thermally activated growth of the cluster size. The evolution of the depolarization as well as the reentrant temperature dependence as a function of Rh content are consistent with the formation of a Kondo-cluster glass below $T_{\rm F1}$ adjacent to a ferromagnetic quantum phase transition at $x^*$.
\end{abstract}

\maketitle

%%%%%%%%%%%%%%%%%%%%%%%%%%%%%%%%
\section{Motivation and Outline}
\label{sec:intro}

Quantum phase transitions (QPTs), defined as zero temperature phase transitions, represent a well-established roadmap in the search for new properties of correlated electron systems\cite{lohneysen2007,senthil2004,si2010, 2003_Vojta_RPP, 2018_Vojta_RepProgPhysb}. The perhaps simplest example of a magnetic QPT is associated with the suppression or the emergence of long-range ferromagnetic order\cite{brando2016} as a function of a non-thermal control parameter such as hydrostatic pressure\cite{2001_Pfleiderer_Naturec,1995_Grosche_PhysicaB,uhlarz2004,saxena2000}, an applied magnetic field\cite{1997_Thessieu_JPhysCondMatter, 2001_Perry_PRL, paglione2003, gegenwart2002}, or chemical composition\cite{fuchs2014,marel2003}. In clean materials a variety of escape routes to ferromagnetic quantum criticality have been identified\cite{brando2016}. For instance, the coupling of the magnetization to electronic soft modes may generically lead to a first-order QPTs \cite{belitz2005, brando2016, 2005_Pfleiderer_JPhysCondensMatterd}, or new forms of order masking the QPT such as spin density wave order\cite{2008_Brando_PhysRevLettb, 2015_Abdul_Jabbar_NaturePhysa, 2018_Friedemann_NaturePhys} or superconductivity\cite{saxena2000,aoki2001,huy2007,aoki2019,2009_Pfleiderer_RevModPhys,2019_Haslbeck_PhysRevB}. In systems featuring defects and disorder, a ferromagnetic QPT may drive the appearance of intermediate phases such as frustrated magnetism or the formation of magnetic clusters\cite{miranda2005,vojta2010}, as well as electronic phase segregation\cite{pfleiderer2010, 2007_Uemura_NaturePhys, 2015_Schmakat_EPJST, 2020_Benka_arXiv}.

An increasing number of studies suggest an intricate interplay of microscopic (atomic) and mesoscopic scales at QPTs. For instance, the temperature dependence of the electrical resistivity at the pressure-tuned quantum phase transition of pure itinerant-electron ferromagnets such as Ni$_3$Al or ZrZn$_2$, for unexplained reasons, is characteristic of disordered ferromagnets \cite{2005_Niklowitz_PhysRevBb,1995_Grosche_PhysicaB}. Related high-pressure studies in the itinerant helimagnet MnSi, connect the anomalous temperature dependence of the resistivity empirically with the presence of strongly fluctuating topological spin textures \cite{2013_Ritz_Naturea, 2013_Ritz_PhysRevBc}. Moreover, microscopic studies of the ferromagnetic QPT in LaCrGe$_3$ suggest the formation of short-range order as an escape route of ferromagnetic quantum criticality \cite{2021_Gati_PhysRevB, 2021_Rana_PhysRevB}. Perhaps most specific so far, recent studies of the transverse field Ising transition in LiHoF$_4$ under small tilted magnetic fields revealed the presence of a line of mesoscale quantum criticality, i.e., purely due to magnetic domains\cite{2022_Wendl_Nature}. 
The additional presence of defects and disorder at QPT, due to the strongly enhanced (or even singular) response functions may readily generate the formation of clusters in the sub-micrometer region or strongly fluctuating ferromagnetically correlated patches akin superparamagnetism. In turn, the interplay of disorder and defects with QPT has attracted considerable theoretical interest as the cause of novel forms of quantum correlations such as quantum Griffiths phases \cite{vojta2010}. 

While it may be intuitive that mesocale textures may be important in the surroundings of QPTs, it is experimentally difficult to prove their existence and to determine their character. Namely, in bulk materials processes on mesoscopic length scales imply correlation lengths that require ultra-small angle scattering. Also, the size of such textures implies large characteristic time-scales that are difficult to determine experimentally in bulk systems. Taken together, this raises the question for experimental methods capable to provide such information. In the study reported here we explore the potential of neutron depolarization measurements to provide such information for the ferromagnetic to intermediate valent transition in the compositional series {\cpr}.

Since the seminal work of Halpern and Holstein in the early 1940s  \cite{halpern1941}, it has been known that the polarization of a neutron beam decreases rapidly when traversing a bulk material with ferromagnetic domains. Numerous studies have demonstrated a great sensitivity of the polarization to the existence of ferromagnetic domains and superconducting flux lines \cite{rauch1968a, drabkin1969, maleev1970, weber1974, rekveldt1979, vandervalk1982, mitsuda1985, mirebeau1986, rosman1990, mirebeau1992, rekveldt1999, 2000_Yusuf_PRB, sato2004, 2006_DeTeresa_PRB, rekveldt2006, pfleiderer2010, treimer2012, treimer2012a, treimer2013, seifert2017, 2021_Kumar_PRM}. Likewise, it is also long known that slow ferromagnetic fluctuations in the paramagnetic state can cause a depolarization  of a neutron beam as well\cite{bakker1968, drabkin1968, takahashi1995}. Moreover, combining neutron depolarization measurements with neutron imaging, spatially resolved information may be obtained \cite{schulz2010, schulz_phd, schulz2016, jorba2019}. This technique known as neutron depolarization imaging (NDI) has been used, e.g., to map out magnetic stray fields \cite{kardjilov2008}. Developing a three-dimensional reconstruction of NDI, neutron depolarization tomography has also been used to determine ferromagnetically polarized regimes \cite{schulz2010}. It is thereby important to note that the spatial resolution of the imaging and tomography is currently limited to a few hundred mikro-meters at best due to beam divergence and detector resolution. Thus, neutron depolarization measurements offer a probe that allows to distinguish the existence of ferromagnetically correlated regimes on microscopic scales (neutron depolarization) from metallurgical inhomogeneities on macroscopic scales (NDI). 

The depolarization of a neutron beam depends on the spatial extent and the average magnetization of the magnetic domains generating the Larmor precession along the trajectory of the neutron beam. Given the wavelength (and hence velocity) of the neutrons this implicitly yields a characteristic time-scale, notably the time required for traversing individual domains, at which the polarization of the neutron beam is affected. In other words, a depolarization is expected when  magnetized patches are (i) sufficiently large, and (ii) sufficiently long-lived, and (iii) the uniform magnetization is sufficiently large. As compared to conventional bulk and transport properties, neutron depolarization measurements provide microscopic information as inferred from the threshold of the depolarization process. This has long been appreciated, though material-specific and experimental set-up-specific details have neither been reported nor exploited. 

To explore the potential of neutron depolarization measurements in studies of ferromagnetic QPTs we decided to study {\cpr} \cite{1988_Kappler_JPC, kappler1991, sereni1993, sereni2005, deppe2006, pikul2006, sereni2006, sereni2007, westerkamp2009, brando2010, schmakat2015a}. The rare earth compound CePd$_{1-x}$Rh$_x$ undergoes a ferromagnetic QPT as a function of Rh content $x$, where ferromagnetism is continuously suppressed above $x=\num{0.6}$ \cite{deppe2006, pikul2006, sereni2005, sereni2006, sereni2007}. In the vicinity of the QPT an exponentially decreasing tail of the onset of ferromagnetic correlations has been observed suggestive of disorder-induced smearing. At high values of $x$ and high temperatures the magnetic properties are characteristic of strong fluctuations that gradually freeze with decreasing temperatures\cite{westerkamp2009}. To determine the inherent length and time-scales of this freezing process and to assess the metallurgical homogeneity of CePd$_{1-x}$Rh$_x$ we performed neutron depolarization measurements in several compositions of intermediate Rh content. Furthermore, we investigated the effect of the magneto-crystalline anisotropy on the neutron depolarization of ferromagnetic CePd$_{1-x}$Rh$_x$ ($x=0.4$)\added{, where we find well-behaved properties of an easy ferromagnetic axis}.

\begin{figure} \centering
	\includegraphics[width=1.0\linewidth]{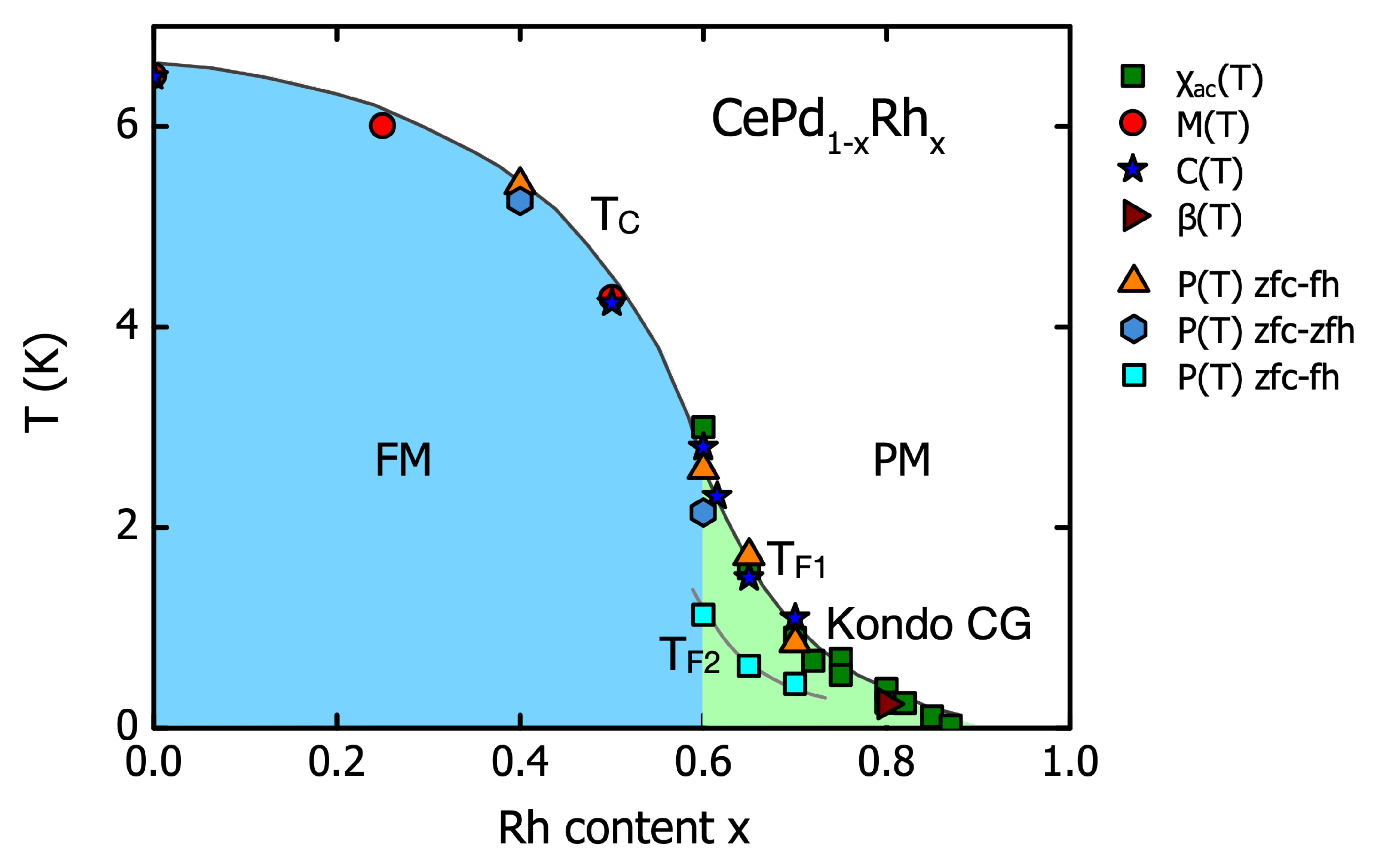}
	\caption{Magnetic phase diagram of CePd$_{1-x}$Rh$_x$. $T_{\rm C}$, $T_{\rm F1}$, and $T_{\rm F2}$ denote the Curie temperature, the spin freezing temperature, and the reentrance temperature, respectively. Shown are values of $T_{\rm C}$, $T_{\rm F1}$, and $T_{\rm F2}$ as inferred from the neutron polarization $P(T)$ recorded in our study, and the ac susceptibility, $\chi'_{\rm ac}(T)$, magnetization, $M(T)$, specific heat, $C(T)$, and thermal volume expansion, $\beta(T)$ of the same samples reported in Refs.\,\onlinecite{westerkamp2008, sereni2007}. With increasing $x$ ferromagnetism (blue shading) is suppressed and spin-frozen behaviour characteristic of Kondo Cluster Glass (green shading) emerges.}
	\label{fig:cepdrh-phasediagram}
\end{figure}

Summarized in Fig.\,\ref{fig:cepdrh-phasediagram} are our main results. For all compositions studied, the onset of neutron depolarzation we observe as a function of temperature is in good agreement with data inferred from the magnetization, ac susceptibility, and specific heat in the same samples as reported previously\cite{westerkamp2008}. For a Rh content up to $x\approx0.6$ all data consistently exhibit the characteristics of long-range ferromagnetic order. This is shown in blue shading in Fig.\,\ref{fig:cepdrh-phasediagram}, where the Curie temperature is denoted as $T_{\rm C}$. In comparison, for {\cpr} with $x>0.6$ all properties consistently exhibit the onset of a spin frozen state below a temperature denoted $T_{\rm F1}$, as shown in green shading in Fig.\,\ref{fig:cepdrh-phasediagram}. The absence of depolarization above $T_{\rm F1}$ is characteristic of a freezing of small, rapidly fluctuating clusters. The temperature and magnetic field history of the neutron depolarization reveals the absence of depolarization below a temperature $T_{\rm F2}$ under zero-field-cooling/field-heating, where $T_{\rm F2}$ roughly tracks $T_{\rm F1}$ for increasing $x$. The pronounced reentrance of the depolarization between $T_{\rm F2}$ and $T_{\rm F1}$ under zero-field-cooling/field-heating reveals a thermally activated increase of the cluster size in the spin-frozen state. Taking into account the rapidly increasing Kondo screening for increasing $x$, our depolarization data of the spin-frozen state reveal microscopic signatures consistent with a Kondo cluster glass (KCG) as proposed before based on the bulk properties\cite{westerkamp2008,westerkamp2009}.

The presentation of our paper is organized as follows. We begin with a pedagogical introduction to neutron depolarization measurements in Sec.\,\ref{sec:ndi}. Starting with the theory of neutron depolarization in Sec.\,\ref{subsec:ndi-theory}, we report estimates of the sensitivity of neutron depolarization on the length and time scales at which ferromagnetic regions begin to generate a noticeable depolarization in Sec.\,\ref{subsec:ndi-scales}. This is followed by key methodological aspects in Sec.\,\ref{subsec:ndi-setups}, where we report an experimental set up that permits to record the neutron depolarization under bipolar field sweeps needed for measurements of different temperature and field histories. In addition, we used spherical neutron polarimetry to discriminate a precessional rotation of the neutron polarization from a generic depolarization in ferromagnetic samples.  

The presentation of our experimental results in Sec.\,\ref{sec:results} proceeds in two subsections. We begin with the dependence of the neutron depolarization on the temperature and field history in Sec.\,\ref{subsec:results_depol}. This is followed by the evidence for a precessional rotation of the polarization in the ferromagnetic state of a single crystal sample in Sec.\,\ref{subsec:results_anisotropy_poli}, where we consider the role of the combination of magnetic anisotropy, temperature and small applied magnetic fields. In Sec.\,\ref{sec:discussion} we discuss our observations in the context of previous studies of the bulk properties and present our arguments in support of a Kondo cluster glass. Our paper closes with a summary of the main conclusions in Sec.\,\ref{sec:conclusion}.

%%%%%%%%%%%%%%%%%%%%%%%%%%%%%%%%
\section{Introduction to $\mathrm{CePd}_{1-x}\mathrm{Rh}_x$ }
\label{sec:cepdrh}

CePd and CeRh represent isotructural siblings that crystallize in the orthorhombic CrB structure. Whereas pure CePd orders ferromagnetically at $T_{\rm C}=\SI{6.5}{\kelvin}$\cite{thornton1998}, CeRh shows the key characteristics of a non-magnetic intermediate valent system down to the lowest temperatures studied. It has long been recognized that the substitutional series {\cpr} permits to explore the evolution from ferromagnetic order to intermediate valent behavior under isostructural conditions \cite{1988_Kappler_JPC, kappler1991, sereni1993} in the additional presence of strong disorder \cite{sereni2005, deppe2006, pikul2006, sereni2007}. Information on the spontaneous magnetic moment as extrapolated to zero temperature is, however, incomplete. In polycrystalline samples ordered moments of 0.87, 0.8 and 0.47\,$\mu_{\rm B}\,{\rm f.u.^{-1}}$ for $x=0$, 0.25 and 0.5, respectively, have been reported\cite{1988_Nieva_ZPhysB, sereni1993}. This compares with easy axis moments along the c-axis of 1.54 and 0.5\,$\mu_{\rm B}\,{\rm f.u.^{-1}}$ in single crystals for $x=0.4$ and 0.6\cite{deppe2006, schmakat2015a}, respectively, where the bulk properties are characteristic of an unchanged easy-axis anisotropy up to large $x$ \cite{deppe2006, brando2010}.

Seminal studies have established that the evolution of the properties of {\cpr} as a function of $x$ share two unusual characteristics with respect to a composition $x^*\approx0.6$ \cite{1988_Kappler_JPC, kappler1991, sereni1993, sereni2005, deppe2006, pikul2006, sereni2006, sereni2007, westerkamp2009, brando2010, schmakat2015a}. On the one hand, with increasing $x$ the lattice constant decreases linearly up to $x^*$, consistent with Vegard's law, followed by a fast superlinear decrease for $x>0.6$. On the other hand, the Weiss temperature, $\Theta_{\rm P}$, as inferred from the inverse susceptibility above $\sim 100\,{\rm K}$ is negative for all $x$. It decreases weakly up to $x^*$ followed by a steep decrease between $x^*$ and $x=1$. 

In the light of the intermediate valent properties of CeRh, it has been argued that the Weiss temperature reflects a mean-value of the Kondo temperature $T_{\rm K}\approx\vert\Theta_{\rm P}\vert$ \cite{1985_Koelling_PRB}. Thus, above $x^*$ the Kondo temperature appears to increase rapidly such that it exceeds the temperature scales on which ferromagnetism is observed. This suggests that the magnetic properties for $x>x^*$ originate from a combination of intermediate valent fluctuations, Kondo screening and disorder.

As a function of increasing $x$ the suppression of ferromagnetism in {\cpr}, notably the transition temperature $T_{\rm C}$ as inferred from the bulk properties, crosses the increase of the average value of $T_{\rm K}$ in the regime of $x^*$ \cite{sereni2007, westerkamp2009}.  A more detailed inspection reveals an unusual concentration dependence of the signatures of ferromagnetism for intermediate values of $x$ \cite{pikul2006, westerkamp2009}. Namely, with increasing $x$ the curvature of $T_{\rm C}$ is initially negative such that $T_{\rm C}$ may be extrapolated to zero approximately around $x^*$. However, as the composition approaches $x^*$ a change of curvature is observed at the onset of ferromagnetic correlations as inferred from pronounced maxima in $\chi'_{\rm ac}(T)$. These maxima  in $\chi'_{\rm ac}(T)$ may be discerned up to $x = 0.87$. The absence of a maximum in $\chi'_{\rm ac}(T)$ down to 20\,mK for $x = 0.9$ suggests the suppression of ferromagnetic correlations for a critical composition close to $x_c = 0.87$. 

Associated with the change of curvature at the onset of ferromagnetic correlations is a change of the character of the processes underlying the maximum in the susceptibility around $x^*$. For $x<x^*$ the bulk properties are consistent with a transition to long-range ferromagnetism. In contrast, for $x \geq 0.6$ the maxima in the susceptibility exhibit a pronounced frequency dependence characteristic of a freezing process similar to that observed in spin glasses \cite{westerkamp2009}. We therefore denote the onset of ferromagnetism and spin freezing with $T_{\rm C}$ and $T_{\rm F1}$, respectively. Interestingly, in the regime of the freezing process at $T_{\rm F1}$ the relative temperature shift between 3\,\% an 10\,\% per decade of the excitation frequency overlaps with the behaviour of canonical metallic spin glasses and superparamagnets, where shifts between $\sim 1\,\%$ and $\sim 30\,\%$ are observed. In addition to the freezing seen in the susceptibility,  the magnetization exhibits a small hysteresis already above $T_{\rm F1}$, where the hysteresis is seen between zero-field-cooled/field-heated and field-cooled/field-heated temperature sweeps. In turn, the hysteresis has been attributed to the formation of ferromagnetic clusters, which freeze at $T_{\rm F1}$.

%%%%%%%

The suppression of ferromagnetic order as well as the suppression of the underlying spontaneous magnetic moment as a function of increasing $x$ in {\cpr} reflect the hybridization of the Ce 4f electrons with the valence electrons of the surrounding ligands. Consistent with the evolution of $T_{\rm K}$ inferred from $\Theta_{\rm P}$, Rh ligands in Ce compounds are known to result in much larger Kondo temperatures than observed for Pd ligands.\cite{1985_Koelling_PRB} This is accompanied by strong local variations of $T_{\rm K}$ due to the random distribution of Rh and Pd atoms. Both aspects are consistent with the entropy and the slope of $\chi'_{\rm ac}(T)$ at 2\,K, which are suggestive of unscreened magnetic moments even when the average of $T_{\rm K}$ exceeds several ten K. 

Taken together, {\cpr} differs distinctly from other strongly disordered Ce-based systems such as CeNi$_{1-x}$Cu$_x$, in which the Ce valence remains nearly trivalent and in which a percolative cluster scenario was proposed\cite{2007_Marcano_PRL}. In recognition of the broad distribution of Kondo temperatures on local scales, which are believed to be responsible for the cluster formation, the low-temperature state in {\cpr} has been denoted a Kondo-cluster glass (KCG). However, this interpretation was so far empirical, without evidence of specific microscopic signatures expected of a Kondo screening.

%%%%%%%

Last but not least, the possible role of quantum correlations in {\cpr} has been addressed in several studies. Namely, putative evidence of a power-law dependence in the specific heat, $C(T)/T \sim T^{\lambda-1}$, with $\lambda= 0.6$ and 0.67 for $x = 0.87$ and 0.9, respectively \cite{sereni2005, pikul2006, sereni2007}, as well as the ac susceptibility, raise the question whether $T_{\rm C}$ near $x = 0.87$ is somehow connected with an additional form of ferromagnetic quantum criticality. This is contrasted by the Gr{\"u}neisen ratio $\Gamma$, which displays a log-divergence, $\Gamma\propto \ln T$, as opposed to a power-law divergence expected for quantum criticality ($\Gamma \propto \beta/C$ where $\beta$ represents the volume thermal expansion coefficient and $C$ the specific heat). Moreover, a power-law form of the magnetic-field dependence of the magnetization at 50\,mK was reported for very large $x$, also consistent with the absence of quantum criticality. Rather, it has been pointed out that these properties are consistent with the scenario of a quantum-Griffiths-phase. However, whereas theory predicts that the values for $\lambda$ inferred from the specific heat and susceptibility should be the same \cite{vojta2009}, they are found to be different experimentally, posing an unresolved inconsistency. 

The properties reported so far for {\cpr} consistently point at an unusual interplay of strong correlations with disorder on multiple scales \cite{1988_Kappler_JPC, kappler1991, sereni1993, sereni2005, deppe2006, pikul2006, sereni2006, sereni2007, westerkamp2009, brando2010, schmakat2015a}. Namely, the bulk properties and limited microscopic information suggest the formation of ferromagnetic clusters that undergo a complex freezing process subject to a distribution of Kondo screening and thermal activation. The properties of {\cpr} reported so far raise questions for the actual size of the underlying microscopic length and time scales as well as the magnetic character of the clusters. In turn this raises the question of the role of the disorder and the distribution of Kondo scales and, last but not least, the relevance of the Kondo screening in the freezing process of the clusters. 

%%%%%%%%%%%%%%%%%%%%%%%%%%%%%%%%
%%%%%%%%%%%%%%%%%%%%%%%%%%%%%%%%
\section{Neutron depolarization }
\label{sec:ndi}

Neutron depolarization measurements are based on the principle of neutron radiography by means of a polarized neutron beam. Since the neutron interacts with magnetic fields via its spin, neutron depolarization measurements allow to identify and determine, within limits, spatially resolved ferromagnetic correlations in bulk materials, i.e., ordered regions such as domains or spin clusters \cite{schlenker1973,schulz2016}. Due to the large penetration depth of neutrons, neutron depolarization measurements allow to use complex sample environment such as cryostats, electro magnets, and pressure cells. 

Examples of neutron depolarization measurements include the three-dimensional imaging of ferromagnetic domains in bulk samples \cite{schulz2010}, as well as the detection of inhomogeneous field distributions \cite{kardjilov2008,piegsa2009,sales2018} as generated, e.g., by screening currents in the vicinity of superconductors or due to the flux lines penetrating superconductors \cite{weber1974,treimer2012, treimer2012a,treimer2013,treimer2014}. By means of neutron depolarization measurements the metallurgical homogeneity of ferromagnetic materials can be characterized when the magnetic properties vary sensitively with chemical composition and internal stress/strain \cite{jorba2019}. Moreover, measurements as a function of temperature allow to map out the distribution of the Curie temperature spatially and hence to infer compositional inhomogeneities across larger sample volumes\cite{pfleiderer2010, schmakat2015a, jorba2019}.

In the following the basic principles of neutron depolarization are reviewed in Sec.\,\ref{subsec:ndi-theory}, followed by material-specific estimates of the spatial and temporal threshold for depolarization to occur in Sec.\,\ref{subsec:ndi-scales}. In Sec.\,\ref{subsec:ndi-setups} the experimental setups used in our study are described.

%%%%%%%%%%%%%%%%%%%%%%%%%%%%%%%%
\subsection{Neutron depolarization in a ferromagnet}
\label{subsec:ndi-theory}

We begin with a summary of the formal description of neutron depolarization measurements of a ferromagnet focussing on a few limiting cases which are required for the interpretation of our results. For the discussion presented in the following we \replaced{consider }{assume} a \added{polarized} neutron beam\added{, the polarization $P$ of which is given as}
\begin{equation}
\added{P=\frac{I_{+}-I_{-}}{I_{+}+I_{-}}}
\end{equation}
\replaced{where $I_{+}$, $I_{-}$ represent the intensities with respect to the polarization axis. In the following we denote the polarization of the incident neutron beam by $P_0$.
The neutron beam }{that} is transmitted through a ferromagnetic sample, \replaced{in which it traverses }{traversing} a series of magnetic domains labelled $i$ with intrinsic fields $\textbf{B}_i$. We assume further that these intrinsic fields are oriented randomly. The classical equation of motion for the neutron spin $\textbf{s}_i$ that couples to the ferromagnetic domain $i$ is given by the Larmor equation
\begin{equation}
\frac{d}{dt}\textbf{s}_i(t) = \gamma \textbf{s}_i(t) \times \textbf{B}_i
\label{eq:larmor}
\end{equation}
where $\gamma$ is the gyromagnetic ratio of the neutron. For constant magnetic field Eq. \ref{eq:larmor} describes a precession of the spin with respect to the direction of the field at the Larmor frequency $\boldsymbol{\omega}_L = -\gamma\textbf{B}_i$. 

Assuming an average field $B_0 = \langle B_i \rangle$ per domain, an average domain length $\delta$ in the direction of flight and infinitesimally thin domain walls the polarization in the $y$-direction may be written as \cite{halpern1941,mitsuda1992}
\begin{equation}
\frac{P}{P_0} = \left[ \left\langle \frac{B_\parallel^2}{B_0^2} \right\rangle_{\!B_0} + \left\langle \frac{B_\perp^2}{B_0^2} \right\rangle_{\!B_0} \left\langle \cos \left( \gamma B_0 \frac{\delta}{v} \right) \right\rangle_{\!\delta} \right]^N
\label{eq:depol}
\end{equation}
where $y$ is perpendicular to the beam direction and parallel to the guide field (representing the quantization axis). Further, \deleted{$P_0$ is the polarization of the incident neutron beam and} $B_\parallel$ and $B_\perp$ represent the components of the magnetic field $B_0$ in each domain parallel and perpendicular to the direction of travel. $N = d/\delta$ represents the average number of magnetic domains on a neutron path through the sample and $v$ is the neutron velocity. Here the brackets $\langle ... \rangle$ denote the average over the magnetic field or the average domain size as denoted by the subscript. The argument of the cosine function corresponds to the Larmor phase collected in each domain $\varphi_L = \omega_L \tau = \gamma B_0 \delta / v$.

Eq. \ref{eq:depol} may be expressed analytically for some limiting cases. First, as proposed by Halpern and Holstein\cite{halpern1941}, averaging over the ensemble of randomly oriented domains for small spin rotations per domain $\omega_L \tau \ll 2\pi$ results in
\begin{equation}
\frac{P}{P_0} = \exp \left( -\frac{1}{3}\gamma^2 B_0^2(T) \frac{d \delta}{v^2} \right)
\label{eq:p_exp}
\end{equation}
where $d$ is the sample thickness in the direction of flight of the neutron and the magnetic flux $B_0$ is assumed to be temperature-dependent. The second case, where $\omega_L \tau \geq 2\pi$, represents a large spin rotation per domain. Eq. \ref{eq:depol} then yields for the polarization
\begin{equation}
\frac{P}{P_0} \approx \exp \left( - N \right).
\label{eq:p_const}
\end{equation}
Assuming that the average domain size is constant, the depolarization is then expected to be constant below the Curie temperature $T_{\rm C}$. 

Based on the solutions described by Eqs.\,\ref{eq:p_exp} and \ref{eq:p_const} it may be concluded that for temperatures below $T_{\rm C}$ the domain configuration leads to a depolarization of the neutron beam, while in the paramagnetic state the polarization is not affected. For the evaluation of our data we assumed that the magnetic field in a single domain may be described using a temperature dependence as follows
\begin{equation}
B_0^2 = \mu_0^2 M^2_0 \left[ \frac{T-T_{\rm C}}{T_{\rm C}} \right]^\beta \left[1- \Theta(T - T_{\rm C}) \right],
\label{eq:bgl}
\end{equation}
where $M_0$ represents the spontaneous magnetization in each domain, $\beta$ is system specific exponent, $\mu_0$ is the vacuum permeability, and $\Theta(x)$ is a Heaviside function that serves to account for an idealized spontaneous symmetry breaking below $T_{\rm C}$. For $\beta=1/2$ Eq.\,\ref{eq:bgl} corresponds to the mean field approximation for ferromagnets. In real systems a smoothed version of the step function such as a Gaussian error function may be used to represent the transition and the emergence of a finite magnetization.

In case the data are recorded with a two-dimensional detector, spatially resolved information may be obtained using Eqs.\,\ref{eq:p_exp} and \ref{eq:p_const}. This provides spatially resolved information on the magnetic ordering temperature $T_{\rm C}$, referred to in the following as a $T_{\rm C}$ map.

Another limit assumes a mono-domain ferromagnetic state, i.e., the sample supports a uniform magnetization without domains\cite{mitsuda1992}. It is important to note that such a state does not cause a depolarization of a monochromatic beam (a polychromatic beam featuring a distribution of wavelengths may depolarize somewhat). Denoting the angle between the polarization and the magnetization by $\alpha$, Eq. \ref{eq:depol} yields
\begin{equation}
P = P_0 \left[ \cos^2(\alpha) + \sin^2(\alpha) \cos \left( \gamma B_0 \frac{d}{v} \right) \right].
\label{eq:p_singledomain}
\end{equation}
If the magnetization is neither parallel nor antiparallel to the polarization ($\alpha \neq n \pi$ for $n=0, 1, 2,\dots$) the polarization vector precesses with respect to the direction of the internal field $B_0$. A seeming decrease of the polarization would then be due to a rotation of the polarization away from the direction for which the polarization is analyzed. Moreover, a change of the magnetization caused, for example, by a variation of the temperature, $T$, or an applied magnetic field, $B = \mu_0 H$, may  result in an oscillation of the polarization due to the cosine term. This has been observed in our study of the ferromagnetic single-crystal {\cpr} ($x=0.4$) as described in Sec.\,\ref{subsec:results_anisotropy_poli}.

%%%%%%%%%%%%%%%%%%%%%%%%%%%%%%%%
\subsection{Sensitivity of neutron depolarization}
\label{subsec:ndi-scales}

In this section we present an estimate of the length- and time-scales of multi-domain ferromagnetic order at which the setup we used for our neutron depolarization study described in Sec.\,\ref{subsec:ndi-setups} was able to detect a depolarization. Several publications have addressed this question\cite{rekveldt1979, vandervalk1982, mirebeau1986, rosman1990, mirebeau1992, mitsuda1992,takahashi1995, rekveldt1999, sato2004, rekveldt2006}. However, to the best of our knowledge quantitative values have not been reported before.

We consider multi-domain ferromagnetic order with an average domain size $\delta$. For decreasing $\delta$ the Larmor phase $\varphi_L$ collected in each domain eventually will no longer be sufficient to depolarize the neutron beam at a level that exceeds the resolution of the setup. This case is described by Eq. \ref{eq:p_exp}. Solving this equation for $\delta$ and replacing $P/P_0=1-\Delta P$ yields
\begin{equation}
\delta = \frac{-3 \log ( 1-\Delta P ) v^2}{\gamma^2 B_0^2 d} .
\label{eq:length-scale}
\end{equation}
where $\Delta P$ is the minimal change of polarization that may be resolved. In our setup the resolution corresponded typically to $\sim\SI{1}{\percent}$ as estimated from the scatter of the data points. 

\begin{figure} 
\centering
	\includegraphics[width=0.9\linewidth]{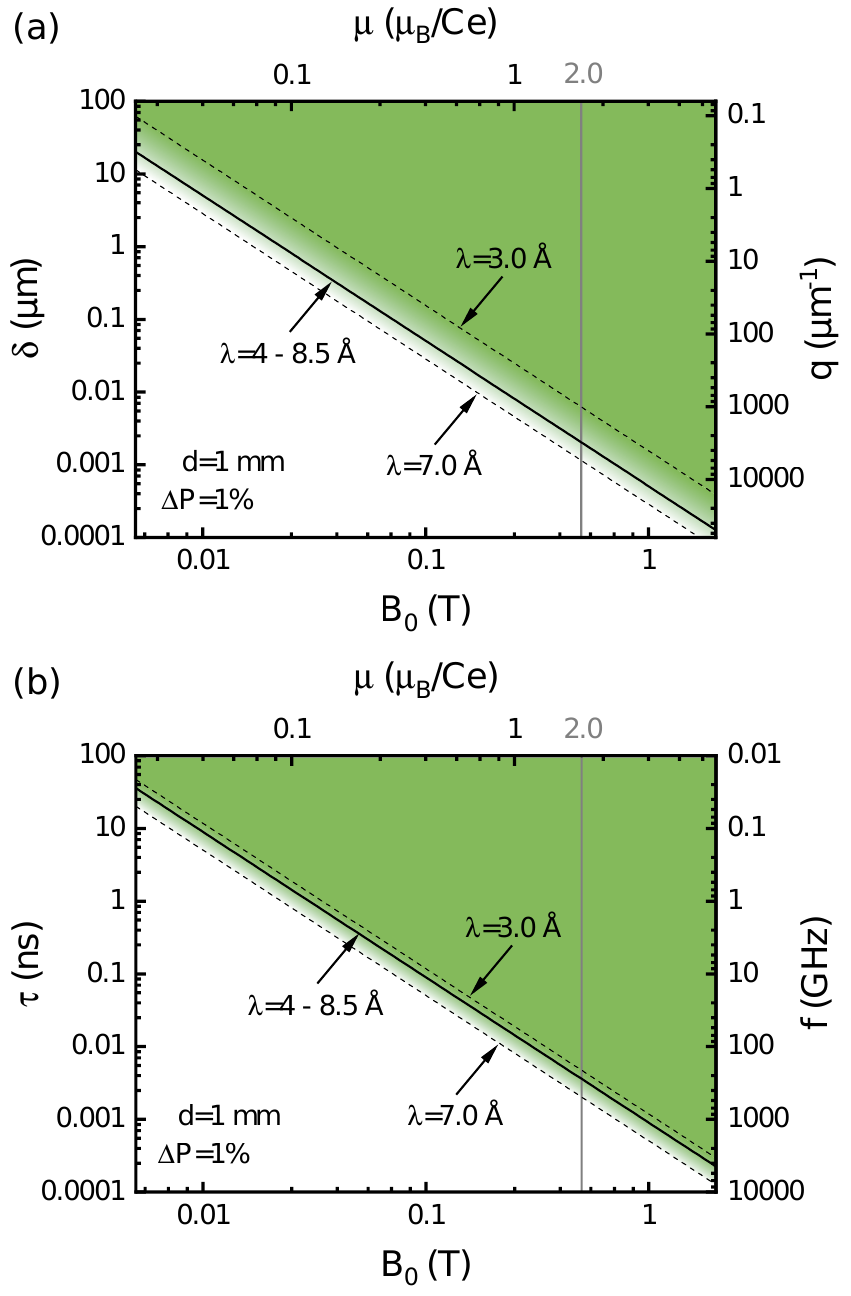} 
\caption{Characteristic length- and time-scales beyond which neutron depolarization is expected in a multi-domain ferromagnet as a function of magnetic flux $B_0$ per domain. Shown in green shading is the parameter range of a noticeable depolarization exceeding $\Delta P=\SI{1}{\percent}$ for a sample thickness of $d=\SI{1}{\milli\meter}$, where the threshold corresponds to monochromatic wavelengths of $\SI{3.0}{\angstrom}$, and $\SI{7.0}{\angstrom}$, as well as a polychromatic spectrum from $\SI{4.0}{\angstrom}$ to $\SI{8.5}{\angstrom}$. The abscissa shown at the top of each panel represent the magnetic flux per domain as expressed in units of the magnetic moment $\mu$ in CePd$_{1-x}$Rh$_x$. Marked in gray is the unscreened moment of {\cpr}. (a) Estimated length scale $\delta$ as a function $B_0$. The ordinate on the right hand side displays a corresponding wave vector $q = 2\pi / \delta$.  (b) Estimated flight time $\tau$ across a domain as a function $B_0$. The ordinate shown on the right hand side displays a corresponding frequency, $f=1/\tau$.}
\label{fig:length-time-scales}
\end{figure}

Shown in Fig.\,\ref{fig:length-time-scales} is an evaluation of Eq. \ref{eq:length-scale} for a typical sample thickness of $d=\SI{1}{\milli\meter}$ and different neutron wavelengths. The minimum average domain length $\delta$ required for a depolarization of \SI{1}{\percent} is shown as a function of the average field $B_0$ in each domain. The abscissa at the top of both panels of Fig.\,\ref{fig:length-time-scales} displays $B_0$ in units of the magnetic moment $\mu$ for CePd$_{1-x}$Rh$_x$ using a unit cell volume of approximately ${187}\,{\rm \AA^{3}}$.\cite{sereni2007} For order of magnitude estimates, $B_0$ may be inferred from the magnetization at sufficiently large magnetic fields. 

Depicted in green shading is the regime, where a notable depolarization is expected. The dashed lines depict the 1\,\% threshold for wavelengths of $\lambda=\SI{3}{\angstrom}$ and $\lambda=\SI{7}{\angstrom}$, corresponding to the wavelengths at the beam-line ANTARES available for monochromatic depolarization measurements. In comparison, the solid black line represents a calculation for a polychromatic spectrum ranging from $\lambda=\SI{4}{\angstrom}$ to \SI{8.5}{\angstrom} as available at the beam-line ANTARES and used in the setup shown in Fig.\,\ref{fig:exp-setup}\,(a).  

With increasing $B_0$ the average size of the domains, $\delta$, that may be detected decreases as shown in Fig.\,\ref{fig:length-time-scales}\,(a). The ordinate on the right hand side of Fig.\,\ref{fig:length-time-scales}\,(a) represents the wave vector $q=2\pi / \delta$  associated with $\delta$. Remarkably, for strong ferromagnets the neutron depolarization is sensitive to ferromagnetic domains down to the sub-nm scale. For instance, the unscreened magnetic moment of CePd$_{1-x}$Rh$_x$ of $2\mu_B$ \cite{pikul2006} as marked in Fig.\ref{fig:length-time-scales}\,(a) by a vertical gray line, implies a spatial sensitivity of \SI{2}{\nano\meter} for the experimental setup shown in Fig.\,\ref{fig:exp-setup}\,(a). In turn, if the average size of ferromagnetic spin clusters exceeds $\sim\SI{2}{\nano\meter}$ a depolarization larger than \SI{1}{\percent} is expected.

It is important to emphasize that the magnetic domains do not need to be static for a depolarization to occur. Rather, the lifetime of the domains must exceed the time, $\tau$, needed by the neutron to travel across the domains given by $\tau=\delta / v$, where $v$ is the neutron velocity. As shown in Fig.\,\ref{fig:length-time-scales}\,(b) typical time scales are of the order of ns, which decreases rapidly with increasing $B_0$. The ordinate on the right hand side of Fig.\,\ref{fig:length-time-scales}\,(b) represents an associated frequency $f=1/\tau$ in the spirit of a fluctuation rate, say, in a superparamagnet.  For large $B_0$ the neutron depolarization measurements are sensitive to ferromagnetically correlated regimes that fluctuate at a time-scale less than pico-seconds. Interestingly, in real materials this corresponds to the time scale of the movement of domain walls at distances that are typical of domain sizes. In turn, this implies that the effective field seen by the neutron does not need to be static on time scales exceeding $\tau$. 

Taken together, these considerations highlight the potential of neutron depolarization measurements as a probe of ferromagnetic textures that fluctuate on a nm scale in the sub-ns regime. The temporal resolution associated with the unscreened moment of CePd$_{1-x}$Rh$_x$ of $2\mu_B$ \cite{pikul2006} for the experimental setup shown in Fig.\,\ref{fig:exp-setup}\,(a) as marked by a gray line in Fig.\ref{fig:length-time-scales}\,(b)  corresponds to $\sim\SI{4}{\pico\second}$.

%%%%%%%%%%%%%%%%%%%%%%%%%%%%%%%%
\subsection{Experimental setups}
\label{subsec:ndi-setups}

For our studies it proved to be essential to record data under different temperature versus field histories. Before turning to an account of the terminology used in our paper, we note that the expression "zero-field" implies the nominal presence of a guide field of roughly 0.5\,mT required to maintain the polarization of the neutron beam. Keeping this in mind, we distinguish the following temperature versus field histories, noting that all data were collected while heating the sample at the same constant rate in order to avoid systematic errors of the sample temperature recorded. First, data recorded after cooling in zero-field (zfc) while heating in zero-field are denoted zfc-zfh (zero-field-cooled - zero-field-heated). Second, data recorded after cooling in zero-field to base temperature and the application of a finite field while heating under this applied magnetic field are denoted zfc-fh (zero-field-cooled - field-heated). Third, after cooling the sample in the presence of a magnetic field applied at sufficiently high temperatures of several K, data were recorded while heating the sample in the same unchanged applied magnetic field. This is denoted fc-fh (field-cooled - field-heated).

\begin{figure}
\includegraphics[width=\linewidth]{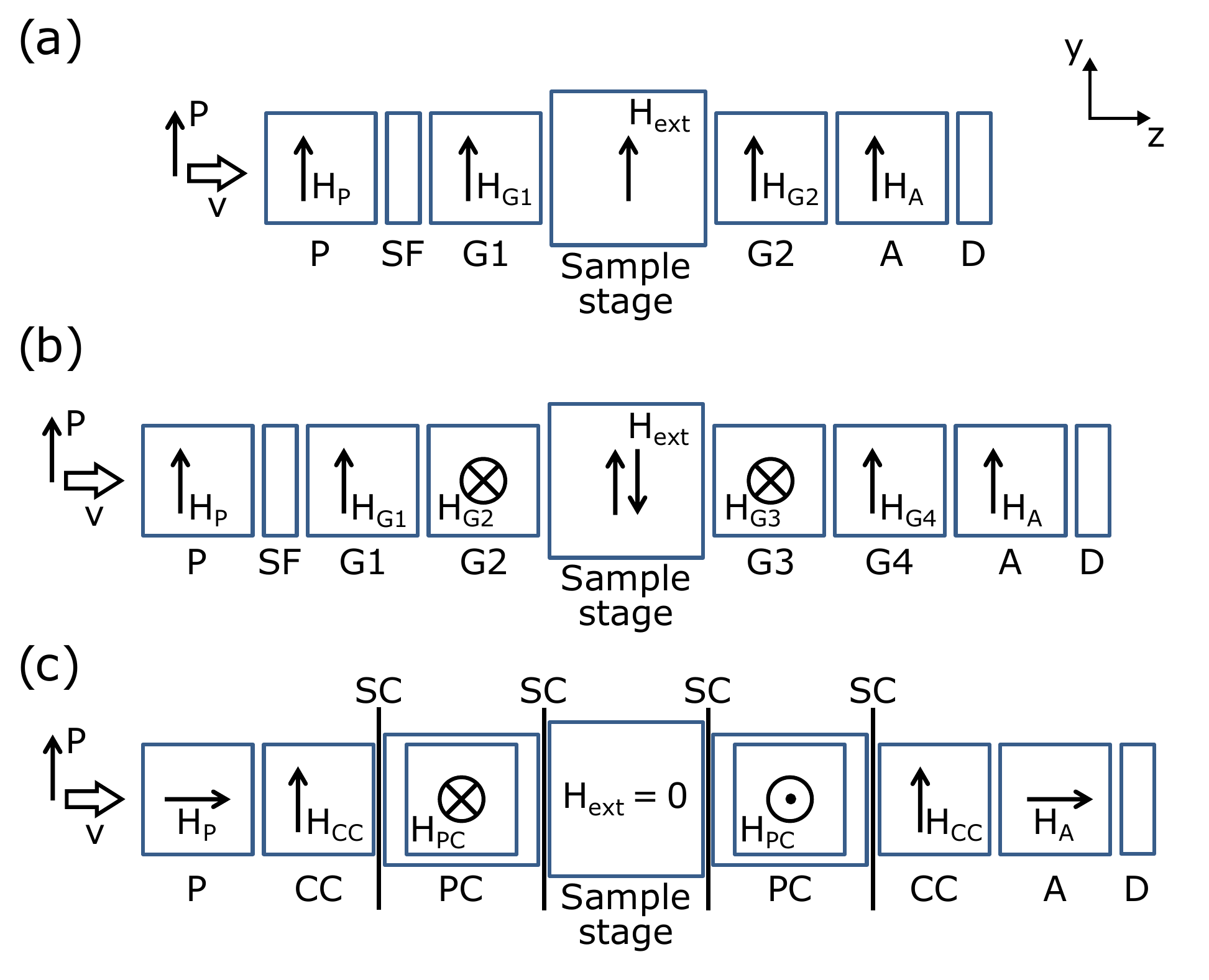}
\caption{Schematic depictions of the three setups used for the neutron depolarization measurements reported in this paper.  (a) The neutron beam passed a polarizer (P) and a spin flipper (SF). The polarization was maintained by guide fields (G1 and G2) in the $y$-direction and hence perpendicular to the neutron path, which was parallel to the $z$-direction across the sample. The sample was placed in a cryostat which was located in a Helmholtz pair of coils generating a magnetic field at the position of the sample. The polarization was analyzed in the $y$-direction using an analyzer (A) and detected by a CCD camera  (D) in combination with a LiF/ZnS converter and scintillator film. (b) Magnetic fields in positive and negative $y$-direction could be applied by adding two horizontal guide fields (G2 and G3) pointing in $x$-direction. (c) Schematic depiction of the setup used for spherical polarization analysis with CryoPAD. Rotatable coupling coils (CC) and two precession coils (PC) were located between two superconducting sheets (SC) that permitted to adjust and to analyze the polarization in arbitrary directions.}
\label{fig:exp-setup}
\end{figure}

The neutron depolarization measurements were carried out at the  beamline ANTARES at FRM II\cite{calzada2009, schulz2015}. A schematic depiction of the setup used initially at ANTARES is shown in Fig. \ref{fig:exp-setup}\,(a). Polarized $^3$He spin filter cells were used for polarizing and analyzing the neutron beam in the $y$-direction which is perpendicular to the neutron beam denoted as $z$-direction. A spin flipper located directly after the polarizer allowed to change the polarization direction from $+y$ to $-y$. This was required for the polarization analysis. Tiny guide fields between the components prevented the loss of polarization, otherwise expected in low-field field regions, where parasitic external magnetic fields dominate. Coarse neutron wavelength selection was achieved by means of a Beryllium filter, resulting in a rather broad wavelength band from $\sim\SI{4}{\angstrom}$ to $\sim\SI{8.5}{\angstrom}$. The neutron detector used was based on a LiF/ZnS scintillator, which converted the neutrons into visible light that was detected by a high resolution CCD camera. A large part of the data reported in Sec.\,\ref{sec:results} were measured using this setup.

Following our first measurements we modified this setup to permit studies of different temperature versus field histories. This required a setup that permitted measurements under arbitrary positive or negative magnetic field strengths. As explained above, true zero-field conditions are very difficult to achive and even small stray fields will cause a severe depolarization of the neutron beam. Therefore, we used small guide fields around the sample position in order to stabilize the polarisation axis.

To satisfy these conditions we installed two additional guide fields pointing in the $x$-direction. As indicated in Fig. \ref{fig:exp-setup}\,(b) these guide fields were placed immediately before and after the Helmholtz coils. This way an adiabatic rotation of the polarization into the horizontal plane was realized that allowed to apply magnetic fields along the positive and the negative $y$-direction by means of the Helmholtz pair without an undefined zero-field transition along the neutron path that would cause a severe depolarization of the neutron beam. Moreover, a neutron velocity selector was installed and the $^3$He polarizers were replaced by polarizing V-cavities. In these experiments we used a neutron wavelength of $\lambda=\SI{4.13}{\angstrom}$ with a wavelength spread of $\Delta\lambda / \lambda=10\%$ given by the velocity selector. Further details of this setup may be found elsewhere\cite{seifert2017, jorba2019, schmakat2015a}.

For all of our neutron depolarization measurements the samples were cooled to temperatures as low as $\sim 0.07\,{\rm K}$ by means of bespoke $^3$He/$^4$He dilution insert as combined with a pulse tube cooler. Magnetic fields were generated with a pair of Helmholtz coils operated at room temperature. 

\begin{figure} \centering
	\includegraphics[width=0.8\linewidth]{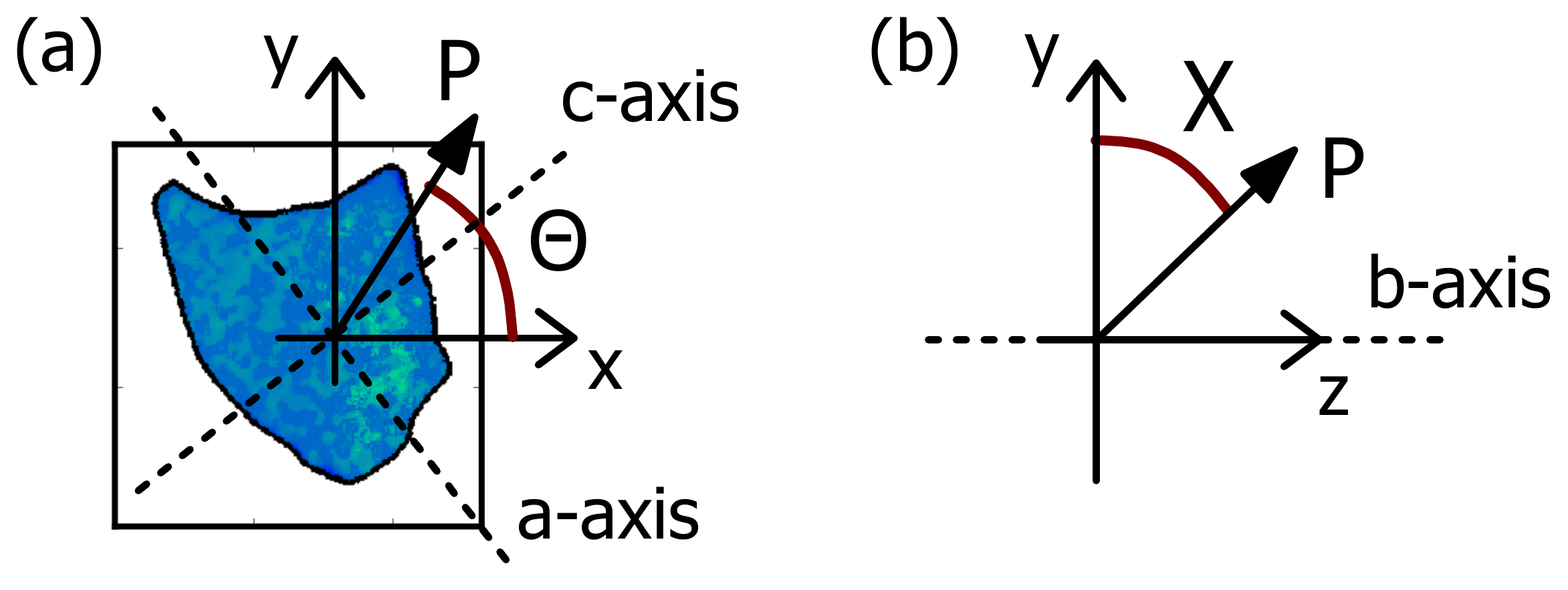} 	   
	\caption{Coordinate systems used in the neutron polarization measurements with CryoPAD, where the crystallographic $a$- and $c$-axis of the CePd$_{1-x}$Rh$_x$ ($x=\num{0.40}$) single crystal were determined by Laue neutron diffraction. (a) Depiction of the sample shape and sample orientation in the $xy$-plane. (b) Definition of the angle $\chi$ in the $yz$-plane. The crystallographic $a$- and $c$-axis of the sample resided in the $xy$-plane perpendicular to the neutron beam. The $b$-axis was parallel to the $z$-direction. CryoPAD allowed to adjust and analyze the polarization in any arbitrary direction defined by the angles $\Theta$ in the $xy$-plane and $\chi$ in the $yz$-plane.}
	\label{fig:cepdrh-anisotropy}
\end{figure}

In addition, spherical neutron polarimetry was carried out using CryoPAD at the beam-line POLI at FRM II \cite{hutanu2016}, where a schematic depiction of the setup is shown in Fig.\,\ref{fig:exp-setup}\,(c). The implementation of coupling coils, precession coils, and magnetic shielding of the sample position permitted a complete determination of the three-dimensional polarization matrix. However, it was not possible to obtain spatially resolved information across the sample because a $^3$He tube had to be used as a neutron detector. A detailed description of the setup may be found elsewhere\cite{hutanu2016}.

Shown schematically in Fig. \ref{fig:cepdrh-anisotropy} is the coordinate system as well as the outline and orientation of the CePd$_{1-x}$Rh$_x$ single crystal as investigated at POLI. In the following the direction of the neutron polarization is denoted by the polar angles $\Theta$ and $\chi$, where the angle $\Theta$ was measured clock-wise in the $xy$-plane and $\Theta=0$ corresponded to the $y$-direction. The angle $\chi$ was measured in the $yz$-plane starting at $\chi=0$ in the $y$-direction. 

In our studies we adjusted and analyzed the polarization always in the same direction to be able to distinguish a generic depolarization from a spherical rotation of the direction of the polarization. Therefore, both nutator angles $\Theta$ and both angles $\chi$ as determined by the precession coils were always kept the same. Finally, the single crystal sample was oriented such that the crystallographic $ac$-plane corresponded to the plane perpendicular to the neutron beam and thus the $xy$-plane in the coordinate system used to account for the polarization. The magnetic easy axis of the system, which corresponded to the crystallographic $c$-axis, hence resided in this plane. %The y-axis of the set up is the vertical axis while the x-axis is horizontal in real space.

The resistivity, ac susceptibility, magnetization, and specific heat of the CePd$_{1-x}$Rh$_x$ samples we investigated in our neutron depolarization measurements were examined rather comprehensively prior to our study as reported elsewhere \cite{deppe2006, westerkamp2008, sereni2007}. Details of the sample preparation may be found in these papers. All samples were poly-crystals with the exception of the sample with $x=\num{0.40}$ which was a single crystal. 

%%%%%%%%%%%%%%%%%%%%%%%%%%%%%%%%
\section{Experimental results}
\label{sec:results}

The presentation of the experimental data is organized in two parts. It begins with the dependence of the neutron depolarization on the temperature and field history in Sec.\,\ref{subsec:results_depol} for a wide range of compositions. This is followed by the variation of the neutron depolarization due to the magnetic anisotropy in single-crystal {\cpr} for the ferromagnetic composition $x=0.4$ in Sec.\,\ref{subsec:results_anisotropy_poli}.

%%%%%%%%%%%%%%%%%%%%%%%%%%%%%%%%
\subsection{Dependence of the neutron depolarization on temperature and field history}
\label{subsec:results_depol}

\begin{figure}
\centering
\includegraphics[width=0.85\linewidth]{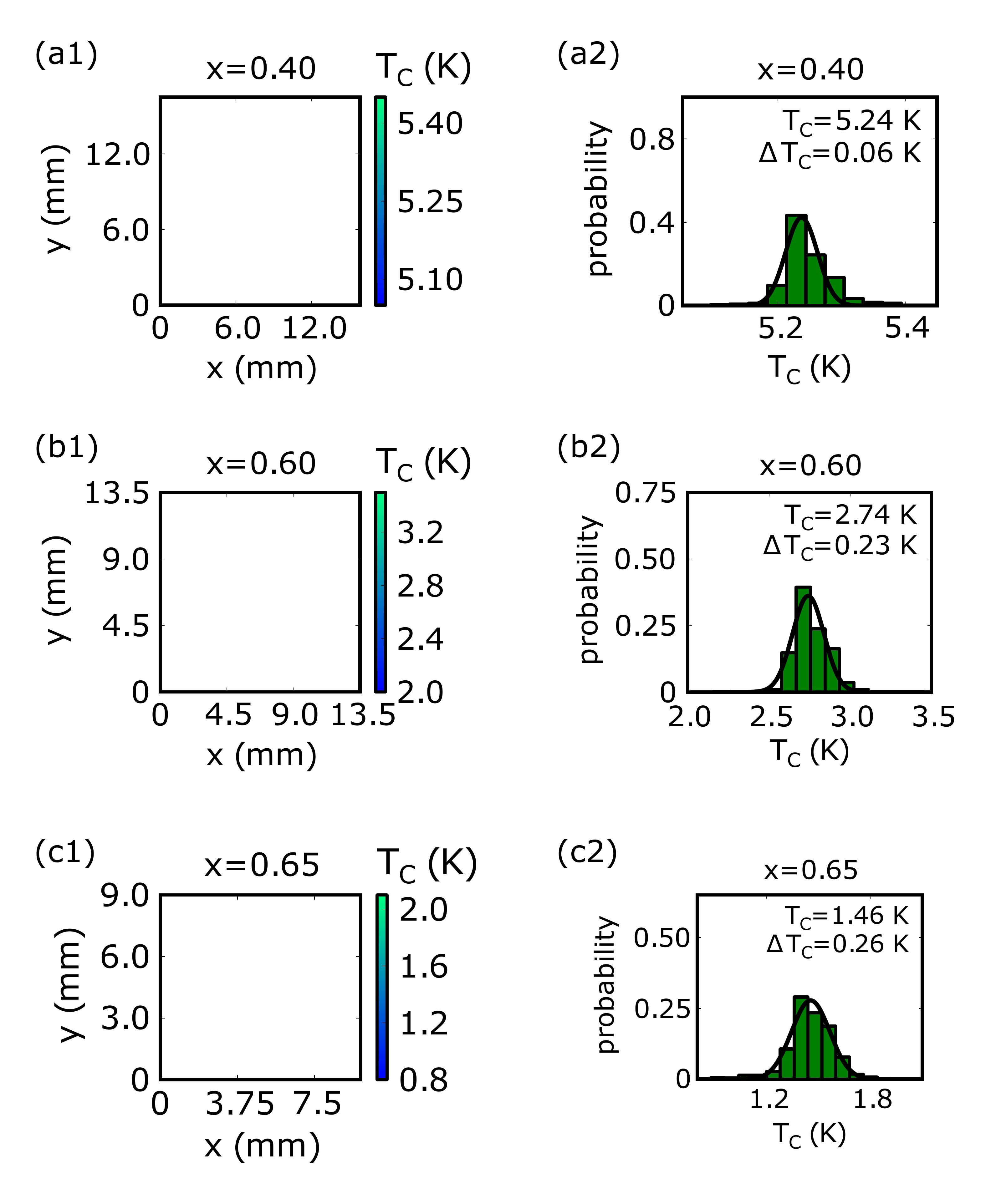}
\caption{Temperature dependence of NDI in {\cpr}. (a1), (b1) and (c1) $T_{\rm C}$ maps of {\cpr} across the sample shape for $x=\num{0.40}$, $x=\num{0.60}$, and $x=\num{0.65}$ as inferred from temperature scans at zero field (see text for details). The color bar denotes the range of the transition temperatures $T_{\rm C}$. (a2), (b2) and (c2) Distribution of transition temperatures inferred from the $T_{\rm C}$ maps shown in panels (a1), (b1), and (c1), where $\Delta T_{\rm C}$ corresponds to the FWHM of the Gaussian fit shown.}
	\label{fig:tc-maps}
\end{figure}

Data reported in the following were recorded at the beam-line ANTARES using the setups shown in Fig. \ref{fig:exp-setup}\,(a) and \ref{fig:exp-setup}\,(b). Shown in Figs.\,\ref{fig:tc-maps}\,(a1), (b1) and (c1) are $T_{\rm C}$ maps across the sample cross-section of the samples with $x=\num{0.40}$, $x=\num{0.60}$, and $x=\num{0.65}$ as inferred from the temperature dependence of neutron depolarization imaging. As reported above, the expression zero-field refers to a very small field $B=\SI{0.5}{\milli\tesla}$ required as a guide field to maintain the neutron polarization. Color bars indicate the transition temperatures while the thin black lines denote the outline of the sample shape. 

Shown in Figs.\,\ref{fig:tc-maps}\,(a2), (b2) and (c2) are the corresponding histograms of the distribution of ordering temperatures across the $T_{\rm C}$ map. The distribution of transition temperatures was fitted with a Gaussian where the values shown in the histograms represented the average value of $T_{\rm C}$, and the asscoiated FWHM, $\Delta T_{\rm C}$. Namely, we found $T_{\rm C}(x=\num{0.40})=5.24\pm0.06\,{\rm K}$, $T_{\rm C}(x=\num{0.60})=2.74\pm0.23\,{\rm K}$, and $T_{\rm C}(x=\num{0.65})=1.46\pm0.26\,{\rm K}$. 

With increasing $x$, values of $T_{\rm C}$ and $T_{\rm F1}$ decrease in excellent agreement with the properties inferred from the bulk properties reported in the literature as shown in Fig.\,\ref{fig:cepdrh-phasediagram}. As $T_{\rm C}$ and $T_{\rm F1}$ decrease with increasing $x$ the FWHM $\Delta T_{\rm C}$ increases. This trend may be explained with an increase of the effects of disorder. Moreover, the sample with $x=\num{0.40}$ was a single-crystal as compared to the poly-crystalline nature of the samples with $x=\num{0.60}$ and $x=\num{0.65}$. It might also reflect the vicinity to the QPT and the concomitant increase of the susceptibility to form ferromagnetic clusters. 

\begin{figure}
	\centering
	\includegraphics[width=1.0\linewidth]{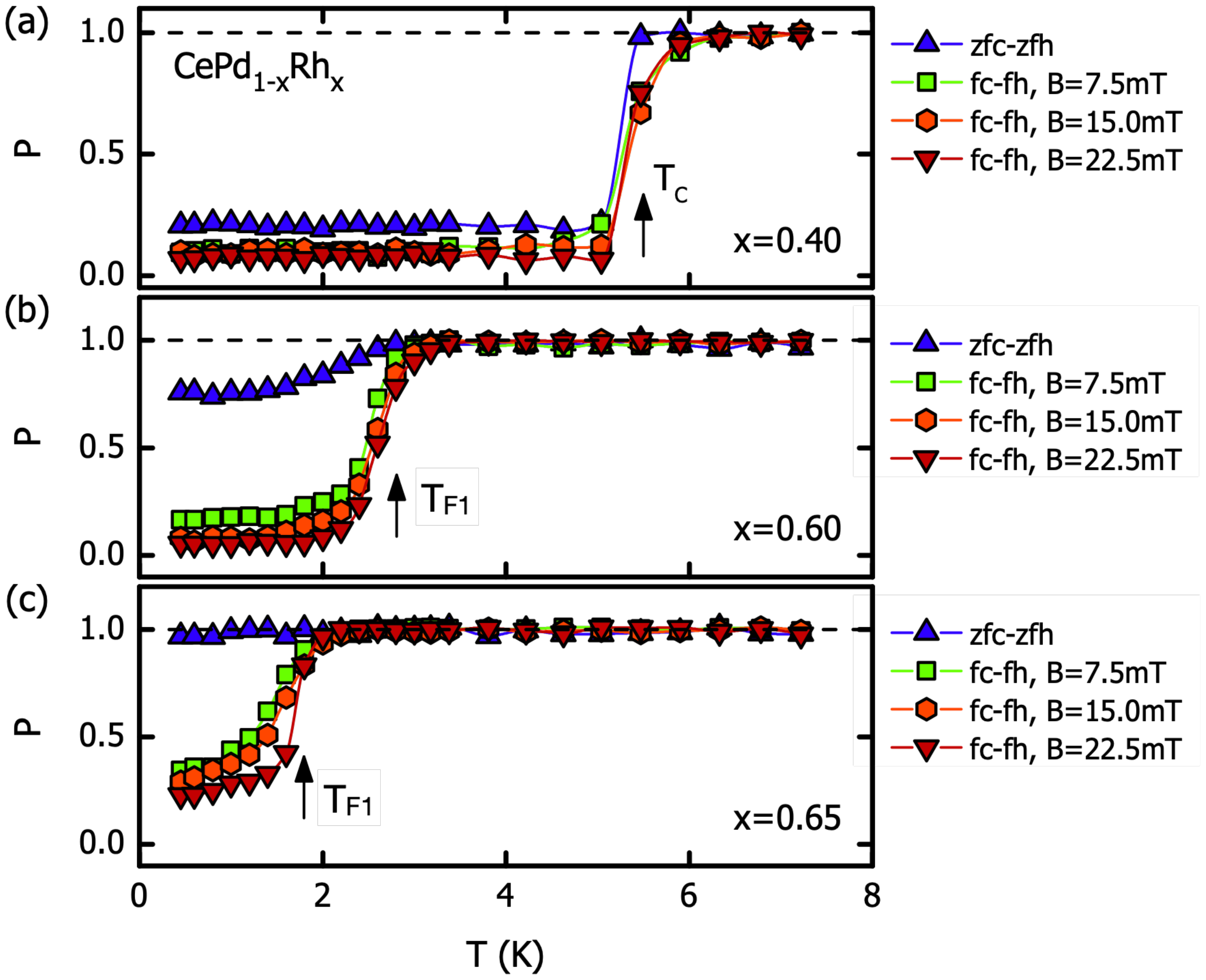}
	\caption{Polarization of CePd$_{1-x}$Rh$_x$ for $x=\num{0.40}$, $x=\num{0.60}$, and $x=\num{0.65}$ as a function of temperature under zfc-zfh and fc-fh at various magnetic fields. Data represent an average over a region of $32\times32$ pixels at the center of each sample. Three data sets were recorded using different external magnetic fields $B=\SI{7.5}{\milli\tesla}$, $B=\SI{15.0}{\milli\tesla}$ and $B=\SI{22.5}{\milli\tesla}$ under field-cooling, respectively. The data recorded under zero-field are shown for better comparision. Below $T_{\rm C}$ the polarization decreases with increasing external field $B$ while the transition broadens as a function of temperature. The arrows indicate the position of $T_{\rm C}$ as determined in zfc-zfh.}
	\label{fig:cepdrh-zfc-vs-fc}
\end{figure}

Shown in Fig.\,\ref{fig:cepdrh-zfc-vs-fc} is the polarization as a function of temperature for $x=\num{0.40}$, $x=\num{0.60}$, and $x=\num{0.65}$ as observed in different temperature versus field histories. The emphasis is here on the effects of the field strength, where the polarization represents an average of a $32\times32$ pixel region in the center of each sample. Data under an applied field were recorded for $B=\SI{7.5}{\milli\tesla}$, $B=\SI{15.0}{\milli\tesla}$, and $B=\SI{22.5}{\milli\tesla}$ (zfc-fh and fc-fh). For comparison also shown are data recorded for zfc-zfh. 

The zfc-zfh data recorded in the sample with $x=\num{0.40}$ shows a sharp drop of the polarization at $T_{\rm C}$ consistent with spontaneous ferromagnetic order forming large domains in zero field or in the presence of small applied magnetic fields as described by Eq. \ref{eq:p_const}. The spontaneous depolarization under zfc-zfh sets in at a well-defined transition temperature and saturates rapidly below $T_{\rm C}$. For increasing applied magnetic field a small broadening is observed at $T_{\rm C}$ while the strength of the depolarization increases slightly. 

In comparison to the sample with $x=\num{0.40}$ the temperature dependence under zfc-zfh for $x=\num{0.60}$ displays only a weak and gradual decrease just below $T_{\rm F1}$ consistent with small domains and/or weak internal fields as described by Eq.\,\ref{eq:p_exp}. Here the size of the depolarization increases remarkably under fc-fh in a small field of $\SI{7.5}{\milli\tesla}$. When further increasing the magnetic field the temperature dependence qualitatively and quantitatively changes only slightly.

For $x=\num{0.65}$ the depolarization below $T_{\rm F1}$ almost vanishes under zfc-zfh. This suggests that spontaneous correlations are virtually suppressed for this Rh concentration on the scales sensitive to neutron depolarization, i.e., the magnetic properties must be featuring very small domains or clusters which are spatially separated from each other. Alternatively, the zfc-zfh data for $x=0.65$ may reflect a strongly fluctuating state. Yet, in the presence of a small applied magnetic field a noteable depolarization is observed consistent with a weak form of ferromagnetism. This suggests that a small applied field stabilizes a ferromagnetic character on length and time scales sensitive to neutron depolarization. We will return to this point in Sec.\,\ref{sec:discussion}.

%%%%%%%%

\begin{figure}
	\centering
	\includegraphics[width=1.0\linewidth]{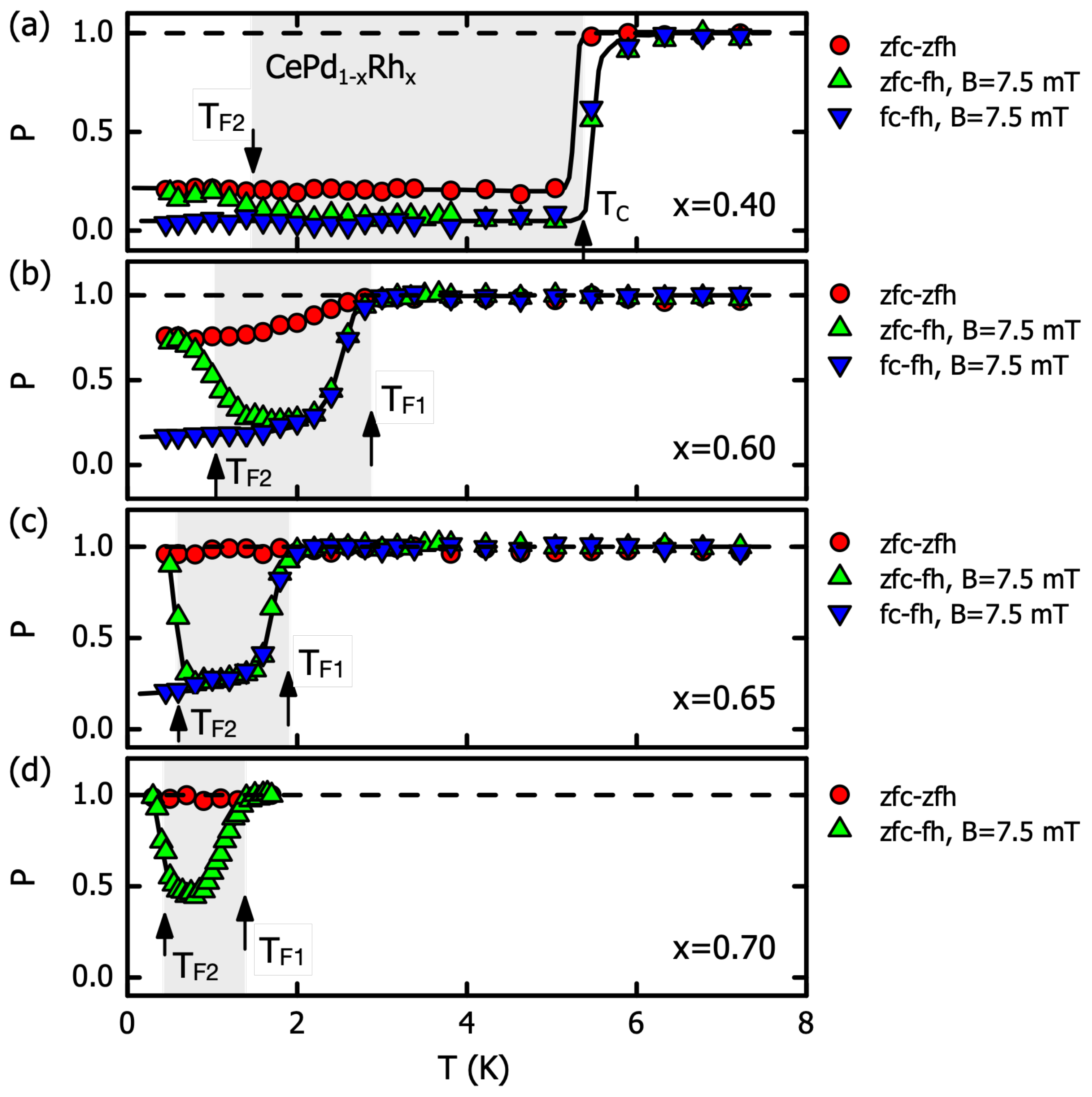} 
	\caption{Polarization of CePd$_{1-x}$Rh$_x$ for $x=\num{0.40}$, $x=\num{0.60}$, $x=\num{0.65}$, and $x=\num{0.70}$ as a function of temperature under zfc-zfh, zfc-fh, and fc-fh. In the zfc-fh and fc-fh measurements a field of $\SI{7.5}{\milli\tesla}$ was applied. The ferromagnetic transition temperature $T_{\rm C}$ is denoted by arrows. The increase of the polarization well below $T_{\rm C}$ under zfc-fh denoted $T_{\rm F2}$ underscores to the formation of a cluster glass below $T_{\rm F1}$ for $x\geq0.6$. The evidence for spontaneous ferromagnetic correlations that are sufficient to depolarize the neutron beam vanishes for $x>\num{0.60}$. However, in this regime a small applied field enhances the depolarizing effects.}
	\label{fig:cepdrh-zfc-fh}
\end{figure}

The pronounced depolarization below a characteristic temperature $T_{\rm F1}$ observed under zfc-zfh and fc-fh is contrasted by a pronounced reentrance of the polarization observed under zfc-fh as illustrated in Fig. \ref{fig:cepdrh-zfc-fh} for $\SI{7.5}{\milli\tesla}$. For ease of comparison also shown in Fig. \ref{fig:cepdrh-zfc-fh} are the data recorded under zfc-zfh and fc-fh at $\SI{7.5}{\milli\tesla}$ shown in Fig.\,\ref{fig:cepdrh-zfc-vs-fc}.  The key signature observed under zfc-fh at $\SI{7.5}{\milli\tesla}$ with decreasing temperature concerns a recovery of the polarization at a temperature $T_{\rm F2}$ well below $T_{\rm F1}$. For the compositions exhibiting a well defined initial decrease of the polarization under fc-fh, the reentrant behaviour is only observed for $x=\num{0.60}$ and $x=\num{0.65}$. For $x=\num{0.40}$ the reentrance is almost absent with a tiny recovery of polarization below $T_{\rm F2}\ll T_{\rm C}$.

Finally, above a critical concentration around $x=\num{0.65}$ no significant spontaneous depolarization is observed under zfc-zfh as a function of temperature. Nonetheless sizeable reentrant behaviour under zfc-fh is still observed, as shown in Fig.\,\ref{fig:cepdrh-zfc-fh} for $x=\num{0.70}$. Thus the reentrant behaviour under zfc-fh prevails as a key signature of the magnetic properties up to high Rh concentrations. Moreover, when taken together with the data recorded for $x=\num{0.60}$ and $x=\num{0.65}$ we observe a decrease of $T_{\rm F2}$ with increasing $x$ that roughly tracks the decreases of $T_{\rm F1}$.

%%%%%%%%%%%%%%%%%%%%%%%%%%%%%%%%
\subsection{Variation of the neutron polarization due to magnetic anisotropy}
\label{subsec:results_anisotropy_poli}

An important facet of the interpretation of the neutron polarization concerns the difference between a generic depolarization and a possible spherical rotation of the direction of the polarization. Such a rotation may be caused by the magnetic anisotropy of the material. Previous studies of the magnetization of {\cpr} are consistent with an easy magnetic $c$-axis \cite{deppe2006}. To follow up on the role of the magnetic anisotropies at zero magnetic field, we tracked the anisotropy of the depolarization as determined at the instrument POLI at FRM II using the 3D polarization analysis device CryoPAD.

\begin{figure}
\centering
	\includegraphics[width=0.7\linewidth]{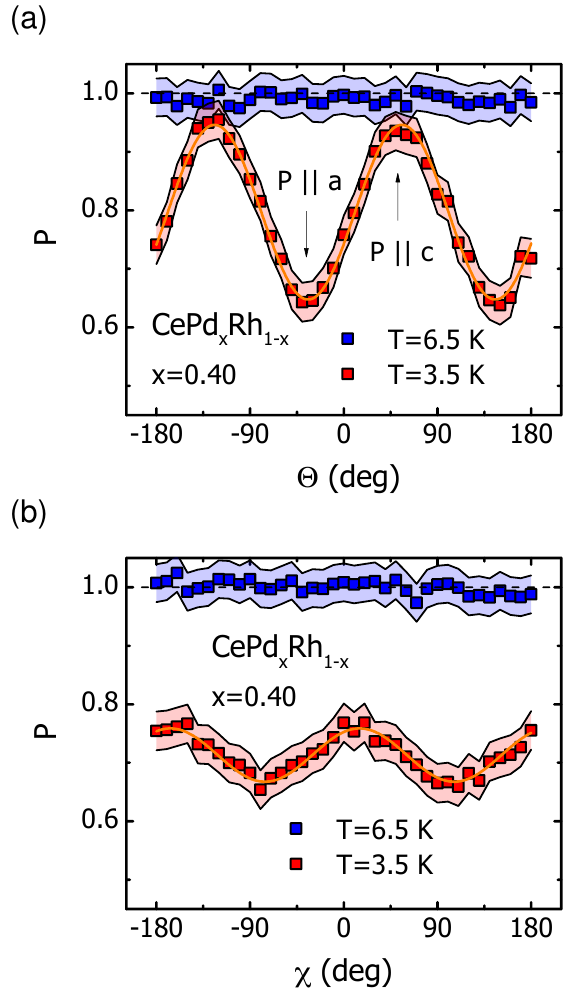} 
%	\subfigimg[width=0.75\linewidth]{~~~~\textplotfont{(a)}}{figure-8a-v1.pdf} \\
%	\subfigimg[width=0.75\linewidth]{~~~~\textplotfont{(b)}}{figure-8b-v1.pdf} 
\caption{Anisotropy of the depolarization as a function of crystallographic orientation for a single crystal of CePd$_{1-x}$Rh$_x$ with Rh concentration $x=\num{0.40}$. For the data shown in (a) and (b) the angles $\Theta$ and $\chi$ were scanned separately, while the polarization was always set and analyzed in the same direction. (a) Polarization dependence as a function of $\Theta$ in the crystallographic ac-plane. The ac-plane was almost perpendicular to the neutron beam. The phase shift in the oscillation suggests that the easy axis of the magnetization is rotated roughly by an angle of $\SI{55}{\deg}$ with respect to the x-axis. (b) Polarization as a function of $\chi$, defining the angle of the polarization in the yz-plane where $\chi\approx\SI{90}{\deg}$ corresponds to the beam direction and roughly to the crystallographic b-axis of the sample.}
\label{fig:cepdrh-anisotropy-depol}
\end{figure}

Our measurements were performed on the single-crystalline sample with $x=\num{0.40}$. This represents a composition for which the bulk properties provide unambiguous evidence of long-range ferromagnetism consistent with the neutron depolarization reported above. The orientation of the sample with respect to the coordinate system of the instrument is illustrated in Fig. \ref{fig:cepdrh-anisotropy}. The ac-plane of the sample was oriented in the xy-plane of the coordinate system of the instrument, with an angle of approximately \SI{55}{\deg} between the crystallographic c-axis and the y-axis of the set-up. The polarization as a function $\Theta$ and $\chi$ are shown in Fig. \ref{fig:cepdrh-anisotropy-depol}\,(a) and \ref{fig:cepdrh-anisotropy-depol}\,(b), respectively. Both angular dependences confirm that the polarization was equal to one and constant above $T_{\rm C}$ as expected for the paramagnetic state. Below the transition temperature a sinusoidal variation was observed as a function of $\Theta$ and $\chi$. The variation in $\chi$ was much weaker than the variation in $\Theta$ which underlines the fact that the easy axis of the magnetization was almost perpendicular to this crystallographic plane.

In the angular scan of $\chi$ there was also a small additional oscillation which suggests that the surface at which the sample was attached to the sample holder, i.e., the crystallographic ac-plane, was not oriented perfectly perpendicular to the neutron beam. Two well-defined minima and maxima were observed for a complete angular scan of the polarization vector. The maxima correspond to the case where the easy axis of the magnetization (in this case the crystallographic c-axis) was almost parallel or anti-parallel to the polarization vector. The minima correspond to a maximum in depolarization where the polarization vector was almost perpendicular to the preferred magnetization axis.

The small amount of depolarization for the case when the polarization was parallel or antiparallel to the anisotropy axis suggests that the magnetic domains were dominantly aligned parallel or antiparallel to the c-axis. For such a regular pattern of domains a depolarizing effect may still be expected due to a finite beam divergence resulting in slightly different neutron path lengths across the sample.

%%%%%%%%%%%%%%%%%%%%%%%%%%%%%%%%
%\subsection{Variation of neutron depolarization due to magnetic anisotropy under finite field}
%\label{subsec:results_anisotropy_antares}

It is in particular instructive to explore the interplay of an applied magnetic field with the effects of magnetic anisotropy, domain populations, neutron depolarisation and spherical precession of the polarization. Namely, the application of a magnetic field results in an effective internal magnetic field $B_\textrm{int}$ that changes the ratio of the domain populations. Moreover, for a small applied magnetic field the internal field will point along the easy axis, even if the easy axis is not parallel to the applied field. 

As the Larmor phase collected by the neutrons depends on their wavelength, a polychromatic beam such as that used in the setup shown in Fig.\,\ref{fig:exp-setup}\,(a) naturally depolarizes when it precesses in a magnetic field. In contrast, for a monochromatic beam featuring a narrow wavelength band the polarization decreases by a few percent only even after several Larmor precessions. If, in addition, the strength of $B_\textrm{int}$ changes due to changes of the magnetization, the total Larmor phase will change also. This finally causes oscillations in the polarization when the magnetization varies monotonically as a function of temperature or the applied magnetic field. In summary, we expect oscillations in the polarization when three conditions are fulfilled: (i) The magnetization of the sample changes, (ii) a monochromatic neutron beam is used, and (iii) the easy axis is not parallel to the polarization.

The polarization observed at POLI in a single-crystalline sample suggests that the magnetic domains in {\cpr} ($x=0.4$) dominantly support a magnetization along the easy magnetic c-axis of the material characteristic of a 3d Ising ferromagnet and consistent with the magnetization \cite{deppe2006}. At zero magnetic field the populations of up and down domains are equal such that the integrated internal field vanishes, $B_\mathrm{int}=\num{0}$. In the presence of an applied magnetic field this ratio changes according to the internal field. The precession of the polarization with respect to the internal field may then be described by Eq. \ref{eq:p_singledomain} where $B_0$ is replaced by $B_\mathrm{int}$. If $B_\mathrm{int}$ changes, e.g., due to changes of the applied magnetic field or changes of the magnetization as a function of temperature, and if the angle between the polarization and the internal field is finite, the cosine part of Eq. \ref{eq:p_singledomain} causes oscillations in the polarization. These oscillations were observed both in temperature and field scans slightly above the Curie temperature $T_{\rm C}=\SI{5.26}{\kelvin}$. At lower temperatures the strong depolarization prevents the appearance of oscillations.

Mathematically this effect may be described by a multiplication of Eq. \ref{eq:p_const}, recognizing that the sample is strongly ferromagnetic, and Eq. \ref{eq:p_singledomain}. To account for the finite distribution of Curie temperatures across the sample, the Heaviside function is replaced by an error function centred at $T_\text{C}$
\begin{equation} 
\begin{split}
\operatorname{erf}_\text{depol}(T, T_{\rm C}, \Delta T_{\rm C}, N) = 
&\frac{1}{2} \left[ \operatorname{erf}  \left( m \cdot (T-T_\text{C}) \right) +1 \right] \\ 
& \cdot \left( 1-P_\text{offset} \right) + P_\text{offset}
\end{split}
\end{equation}
which varies between $P_\text{offset}=3^{-N}$ and 1. The parameter $m$ represents the slope at $T=T_\text{C}$ such that 
\begin{equation} \label{eq:erf-depol-Tc}
\Delta T_\text{C} = \frac{2 \sqrt{\ln 2}}{m}
\end{equation}
represents the half-width, $\Delta T_\text{C}$, of the distribution of $T_\text{C}$.
The polarization in the presence of a magnetic anisotropy, i.e., an angle between the easy axis and the crystallographic c-axis, is then given by
\begin{equation} \label{eq:depol-rotation}
\begin{split}
P = P_0 & \cdot \left[ \cos^2(\alpha) + \sin^2(\alpha) \cos \left( \gamma \mu_0 M(B,T) \frac{d}{v} \right) \right] \\
& \cdot \operatorname{erf}_\textrm{depol} \left( T, T_{\rm C}, \Delta T_{\rm C}, N \right).
\end{split}
\end{equation}

Experimental evidence for this behavior may be observed in the single-crystal with $x=\num{0.40}$ using the setup illustrated in Fig. \ref{fig:exp-setup}\,(b). Data were recorded at the beam-line ANTARES after a major instrument upgrade\cite{calzada2009,schulz2015} in which an additional neutron velocity selector was installed to monochromatize the beam. For the measurements reported here the crystal was oriented with the crystallographic c-axis under an angle of approximately $\alpha=\SI{45}{\deg}$ with respect to the polarization of the incident beam.

%%%%%

\begin{figure} \centering
\includegraphics[width=0.8\linewidth]{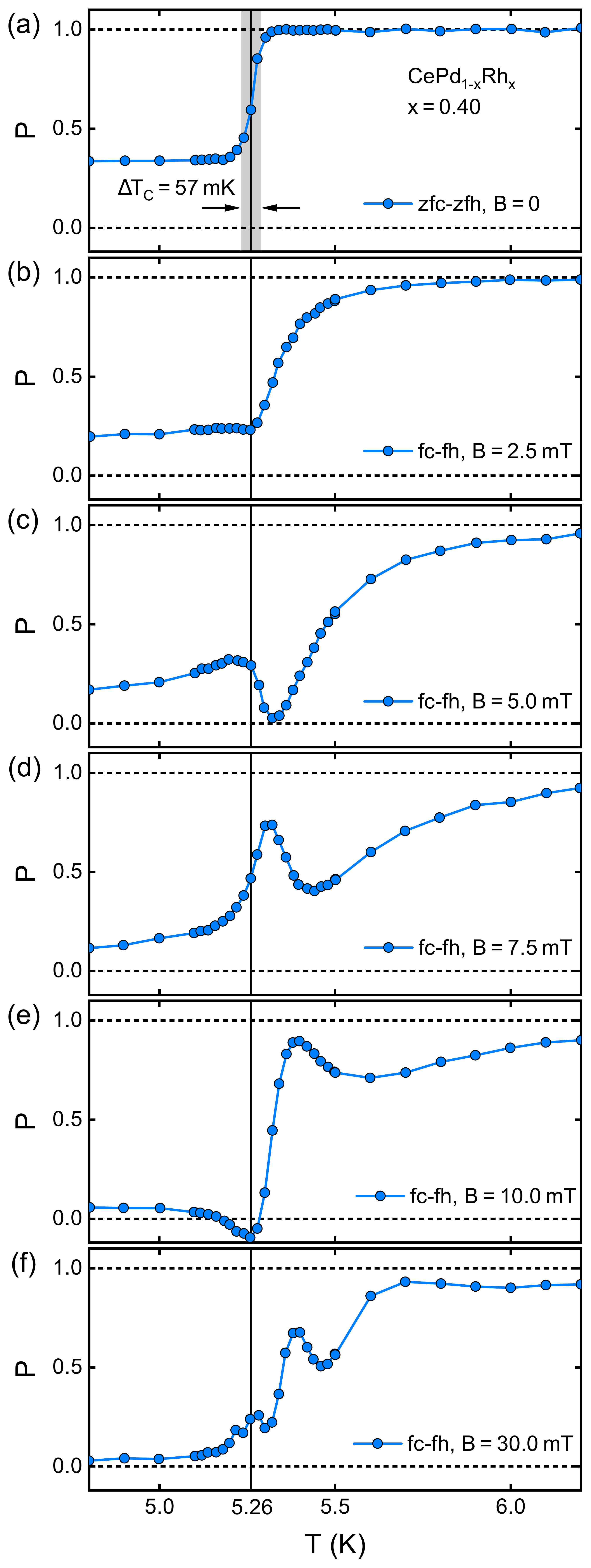}
\caption{Comparison of several fc-fh temperature scans for different magnetic fields of a CePd$_{1-x}$Rh$_x$, $x=\num{0.40}$ single crystal. Applying small fields leads to a decrease of the polarization well above the ordering temperature $T_{\rm C}=\SI{5.26}{\kelvin}$ determined from the zfc-zfh scan. Additionally, oscillations close to $T_{\rm C}$ appear in the polarization signal if small fields are applied. The width of the transition as determined from a fit of Eq. \ref{eq:p_const} to the zf data set is $\Delta T_{\rm C} = \SI{57}{\milli\kelvin}$.}
	\label{fig:cepdrh-tcluster}
\end{figure}

Shown in Fig.\,\ref{fig:cepdrh-tcluster} is the polarization as a function of temperature across the phase transition at $T_{\rm C}=\SI{5.26}{\kelvin}$. Under zfc-zfh, shown in Fig. \ref{fig:cepdrh-tcluster}\,(a), a sharp drop of the polarization was observed at $T_{\rm C}$ in agreement with the data shown in Fig.\,\ref{fig:cepdrh-zfc-vs-fc}. Also, the width of the transition of $\Delta T_{\rm C} = \SI{57}{\milli\kelvin}$ compares well with the  $T_{\rm C}$ maps shown in Fig. \ref{fig:tc-maps}. 

In comparison, the temperature dependence of the polarization became more complex under fc-fh as recored in various applied magnetic fields between \num{0} and \SI{30}{\milli\tesla}, shown in Figs.\,\ref{fig:cepdrh-tcluster}\,(b) through \ref{fig:cepdrh-tcluster}\,(f). In a small applied field of \SI{2.5}{\milli\tesla} the onset of the decrease of the polarization shifted to higher temperatures and exhibited considerable broadening. Similar behavior was observed in several neutron depolarization measurements reported in the literature \cite{bakker1968, drabkin1968, rauch1968a, drabkin1969, maleev1970, mitsuda1985, endoh1986}, which were attributed to a change of the temperature dependence to the slowing down of ferromagnetic fluctuations close to $T_{\rm C}$, such that they satisfy the conditions for depolarizing the neutron beam.

Further, when increasing the applied magnetic field oscillations in the polarization as a function of temperature emerged close to $T_{\rm C}$. These oscillations may be explained by a precession of the polarization with respect to the effective internal field comprising the interplay of the magnetization under the applied magnetic field and the easy magnetic axis of the sample. This interpretation is corroborated by the observation of a polarization that is nominally negative for $B=\SI{10}{\milli\tesla}$ which cannot be accounted for by a depolarization alone.

\begin{figure} \centering
\includegraphics[width=0.8\linewidth]{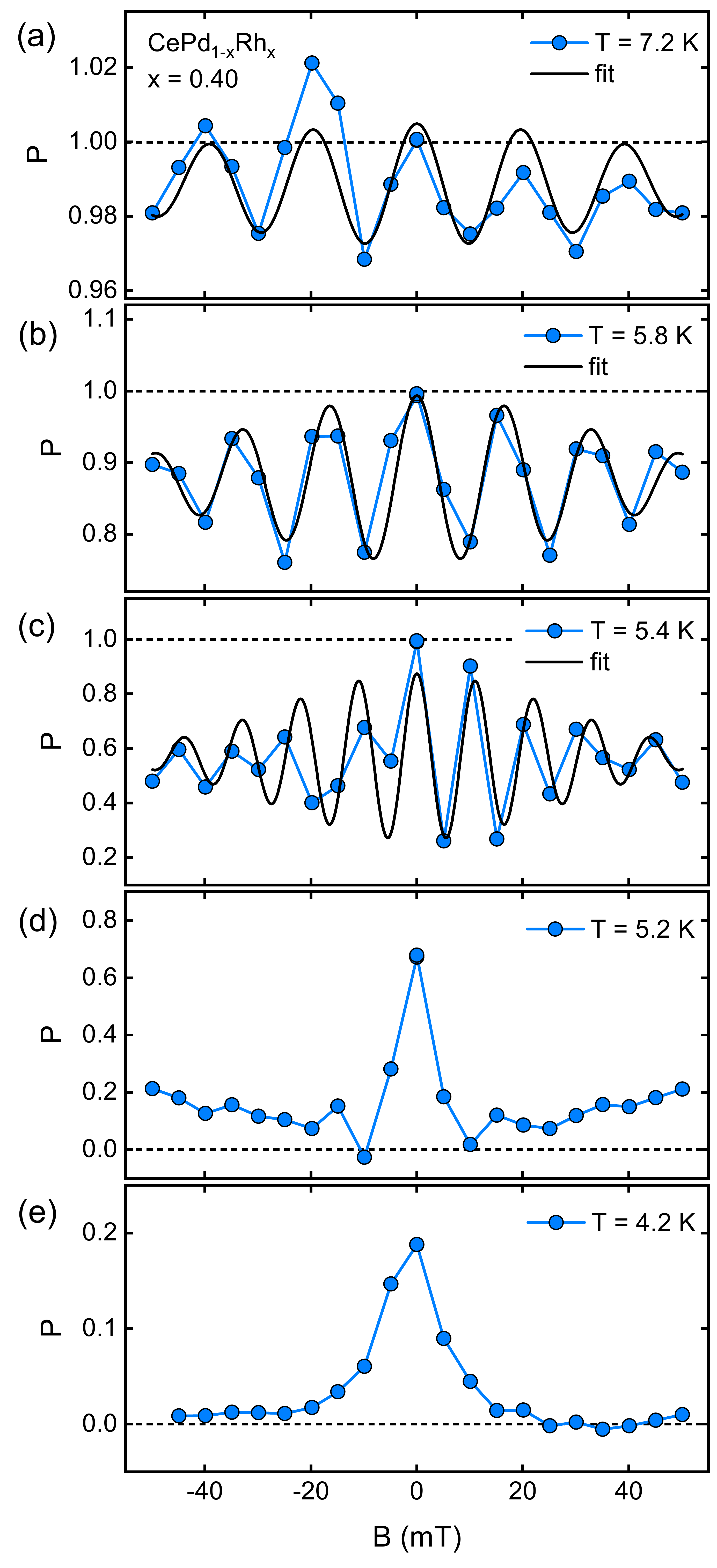}
\caption{Polarization as a function of magnetic field of single-crystal CePd$_{1-x}$Rh$_x$ with $x=\num{0.40}$ for various temperatures. An oscillation in the polarization is already present at temperatures well above $T_{\rm C}=\SI{5.26}{\kelvin}$. To illustrate key characteristics of the oscillations a fit using a gaussian-damped cosine function is shown.}
\label{fig:cepdrh-bstack-0.40}
\end{figure}

Similar oscillations of the polarization were also observed in magnetic field sweeps. As explained above, the setup shown in Fig. \ref{fig:exp-setup}\,(b) allowed to measure the polarization in bipolar field cycles. Shown in Fig.\,\ref{fig:cepdrh-bstack-0.40} is the polarization as a function of applied field between $+\SI{50}{\milli\tesla}$ and $-\SI{50}{\milli\tesla}$ at various temperatures below and above $T_{\rm C}=\SI{5.26}{\kelvin}$ for the single-crystal sample with $x=\num{0.40}$. Rough fits of the data illustrating the frequency and amplitude of the oscillations were extracted using a gaussian-damped cosine function.

Already at \SI{7.2}{\kelvin}, shown in Fig.\,\ref{fig:cepdrh-bstack-0.40}\,(a), a pronounced oscillation may be discerned, i.e., well above $T_{\rm C}$. The frequency of the oscillation increases with decreasing temperature, consistent with an increase of the internal field. The amplitude of the oscillation is damped for increasing magnitude of the applied field due to the finite wavelength spread of the neutron beam. This effect compares with similar behaviour seen, e.g., in neutron spin-echo measurements \cite{2019_Franz_JPSJ}. The oscillation is smeared out below $T_{\rm C}$ where a small applied magnetic field of \SI{20}{\milli\tesla} is already sufficient to completely depolarize the neutron beam, i.e.,  $P=\num{0}$, as shown in Fig.\,\ref{fig:cepdrh-bstack-0.40}\,(e).

%%%%%%%%%%%%%%%%%%%%%%%%%%%%%%%%
%\subsection{Anisotropy effects on the polarization signal}
%\label{subsec:discussion_anisotropy}

\begin{figure}
	\centering
	\includegraphics[width=0.9\linewidth]{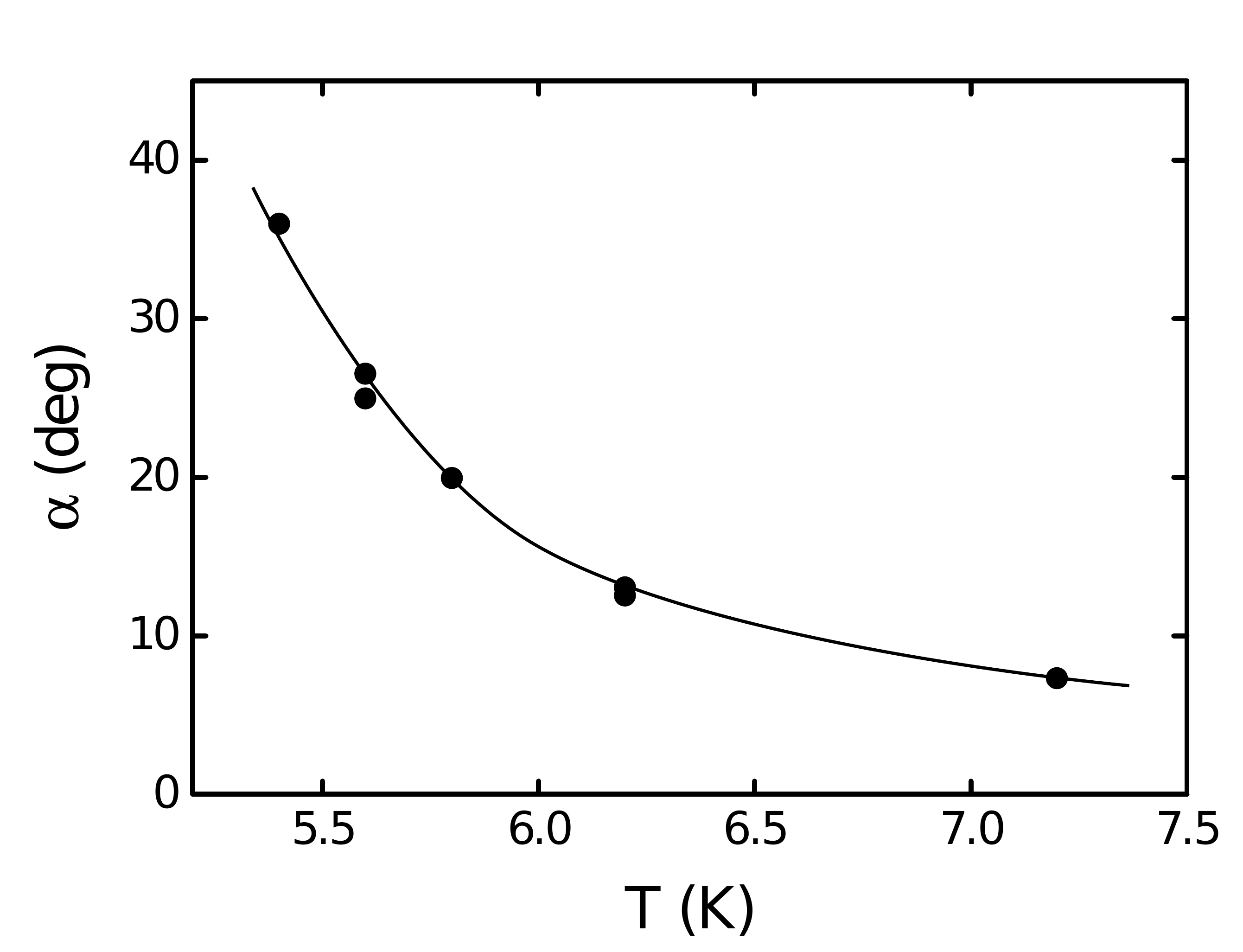}
	\caption{Angle $\alpha$ between the c-axis of a CePd$_{1-x}$Rh$_x$ single-crystal with $x=\num{0.40}$ and the direction of the neutron polarization $P$ above the ferromagnetic transition temperature $T_{\rm C}=5.26\,{\rm K}$. Data points were inferred from the field sweeps shown in Fig. \ref{fig:cepdrh-bstack-0.40} using Eq. \ref{eq:depol-rotation}. A continuous decrease of $\alpha$ is observed with increasing temperature. The black line is a guide to the eye.}
	\label{fig:alpha}
\end{figure}

A direct comparison of Eq. \ref{eq:depol-rotation} with the temperature dependence observed experimentally is not satisfactory due to the large number of parameters and the lack of information of the precise temperature dependence of the magnetization $M(T)=B_\textrm{int}(T)/\mu_0$. In contrast, an evaluation of the magnetic field dependence is possible as the depolarization term stays constant and the field-dependence of the magnetization $M(B)$ is roughly linear in small applied magnetic fields. In turn, this permits to infer the angle $\alpha$ between $B_\textrm{int}$ and the polarization as a function of temperature as shown in Fig. \ref{fig:alpha}.

With increasing temperature above $T_{\rm C}=5.26\,{\rm K}$ the angle $\alpha$ decreases. As the magnetic anisotropy inferred from the bulk properties is unchanged up to \SI{100}{\kelvin}\cite{westerkamp2008} the temperature dependence may be attributed to the decrease of the easy-axis susceptibility and the associated decrease of the life-time of fluctuations in the paramagnetic state vis a vis with the time needed of a neutron to traverse the regime of a fluctuation. As depolarizing effects decrease with increasing temperature above $T_{\rm C}$, the combination of these time-scales may be effectively viewed in terms of a magnetic field causing dominantly a rotation of the polarization direction. 

%%%%%%%%%%%%%%%%%%%%%%%%%%%%%%%%
\section{Discussion}
\label{sec:discussion}

%%%%%%%%%%%%%%%%%%%%%%%%%%

For the interpretation of our experimental results it is helpful to recall at first the notions and terminology used in conventional spin freezing processes starting from a paramagnetic state. In so-called canonical spin glasses the separation and interactions between the spins are sufficiently small, such that the frozen state is characterized by an ensemble of essentially randomly oriented, uncorrelated microscopic spins. In systems with a larger density of spins and larger interactions, clusters of correlated spins may form under decreasing temperature. The associated frozen state is commonly dubbed a cluster glass. Finally, in the limit of strongly interacting densely packed spins, correlated regimes may form that behave essentially like very large macroscopic spins and the behaviour is referred to as superparamagnetism. For completeness we note that spin-frozen states which emerge from long-range ordered states are referred to as a reentrant spin glass -- a misleading expression as the long-range ordered state actually exhibits the reentrant temperature dependence. 

Regardless of the precise character of the spin-frozen state, neutron depolarization is expected if the same threshold conditions are satisfied as in a multi-domain ferromagnet. To distinguish long-range ferromagnetism from a spin-frozen state is, in turn, not straight-forward and requires consideration of further information such as the bulk properties. In particular, apart from quantitative differences of the size of the depolarization, the onset of the depolarization will be insensitive to the precise magnetic field and temperature history. The former depends on the precise alignment of the domains, whereas the latter depends on the interactions and the size of the correlated regimes. 

Further, it is also important to distinguish a generic depolarization from the rotation of the polarization axis. Both effects may be present simultaneously as reported in the literature, e.g., for Ni and observed in our study of CePd$_{1-x}$Rh$_x$ for $x=0.4$. This compares with Fe$_{1-x}$Cr$_x$ \cite{2020_Benka_arXiv}, where a pronounced depolarization was recently reported deep in the paramagnetic state far above the freezing temperature observed in the bulk properties.  In this context we also wish to note that reentrant spin glasses may exhibit a depolarization for all temperatures below the onset of sufficiently strong ferromagnetic correlations. If the long range ordered state is ferromagnetic the depolarization may start below the Curie temperature and prevail unchanged into the spin glass regime\cite{2020_Benka_arXiv}. If, in contrast, the long-range ordered state is antiferromagnetic, a depolarization may only be expected below the spin-glass temperature. 

%%%%%

We turn now to the evolution of the nature of ferromagnetic correlations in {\cpr} as a function of increasing Rh content $x$, which is the result of several microscopic interactions. Notably, magnetic moments develop with decreasing temperature that interact by virtue of an exchange coupling. Crystal electric fields and spin-orbit coupling partly quench the magnetic moments and introduce magnetic anisotropies. The Kondo effect results in an additional screening of these moments and changes of the concomitant interactions. As a function of increasing Rh content the spontaneous moment at zero temperature decreases from an almost unscreened large value for $x=0$. Both the change of the lattice constant and the concomitant decrease of the density of states at the Fermi level, as well as the increase of the Kondo screening under increasing Rh content control the suppression of ferromagnetism. 

%%%%%

For all {\cpr} samples studied we observe neutron depolarization. Up to $x=0.65$ even a spontaneous depolarization under zfc-zfh conditions is observed.  This provides unambiguous microscopic evidence of the ferromagnetic character of the spin correlations up to $x=\num{0.70}$, the largest value studied. In addition, the $T_{\rm C}$ maps depicted in Fig.\,\ref{fig:tc-maps} show that the distribution of the ordering temperature varies only slightly over the sample cross-section with well-defined ordering and/or freezing temperatures. The samples are hence metallurgically homogeneous on macroscopic scales. This confirms that bulk properties like the magnetization reflect intrinsic behavior.

The onset of the neutron depolarization at $T_{\rm C}$/$T_{\rm F1}$ is in excellent agreement with the ordering/freezing temperature observed in the bulk properties. The behaviour at $T_{\rm C}$/$T_{\rm F1}$ does not depend on the temperature and field history, apart from a small broadening at $T_{\rm C}$ for $x=0.4$. Indeed, for $x=0.4$ the strong easy-axis ferromagnetism causes a Larmor rotation of the polarization even well above the Curie temperature, whereas a pronounced depolarization is observed below $T_{\rm C}$. The properties of the ferromagnetically ordered compositions are, hence, perfectly consistent with an Ising ferromagnet without noticeable evidence of disorder, e.g., such as depolarization above $T_{\rm C}$. \added{This provides an important point of reference for the emergence of the reentrant and glassy behaviour near quantum criticality.}

For $x>0.6$ the agreement of $T_{\rm F1}$ observed in the depolarization and the bulk properties contrasts the observation of a small but finite hysteresis in the magnetization for $T>T_{\rm F1}$ that has been attributed to the formation of clusters\cite{westerkamp2009}. The absence of depolarization above $T_{\rm F1}$  thus shows that the size and the life-time of the clusters inferred from the magnetization must be tiny and below the threshold of depolarization, consistent with a decrease of the ordered moment and the interactions under increasing Rh content. The behaviour we observe in {\cpr} contrasts that observed in the superparamagnetic regime of Fe$_{1-x}$Cr$_x$ where a sizeable depolarization is observed well above the freezing temperature \cite{2020_Benka_arXiv}. It underscores the formation of a cluster glass at $T_{\rm F1}$ in {\cpr}, however, consisting of tiny clusters. 

%%%%%

A highly unconventional property emerges, finally, under zero-field-cooling/field-heating. All samples with $x>0.6$ exhibit a pronounced reentrance of the depolarization between $T_{\rm F1}$ and $T_{\rm F2}$. In fact, even the ferromagnetic single-crystal with $x=0.4$ displayed a reentrance at $T_{\rm F2}$, though barely noticeable. Here it may be helpful to note that the reentrant temperature dependence of the depolarzation under zfc-fh cannot be the signature of a reentrant spin-glass. Rather, the lack of depolarization up to $T_{\rm F2}$ under zero-field-cooling/field-heating implies that the clusters which undergo a freezing at $T_{\rm F1}$ must be tiny in the absence of an applied field. This is consistent with the absence of a depolarization above $T_{\rm F1}$ despite the presence of hysteresis in the magnetization. 

Yet, the applied magnetic field of 7.5\,mT under which reentrance is observed is small, and the energy scale associated with the applied field corresponds to a temperature of several hundred milli-Kelvin. The Zeeman energy of the applied magnetic field is hence roughly consistent with the values of $T_{\rm F2}$. Thus, when heating the sample in a small applied magnetic field after zero-field-cooling, a thermally activated formation of clusters may take place at $T_{\rm F2}$, where the resulting clusters are sufficiently large to generate a sizeable depolarization. 

Considering the combination of energy scales in {\cpr} mentioned above, it is instructive to discuss the possible origin of the small size of the clusters at zero magnetic field that undergo the spin freezing at $T_{\rm F1}$. Experimentally we find that the freezing temperature $T_{\rm F1}$  and the reentrance temperature $T_{\rm F2}$ decrease with increasing $x$ and roughly track each other.  As a function of increasing Rh content, this is consistent with the reduction of the spontaneous magnetic moment and the strength of the interactions, as well as the steep increase of the Kondo screening and the distribution of Kondo temperatures for $x>x^*\approx0.6$. Namely, as the moment decreases the freezing temperatures decrease, empirically suggesting that the additional Kondo screening above $x^*$ controls the small size of the clusters.

%%%%%%%%%%%%%%%%%%%%%%%%%%%%%%%%
%%%%%%%%%%%%%%%%%%%%%%%%%%%%%%%%

The setup shown in Fig.\ref{fig:exp-setup}\,(a) yields detection thresholds of $\delta > \SI{2}{\nano\meter}$ and $\tau > \SI{4}{\pico\second}$ for an unscreened moment of $2\,\mu_{\rm B}{\rm f.u.^{-1}}$ in CePd$_{1-x}$Rh$_x$ as discussed in Sec. \ref{subsec:ndi-scales}. However, ferromagnetic fluctuations in this parameter regime and on this time scale are not plausible based on the bulk properties and the value of $T_{F1}$. Therefore, we attribute the change in polarization at $T_{F2}$ to an increase of the average size of the clusters. 

\begin{figure} \centering
\includegraphics[width=0.9\linewidth]{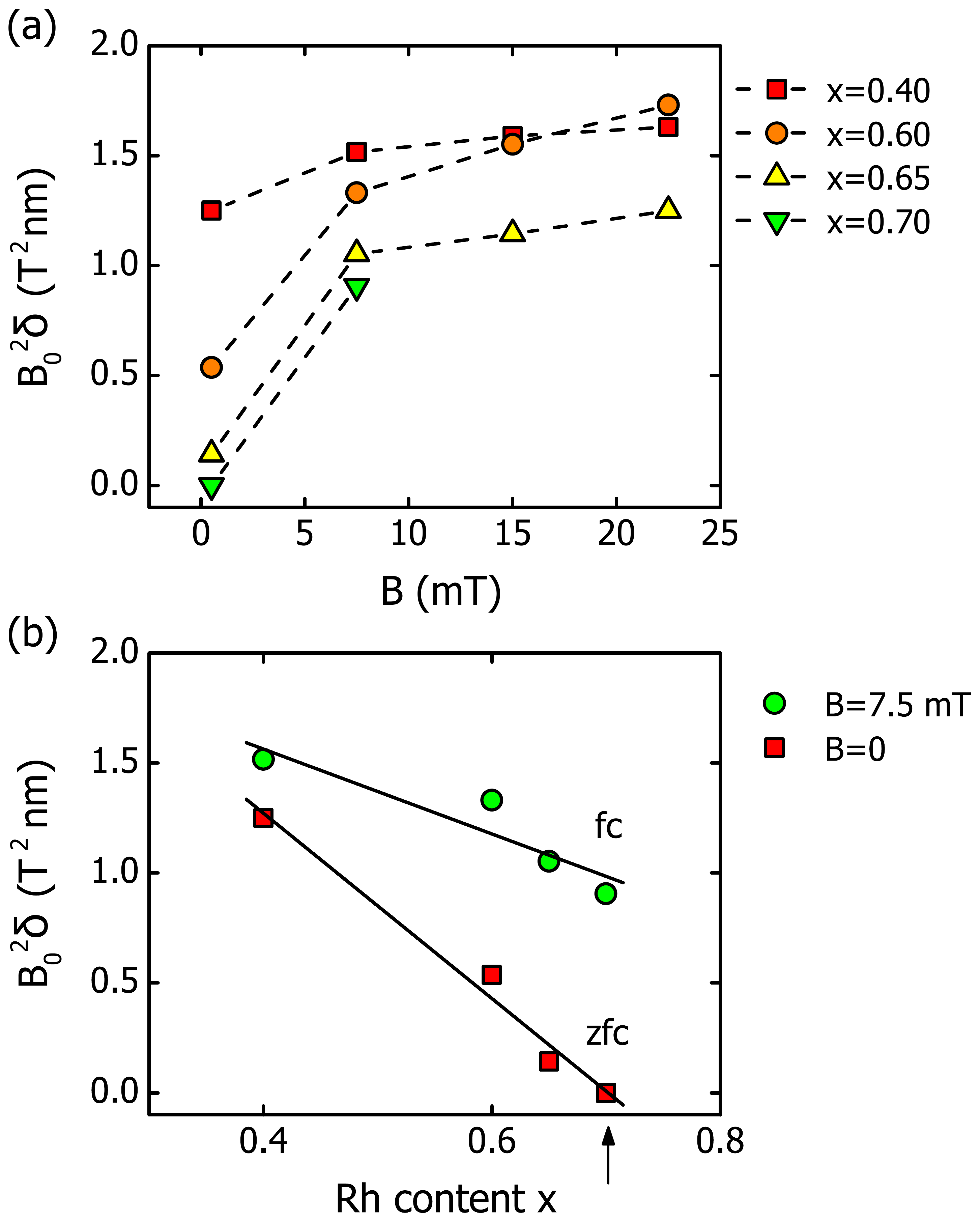}
\caption{(a) Product $B_0^2\delta$ of average flux density per domain $B_0$ and average domain length $\delta$ in transmission direction as a function of external magnetic field $B=\mu_0 H$. The behavior is monotonic, however with increasing Rh content $x$ the initial value at zero applied field vanishes for $x>\num{0.65}$. (b) The product $B_0^2\delta$ shown as function of Rh concentration $x$ derived from zero-field measurements and under an applied field $B=\SI{7.5}{\milli\tesla}$ for each concentration. The signature after zero-field-cooling vanishes at $x=\num{0.70}$ which implies that $B_0^2\delta\rightarrow 0$.}
\label{fig:cepdrh-b0-vs}
\end{figure}

Fitting the depolarization shown in Fig.\,\ref{fig:cepdrh-zfc-vs-fc} with Eq.\,\ref{eq:p_exp}, typical values of $B_0^2\delta$ may be inferred, i.e., the average field per domain $B_0$ squared times the average domain size $\delta$. Shown in Figs. \ref{fig:cepdrh-b0-vs}\,(a) and \ref{fig:cepdrh-b0-vs}\,(b) is the extrapolated zero-temperature limit of $B_0^2\delta$ as a function of the applied field $B=\mu_0 H$ and the Rh concentration $x$, respectively. Under small applied magnetic fields $B_0^2\delta$ increases significantly. The absolute values of $T_{\rm F2}$ as compared to the applied magnetic field suggest that this may be attributed to a thermally activated increase of the average cluster size. The broad distribution of Kondo temperatures at large values of $x$ support this suggestion. Likewise, $B_0^2\delta$ decreases as a function of increasing Rh content, as expected when approaching the intermediate valent properties of CeRh. 

The average size $\delta$ of the clusters may, finally, be estimated when taking into account the magnetic moment of CePd$_{1-x}$Rh$_x$\cite{1988_Nieva_ZPhysB,sereni1993, schmakat2015a, deppe2006}. In the field-cooled state the cluster size decreases continuously from a value exceeding \SI{6}{\nano\meter} to \SI{4}{\nano\meter} when the Rh concentration increases from $x=0.4$ to 0.7. Under zero-field-cooling $\delta$ decreases from \SI{5}{\nano\meter} to a value below the detection threshold. This is consistent with basic estimates of the cluster size of approximately 5 spins in the tail region of the phase diagram\cite{westerkamp2008}.

%%%%%%%%%%%%%%%%%%%%%%%%%%%%%%%%

%%%%%%%%%%%%%%%%%%%%%%%%%%%%%%%%
\section{Conclusions}
\label{sec:conclusion}

In conclusion, we have carried out neutron depolarization measurements of ferromagnetism and spin freezing in {\cpr}. We find clear signatures of ferromagnetic correlations up to a Rh concentration of $x=\num{0.70}$\added{, where the 3D polarization analysis of a single crystal with $x=0.4$ underscores well behaved long-range ferromagnetic order}. The ordering and freezing temperatures are in good agreement with the bulk properties.  A reentrant temperature dependence of the depolarization under zero-field-cooling/field-heating of the Rh compositions featuring spin freezing reveals thermally activated cluster growth in the spin-frozen state. The sensitivity of our setup and the estimated size of the ferromagnetic correlations provide microscopic information consistent with the formation of a Kondo cluster glass, initially proposed on the basis of the bulk properties.\cite{westerkamp2009,westerkamp2008} The Kondo cluster glass emerges adjacent to a ferromagnetic QPT. Taken together, our observations in {\cpr} underscore the potential of  neutron depolarization as a microscopic probe of ferromagnetic quantum phase transitions and concomitant escape routes. 

%%%%%%%%%%%%%%%%%%%%%%%%%%%%%%%%
\acknowledgments
We wish to thank Peter B\"oni for fruitful discussions and the team at the beamline ANTARES for support with our experiments. This work has been funded by the Deutsche Forschungsgemeinschaft (DFG, German Research Foundation) under FOR960 (Quantum Phase Transitions), TRR80 (From Electronic Correlations to Functionality, Project No. 107745057) and the excellence cluster MCQST under Germany's Excellence Strategy EXC-2111 (Project No. 390814868). Financial support by the European Research Council (ERC) through Advanced Grants No. 291079 (TOPFIT) and No. 788031 (ExQuiSid) is gratefully acknowledged. M.S., P.S., and P.J. acknowledge financial support through the TUM Graduate School.

%%%%%%%%%%%%%%%%%%%%%%%%%%%%%%%%
%\bibliography{CePdRh_paper}

\begin{thebibliography}{94}%
\makeatletter
\providecommand \@ifxundefined [1]{%
 \@ifx{#1\undefined}
}%
\providecommand \@ifnum [1]{%
 \ifnum #1\expandafter \@firstoftwo
 \else \expandafter \@secondoftwo
 \fi
}%
\providecommand \@ifx [1]{%
 \ifx #1\expandafter \@firstoftwo
 \else \expandafter \@secondoftwo
 \fi
}%
\providecommand \natexlab [1]{#1}%
\providecommand \enquote  [1]{``#1''}%
\providecommand \bibnamefont  [1]{#1}%
\providecommand \bibfnamefont [1]{#1}%
\providecommand \citenamefont [1]{#1}%
\providecommand \href@noop [0]{\@secondoftwo}%
\providecommand \href [0]{\begingroup \@sanitize@url \@href}%
\providecommand \@href[1]{\@@startlink{#1}\@@href}%
\providecommand \@@href[1]{\endgroup#1\@@endlink}%
\providecommand \@sanitize@url [0]{\catcode `\\12\catcode `\$12\catcode
  `\&12\catcode `\#12\catcode `\^12\catcode `\_12\catcode `\%12\relax}%
\providecommand \@@startlink[1]{}%
\providecommand \@@endlink[0]{}%
\providecommand \url  [0]{\begingroup\@sanitize@url \@url }%
\providecommand \@url [1]{\endgroup\@href {#1}{\urlprefix }}%
\providecommand \urlprefix  [0]{URL }%
\providecommand \Eprint [0]{\href }%
\providecommand \doibase [0]{http://dx.doi.org/}%
\providecommand \selectlanguage [0]{\@gobble}%
\providecommand \bibinfo  [0]{\@secondoftwo}%
\providecommand \bibfield  [0]{\@secondoftwo}%
\providecommand \translation [1]{[#1]}%
\providecommand \BibitemOpen [0]{}%
\providecommand \bibitemStop [0]{}%
\providecommand \bibitemNoStop [0]{.\EOS\space}%
\providecommand \EOS [0]{\spacefactor3000\relax}%
\providecommand \BibitemShut  [1]{\csname bibitem#1\endcsname}%
\let\auto@bib@innerbib\@empty
%</preamble>
\bibitem [{\citenamefont {L{\"o}hneysen}\ \emph {et~al.}(2007)\citenamefont
  {L{\"o}hneysen}, \citenamefont {Rosch}, \citenamefont {Vojta},\ and\
  \citenamefont {W{\"o}lfle}}]{lohneysen2007}%
  \BibitemOpen
  \bibfield  {author} {\bibinfo {author} {\bibfnamefont {v.~Hilbert}\
  \bibnamefont {L{\"o}hneysen}}, \bibinfo {author} {\bibfnamefont {Achim}\
  \bibnamefont {Rosch}}, \bibinfo {author} {\bibfnamefont {Matthias}\
  \bibnamefont {Vojta}}, \ and\ \bibinfo {author} {\bibfnamefont {Peter}\
  \bibnamefont {W{\"o}lfle}},\ }\bibfield  {title} {{\selectlanguage
  {english}\enquote {\bibinfo {title} {Fermi-liquid instabilities at magnetic
  quantum phase transitions},}\ }}\href {\doibase 10.1103/RevModPhys.79.1015}
  {\bibfield  {journal} {\bibinfo  {journal} {Reviews of Modern Physics}\
  }\textbf {\bibinfo {volume} {79}},\ \bibinfo {pages} {1015--1075} (\bibinfo
  {year} {2007})}\BibitemShut {NoStop}%
\bibitem [{\citenamefont {Senthil}\ \emph {et~al.}(2004)\citenamefont
  {Senthil}, \citenamefont {Vojta},\ and\ \citenamefont
  {Sachdev}}]{senthil2004}%
  \BibitemOpen
  \bibfield  {author} {\bibinfo {author} {\bibfnamefont {T.}~\bibnamefont
  {Senthil}}, \bibinfo {author} {\bibfnamefont {M.}~\bibnamefont {Vojta}}, \
  and\ \bibinfo {author} {\bibfnamefont {S.}~\bibnamefont {Sachdev}},\
  }\bibfield  {title} {{\selectlanguage {english}\enquote {\bibinfo {title}
  {Weak magnetism and non-{{Fermi}} liquids near heavy-fermion critical
  points},}\ }}\href {\doibase 10.1103/PhysRevB.69.035111} {\bibfield
  {journal} {\bibinfo  {journal} {Physical Review B}\ }\textbf {\bibinfo
  {volume} {69}},\ \bibinfo {pages} {035111} (\bibinfo {year}
  {2004})}\BibitemShut {NoStop}%
\bibitem [{\citenamefont {Si}\ and\ \citenamefont {Steglich}(2010)}]{si2010}%
  \BibitemOpen
  \bibfield  {author} {\bibinfo {author} {\bibfnamefont {Q.}~\bibnamefont
  {Si}}\ and\ \bibinfo {author} {\bibfnamefont {F.}~\bibnamefont {Steglich}},\
  }\bibfield  {title} {{\selectlanguage {english}\enquote {\bibinfo {title}
  {Heavy {{Fermions}} and {{Quantum Phase Transitions}}},}\ }}\href {\doibase
  10.1126/science.1191195} {\bibfield  {journal} {\bibinfo  {journal}
  {Science}\ }\textbf {\bibinfo {volume} {329}},\ \bibinfo {pages} {1161--1166}
  (\bibinfo {year} {2010})}\BibitemShut {NoStop}%
\bibitem [{\citenamefont {Vojta}(2003)}]{2003_Vojta_RPP}%
  \BibitemOpen
  \bibfield  {author} {\bibinfo {author} {\bibfnamefont {M.}~\bibnamefont
  {Vojta}},\ }\bibfield  {title} {\enquote {\bibinfo {title} {Quantum phase
  transitions},}\ }\href@noop {} {\bibfield  {journal} {\bibinfo  {journal}
  {Rep. Prog. Phys.}\ }\textbf {\bibinfo {volume} {66}},\ \bibinfo {pages}
  {2069} (\bibinfo {year} {2003})}\BibitemShut {NoStop}%
\bibitem [{\citenamefont {Vojta}(2018)}]{2018_Vojta_RepProgPhysb}%
  \BibitemOpen
  \bibfield  {author} {\bibinfo {author} {\bibfnamefont {M.}~\bibnamefont
  {Vojta}},\ }\bibfield  {title} {\enquote {\bibinfo {title} {Frustration and
  quantum criticality},}\ }\href {\doibase 10.1088/1361-6633/aab6be} {\bibfield
   {journal} {\bibinfo  {journal} {Rep. Prog. Phys.}\ }\textbf {\bibinfo
  {volume} {81}},\ \bibinfo {pages} {064501} (\bibinfo {year}
  {2018})}\BibitemShut {NoStop}%
\bibitem [{\citenamefont {Brando}\ \emph {et~al.}(2016)\citenamefont {Brando},
  \citenamefont {Belitz}, \citenamefont {Grosche},\ and\ \citenamefont
  {Kirkpatrick}}]{brando2016}%
  \BibitemOpen
  \bibfield  {author} {\bibinfo {author} {\bibfnamefont {M.}~\bibnamefont
  {Brando}}, \bibinfo {author} {\bibfnamefont {D.}~\bibnamefont {Belitz}},
  \bibinfo {author} {\bibfnamefont {F.~M.}\ \bibnamefont {Grosche}}, \ and\
  \bibinfo {author} {\bibfnamefont {T.~R.}\ \bibnamefont {Kirkpatrick}},\
  }\bibfield  {title} {{\selectlanguage {english}\enquote {\bibinfo {title}
  {Metallic quantum ferromagnets},}\ }}\href {\doibase
  10.1103/RevModPhys.88.025006} {\bibfield  {journal} {\bibinfo  {journal}
  {Reviews of Modern Physics}\ }\textbf {\bibinfo {volume} {88}},\ \bibinfo
  {pages} {025006} (\bibinfo {year} {2016})}\BibitemShut {NoStop}%
\bibitem [{\citenamefont {Pfleiderer}\ \emph {et~al.}(2001)\citenamefont
  {Pfleiderer}, \citenamefont {Julian},\ and\ \citenamefont
  {Lonzarich}}]{2001_Pfleiderer_Naturec}%
  \BibitemOpen
  \bibfield  {author} {\bibinfo {author} {\bibfnamefont {C.}~\bibnamefont
  {Pfleiderer}}, \bibinfo {author} {\bibfnamefont {S.~R.}\ \bibnamefont
  {Julian}}, \ and\ \bibinfo {author} {\bibfnamefont {G.~G.}\ \bibnamefont
  {Lonzarich}},\ }\bibfield  {title} {\enquote {\bibinfo {title}
  {Non-{{Fermi-liquid}} nature of the normal state of itinerant-electron
  ferromagnets},}\ }\href {\doibase 10.1038/35106527} {\bibfield  {journal}
  {\bibinfo  {journal} {Nature}\ }\textbf {\bibinfo {volume} {414}},\ \bibinfo
  {pages} {427} (\bibinfo {year} {2001})}\BibitemShut {NoStop}%
\bibitem [{\citenamefont {Grosche}\ \emph {et~al.}(1995)\citenamefont
  {Grosche}, \citenamefont {Pfleiderer}, \citenamefont {McMullan},
  \citenamefont {Lonzarich},\ and\ \citenamefont
  {Bernhoeft}}]{1995_Grosche_PhysicaB}%
  \BibitemOpen
  \bibfield  {author} {\bibinfo {author} {\bibfnamefont {F.M.}\ \bibnamefont
  {Grosche}}, \bibinfo {author} {\bibfnamefont {C.}~\bibnamefont {Pfleiderer}},
  \bibinfo {author} {\bibfnamefont {G.J.}\ \bibnamefont {McMullan}}, \bibinfo
  {author} {\bibfnamefont {G.G.}\ \bibnamefont {Lonzarich}}, \ and\ \bibinfo
  {author} {\bibfnamefont {N.R.}\ \bibnamefont {Bernhoeft}},\ }\bibfield
  {title} {\enquote {\bibinfo {title} {Critical behaviour of {{ZrZn$_2$}}},}\
  }\href {\doibase 10.1016/0921-4526(94)00356-Z} {\bibfield  {journal}
  {\bibinfo  {journal} {Physica B: Condensed Matter}\ }\textbf {\bibinfo
  {volume} {206--207}},\ \bibinfo {pages} {20} (\bibinfo {year}
  {1995})}\BibitemShut {NoStop}%
\bibitem [{\citenamefont {Uhlarz}\ \emph {et~al.}(2004)\citenamefont {Uhlarz},
  \citenamefont {Pfleiderer},\ and\ \citenamefont {Hayden}}]{uhlarz2004}%
  \BibitemOpen
  \bibfield  {author} {\bibinfo {author} {\bibfnamefont {M.}~\bibnamefont
  {Uhlarz}}, \bibinfo {author} {\bibfnamefont {C.}~\bibnamefont {Pfleiderer}},
  \ and\ \bibinfo {author} {\bibfnamefont {S.~M.}\ \bibnamefont {Hayden}},\
  }\bibfield  {title} {{\selectlanguage {english}\enquote {\bibinfo {title}
  {Quantum {{Phase Transitions}} in the {{Itinerant Ferromagnet ZrZn$_2$}}},}\
  }}\href {\doibase 10.1103/PhysRevLett.93.256404} {\bibfield  {journal}
  {\bibinfo  {journal} {Physical Review Letters}\ }\textbf {\bibinfo {volume}
  {93}},\ \bibinfo {pages} {256404} (\bibinfo {year} {2004})}\BibitemShut
  {NoStop}%
\bibitem [{\citenamefont {Saxena}\ \emph {et~al.}(2000)\citenamefont {Saxena},
  \citenamefont {Agarwal}, \citenamefont {Ahilan}, \citenamefont {Grosche},
  \citenamefont {Haselwimmer}, \citenamefont {Steiner}, \citenamefont {Pugh},
  \citenamefont {Walker}, \citenamefont {Julian}, \citenamefont {Monthoux},
  \citenamefont {Lonzarich}, \citenamefont {Huxley}, \citenamefont {Sheikin},
  \citenamefont {Braithwaite},\ and\ \citenamefont {Flouquet}}]{saxena2000}%
  \BibitemOpen
  \bibfield  {author} {\bibinfo {author} {\bibfnamefont {S.~S.}\ \bibnamefont
  {Saxena}}, \bibinfo {author} {\bibfnamefont {P.}~\bibnamefont {Agarwal}},
  \bibinfo {author} {\bibfnamefont {K.}~\bibnamefont {Ahilan}}, \bibinfo
  {author} {\bibfnamefont {F.~M.}\ \bibnamefont {Grosche}}, \bibinfo {author}
  {\bibfnamefont {R.~K.~W.}\ \bibnamefont {Haselwimmer}}, \bibinfo {author}
  {\bibfnamefont {M.~J.}\ \bibnamefont {Steiner}}, \bibinfo {author}
  {\bibfnamefont {E.}~\bibnamefont {Pugh}}, \bibinfo {author} {\bibfnamefont
  {I.~R.}\ \bibnamefont {Walker}}, \bibinfo {author} {\bibfnamefont {S.~R.}\
  \bibnamefont {Julian}}, \bibinfo {author} {\bibfnamefont {P.}~\bibnamefont
  {Monthoux}}, \bibinfo {author} {\bibfnamefont {G.~G.}\ \bibnamefont
  {Lonzarich}}, \bibinfo {author} {\bibfnamefont {A.}~\bibnamefont {Huxley}},
  \bibinfo {author} {\bibfnamefont {I.}~\bibnamefont {Sheikin}}, \bibinfo
  {author} {\bibfnamefont {D.}~\bibnamefont {Braithwaite}}, \ and\ \bibinfo
  {author} {\bibfnamefont {J.}~\bibnamefont {Flouquet}},\ }\bibfield  {title}
  {{\selectlanguage {english}\enquote {\bibinfo {title} {Superconductivity on
  the border of itinerant-electron ferromagnetism in {{UGe$_2$}}},}\ }}\href
  {\doibase 10.1038/35020500} {\bibfield  {journal} {\bibinfo  {journal}
  {Nature}\ }\textbf {\bibinfo {volume} {406}},\ \bibinfo {pages} {587--592}
  (\bibinfo {year} {2000})}\BibitemShut {NoStop}%
\bibitem [{\citenamefont {Thessieu}\ \emph {et~al.}(1997)\citenamefont
  {Thessieu}, \citenamefont {Pfleiderer}, \citenamefont {Stepanov},\ and\
  \citenamefont {Flouquet}}]{1997_Thessieu_JPhysCondMatter}%
  \BibitemOpen
  \bibfield  {author} {\bibinfo {author} {\bibfnamefont {C.}~\bibnamefont
  {Thessieu}}, \bibinfo {author} {\bibfnamefont {C.}~\bibnamefont
  {Pfleiderer}}, \bibinfo {author} {\bibfnamefont {A.~N.}\ \bibnamefont
  {Stepanov}}, \ and\ \bibinfo {author} {\bibfnamefont {J.}~\bibnamefont
  {Flouquet}},\ }\bibfield  {title} {\enquote {\bibinfo {title} {Field
  dependence of the magnetic quantum phase transition in {{MnSi}}},}\
  }\href@noop {} {\bibfield  {journal} {\bibinfo  {journal} {J. Phys. Cond.
  Matter}\ }\textbf {\bibinfo {volume} {9}},\ \bibinfo {pages} {6677} (\bibinfo
  {year} {1997})}\BibitemShut {NoStop}%
\bibitem [{\citenamefont {Perry}\ \emph {et~al.}(2001)\citenamefont {Perry},
  \citenamefont {Galvin}, \citenamefont {Grigera}, \citenamefont {Capogna},
  \citenamefont {Schofield}, \citenamefont {Mackenzie}, \citenamefont {Chiao},
  \citenamefont {Julian}, \citenamefont {Ikeda}, \citenamefont {Nakatsuji},
  \citenamefont {Maeno},\ and\ \citenamefont {Pfleiderer}}]{2001_Perry_PRL}%
  \BibitemOpen
  \bibfield  {author} {\bibinfo {author} {\bibfnamefont {R.~S.}\ \bibnamefont
  {Perry}}, \bibinfo {author} {\bibfnamefont {L.~M.}\ \bibnamefont {Galvin}},
  \bibinfo {author} {\bibfnamefont {S.~A.}\ \bibnamefont {Grigera}}, \bibinfo
  {author} {\bibfnamefont {L.}~\bibnamefont {Capogna}}, \bibinfo {author}
  {\bibfnamefont {A.~J.}\ \bibnamefont {Schofield}}, \bibinfo {author}
  {\bibfnamefont {A.~P.}\ \bibnamefont {Mackenzie}}, \bibinfo {author}
  {\bibfnamefont {M.}~\bibnamefont {Chiao}}, \bibinfo {author} {\bibfnamefont
  {S.~R.}\ \bibnamefont {Julian}}, \bibinfo {author} {\bibfnamefont {S.~I.}\
  \bibnamefont {Ikeda}}, \bibinfo {author} {\bibfnamefont {S.}~\bibnamefont
  {Nakatsuji}}, \bibinfo {author} {\bibfnamefont {Y.}~\bibnamefont {Maeno}}, \
  and\ \bibinfo {author} {\bibfnamefont {C.}~\bibnamefont {Pfleiderer}},\
  }\bibfield  {title} {\enquote {\bibinfo {title} {Metamagnetism and critical
  fluctuations in high quality single crystals of the bilayer ruthenate
  {Sr$_3$Ru$_2$O$_7$}},}\ }\href {\doibase 10.1103/PhysRevLett.86.2661}
  {\bibfield  {journal} {\bibinfo  {journal} {Phys. Rev. Lett.}\ }\textbf
  {\bibinfo {volume} {86}},\ \bibinfo {pages} {2661} (\bibinfo {year}
  {2001})}\BibitemShut {NoStop}%
\bibitem [{\citenamefont {Paglione}\ \emph {et~al.}(2003)\citenamefont
  {Paglione}, \citenamefont {Tanatar}, \citenamefont {Hawthorn}, \citenamefont
  {Boaknin}, \citenamefont {Hill}, \citenamefont {Ronning}, \citenamefont
  {Sutherland}, \citenamefont {Taillefer}, \citenamefont {Petrovic},\ and\
  \citenamefont {Canfield}}]{paglione2003}%
  \BibitemOpen
  \bibfield  {author} {\bibinfo {author} {\bibfnamefont {J.}~\bibnamefont
  {Paglione}}, \bibinfo {author} {\bibfnamefont {M.~A.}\ \bibnamefont
  {Tanatar}}, \bibinfo {author} {\bibfnamefont {D.~G.}\ \bibnamefont
  {Hawthorn}}, \bibinfo {author} {\bibfnamefont {Etienne}\ \bibnamefont
  {Boaknin}}, \bibinfo {author} {\bibfnamefont {R.~W.}\ \bibnamefont {Hill}},
  \bibinfo {author} {\bibfnamefont {F.}~\bibnamefont {Ronning}}, \bibinfo
  {author} {\bibfnamefont {M.}~\bibnamefont {Sutherland}}, \bibinfo {author}
  {\bibfnamefont {Louis}\ \bibnamefont {Taillefer}}, \bibinfo {author}
  {\bibfnamefont {C.}~\bibnamefont {Petrovic}}, \ and\ \bibinfo {author}
  {\bibfnamefont {P.~C.}\ \bibnamefont {Canfield}},\ }\bibfield  {title}
  {{\selectlanguage {english}\enquote {\bibinfo {title} {Field-{{Induced
  Quantum Critical Point}} in {{CeCoIn$_5$}}},}\ }}\href {\doibase
  10.1103/PhysRevLett.91.246405} {\bibfield  {journal} {\bibinfo  {journal}
  {Physical Review Letters}\ }\textbf {\bibinfo {volume} {91}},\ \bibinfo
  {pages} {246405} (\bibinfo {year} {2003})}\BibitemShut {NoStop}%
\bibitem [{\citenamefont {Gegenwart}\ \emph {et~al.}(2002)\citenamefont
  {Gegenwart}, \citenamefont {Custers}, \citenamefont {Geibel}, \citenamefont
  {Neumaier}, \citenamefont {Tayama}, \citenamefont {Tenya}, \citenamefont
  {Trovarelli},\ and\ \citenamefont {Steglich}}]{gegenwart2002}%
  \BibitemOpen
  \bibfield  {author} {\bibinfo {author} {\bibfnamefont {P.}~\bibnamefont
  {Gegenwart}}, \bibinfo {author} {\bibfnamefont {J.}~\bibnamefont {Custers}},
  \bibinfo {author} {\bibfnamefont {C.}~\bibnamefont {Geibel}}, \bibinfo
  {author} {\bibfnamefont {K.}~\bibnamefont {Neumaier}}, \bibinfo {author}
  {\bibfnamefont {T.}~\bibnamefont {Tayama}}, \bibinfo {author} {\bibfnamefont
  {K.}~\bibnamefont {Tenya}}, \bibinfo {author} {\bibfnamefont
  {O.}~\bibnamefont {Trovarelli}}, \ and\ \bibinfo {author} {\bibfnamefont
  {F.}~\bibnamefont {Steglich}},\ }\bibfield  {title} {{\selectlanguage
  {english}\enquote {\bibinfo {title} {Magnetic-{{Field Induced Quantum
  Critical Point}} in {{YbRh$_2$Si$_2$}}},}\ }}\href {\doibase
  10.1103/PhysRevLett.89.056402} {\bibfield  {journal} {\bibinfo  {journal}
  {Physical Review Letters}\ }\textbf {\bibinfo {volume} {89}},\ \bibinfo
  {pages} {056402} (\bibinfo {year} {2002})}\BibitemShut {NoStop}%
\bibitem [{\citenamefont {Fuchs}\ \emph {et~al.}(2014)\citenamefont {Fuchs},
  \citenamefont {Wissinger}, \citenamefont {Schmalian}, \citenamefont {Huang},
  \citenamefont {Fromknecht}, \citenamefont {Schneider},\ and\ \citenamefont
  {v.~L{\"o}hneysen}}]{fuchs2014}%
  \BibitemOpen
  \bibfield  {author} {\bibinfo {author} {\bibfnamefont {D.}~\bibnamefont
  {Fuchs}}, \bibinfo {author} {\bibfnamefont {M.}~\bibnamefont {Wissinger}},
  \bibinfo {author} {\bibfnamefont {J.}~\bibnamefont {Schmalian}}, \bibinfo
  {author} {\bibfnamefont {C.-L.}\ \bibnamefont {Huang}}, \bibinfo {author}
  {\bibfnamefont {R.}~\bibnamefont {Fromknecht}}, \bibinfo {author}
  {\bibfnamefont {R.}~\bibnamefont {Schneider}}, \ and\ \bibinfo {author}
  {\bibfnamefont {H.}~\bibnamefont {v.~L{\"o}hneysen}},\ }\bibfield  {title}
  {{\selectlanguage {english}\enquote {\bibinfo {title} {Critical scaling
  analysis of the itinerant ferromagnet {Sr$_{1-x}$Ca$_x$RuO$_3$}},}\ }}\href
  {\doibase 10.1103/PhysRevB.89.174405} {\bibfield  {journal} {\bibinfo
  {journal} {Physical Review B}\ }\textbf {\bibinfo {volume} {89}},\ \bibinfo
  {pages} {174405} (\bibinfo {year} {2014})}\BibitemShut {NoStop}%
\bibitem [{\citenamefont {van~der Marel}\ \emph {et~al.}(2003)\citenamefont
  {van~der Marel}, \citenamefont {Molegraaf}, \citenamefont {Zaanen},
  \citenamefont {Nussinov}, \citenamefont {Carbone}, \citenamefont
  {Damascelli}, \citenamefont {Eisaki}, \citenamefont {Greven}, \citenamefont
  {Kes},\ and\ \citenamefont {Li}}]{marel2003}%
  \BibitemOpen
  \bibfield  {author} {\bibinfo {author} {\bibfnamefont {D.}~\bibnamefont
  {van~der Marel}}, \bibinfo {author} {\bibfnamefont {H.~J.~A.}\ \bibnamefont
  {Molegraaf}}, \bibinfo {author} {\bibfnamefont {J.}~\bibnamefont {Zaanen}},
  \bibinfo {author} {\bibfnamefont {Z.}~\bibnamefont {Nussinov}}, \bibinfo
  {author} {\bibfnamefont {F.}~\bibnamefont {Carbone}}, \bibinfo {author}
  {\bibfnamefont {A.}~\bibnamefont {Damascelli}}, \bibinfo {author}
  {\bibfnamefont {H.}~\bibnamefont {Eisaki}}, \bibinfo {author} {\bibfnamefont
  {M.}~\bibnamefont {Greven}}, \bibinfo {author} {\bibfnamefont {P.~H.}\
  \bibnamefont {Kes}}, \ and\ \bibinfo {author} {\bibfnamefont
  {M.}~\bibnamefont {Li}},\ }\bibfield  {title} {{\selectlanguage
  {english}\enquote {\bibinfo {title} {Quantum critical behaviour in a
  high-{{Tc}} superconductor},}\ }}\href {\doibase 10.1038/nature01978}
  {\bibfield  {journal} {\bibinfo  {journal} {Nature}\ }\textbf {\bibinfo
  {volume} {425}},\ \bibinfo {pages} {271--274} (\bibinfo {year}
  {2003})}\BibitemShut {NoStop}%
\bibitem [{\citenamefont {Belitz}\ \emph {et~al.}(2005)\citenamefont {Belitz},
  \citenamefont {Kirkpatrick},\ and\ \citenamefont
  {Rollb{\"u}hler}}]{belitz2005}%
  \BibitemOpen
  \bibfield  {author} {\bibinfo {author} {\bibfnamefont {D.}~\bibnamefont
  {Belitz}}, \bibinfo {author} {\bibfnamefont {T.~R.}\ \bibnamefont
  {Kirkpatrick}}, \ and\ \bibinfo {author} {\bibfnamefont {J{\"o}rg}\
  \bibnamefont {Rollb{\"u}hler}},\ }\bibfield  {title} {{\selectlanguage
  {english}\enquote {\bibinfo {title} {Tricritical {{Behavior}} in {{Itinerant
  Quantum Ferromagnets}}},}\ }}\href {\doibase 10.1103/PhysRevLett.94.247205}
  {\bibfield  {journal} {\bibinfo  {journal} {Physical Review Letters}\
  }\textbf {\bibinfo {volume} {94}},\ \bibinfo {pages} {247205} (\bibinfo
  {year} {2005})}\BibitemShut {NoStop}%
\bibitem [{\citenamefont
  {Pfleiderer}(2005)}]{2005_Pfleiderer_JPhysCondensMatterd}%
  \BibitemOpen
  \bibfield  {author} {\bibinfo {author} {\bibfnamefont {C.}~\bibnamefont
  {Pfleiderer}},\ }\bibfield  {title} {\enquote {\bibinfo {title} {Why first
  order quantum phase transitions are interesting},}\ }\href {\doibase
  10.1088/0953-8984/17/11/031} {\bibfield  {journal} {\bibinfo  {journal} {J.
  Phys.: Condens. Matter}\ }\textbf {\bibinfo {volume} {17}},\ \bibinfo {pages}
  {S987} (\bibinfo {year} {2005})}\BibitemShut {NoStop}%
\bibitem [{\citenamefont {Brando}\ \emph {et~al.}(2008)\citenamefont {Brando},
  \citenamefont {Duncan}, \citenamefont {{Moroni-Klementowicz}}, \citenamefont
  {Albrecht}, \citenamefont {Gr{\"u}ner}, \citenamefont {Ballou},\ and\
  \citenamefont {Grosche}}]{2008_Brando_PhysRevLettb}%
  \BibitemOpen
  \bibfield  {author} {\bibinfo {author} {\bibfnamefont {M.}~\bibnamefont
  {Brando}}, \bibinfo {author} {\bibfnamefont {W.~J.}\ \bibnamefont {Duncan}},
  \bibinfo {author} {\bibfnamefont {D.}~\bibnamefont {{Moroni-Klementowicz}}},
  \bibinfo {author} {\bibfnamefont {C.}~\bibnamefont {Albrecht}}, \bibinfo
  {author} {\bibfnamefont {D.}~\bibnamefont {Gr{\"u}ner}}, \bibinfo {author}
  {\bibfnamefont {R.}~\bibnamefont {Ballou}}, \ and\ \bibinfo {author}
  {\bibfnamefont {F.~M.}\ \bibnamefont {Grosche}},\ }\bibfield  {title}
  {\enquote {\bibinfo {title} {Logarithmic {{Fermi-Liquid Breakdown}} in
  {{NbFe$_2$}}},}\ }\href {\doibase 10.1103/PhysRevLett.101.026401} {\bibfield
  {journal} {\bibinfo  {journal} {Phys. Rev. Lett.}\ }\textbf {\bibinfo
  {volume} {101}},\ \bibinfo {pages} {026401} (\bibinfo {year}
  {2008})}\BibitemShut {NoStop}%
\bibitem [{\citenamefont {{Abdul-Jabbar}}\ \emph {et~al.}(2015)\citenamefont
  {{Abdul-Jabbar}}, \citenamefont {Sokolov}, \citenamefont {O'Neill},
  \citenamefont {Stock}, \citenamefont {Wermeille}, \citenamefont {Demmel},
  \citenamefont {Kr{\"u}ger}, \citenamefont {Green}, \citenamefont
  {{L{\'e}vy-Bertrand}}, \citenamefont {Grenier},\ and\ \citenamefont
  {Huxley}}]{2015_Abdul_Jabbar_NaturePhysa}%
  \BibitemOpen
  \bibfield  {author} {\bibinfo {author} {\bibfnamefont {G.}~\bibnamefont
  {{Abdul-Jabbar}}}, \bibinfo {author} {\bibfnamefont {D.~A.}\ \bibnamefont
  {Sokolov}}, \bibinfo {author} {\bibfnamefont {C.~D.}\ \bibnamefont
  {O'Neill}}, \bibinfo {author} {\bibfnamefont {C.}~\bibnamefont {Stock}},
  \bibinfo {author} {\bibfnamefont {D.}~\bibnamefont {Wermeille}}, \bibinfo
  {author} {\bibfnamefont {F.}~\bibnamefont {Demmel}}, \bibinfo {author}
  {\bibfnamefont {F.}~\bibnamefont {Kr{\"u}ger}}, \bibinfo {author}
  {\bibfnamefont {A.~G.}\ \bibnamefont {Green}}, \bibinfo {author}
  {\bibfnamefont {F.}~\bibnamefont {{L{\'e}vy-Bertrand}}}, \bibinfo {author}
  {\bibfnamefont {B.}~\bibnamefont {Grenier}}, \ and\ \bibinfo {author}
  {\bibfnamefont {A.~D.}\ \bibnamefont {Huxley}},\ }\bibfield  {title}
  {\enquote {\bibinfo {title} {Modulated magnetism in {{PrPtAl}}},}\ }\href
  {\doibase 10.1038/nphys3238} {\bibfield  {journal} {\bibinfo  {journal}
  {Nature Phys}\ }\textbf {\bibinfo {volume} {11}},\ \bibinfo {pages} {321}
  (\bibinfo {year} {2015})}\BibitemShut {NoStop}%
\bibitem [{\citenamefont {Friedemann}\ \emph {et~al.}(2018)\citenamefont
  {Friedemann}, \citenamefont {Duncan}, \citenamefont {Hirschberger},
  \citenamefont {Bauer}, \citenamefont {K{\"u}chler}, \citenamefont {Neubauer},
  \citenamefont {Brando}, \citenamefont {Pfleiderer},\ and\ \citenamefont
  {Grosche}}]{2018_Friedemann_NaturePhys}%
  \BibitemOpen
  \bibfield  {author} {\bibinfo {author} {\bibfnamefont {S.}~\bibnamefont
  {Friedemann}}, \bibinfo {author} {\bibfnamefont {W.~J.}\ \bibnamefont
  {Duncan}}, \bibinfo {author} {\bibfnamefont {M.}~\bibnamefont
  {Hirschberger}}, \bibinfo {author} {\bibfnamefont {T.~W.}\ \bibnamefont
  {Bauer}}, \bibinfo {author} {\bibfnamefont {R.}~\bibnamefont {K{\"u}chler}},
  \bibinfo {author} {\bibfnamefont {A.}~\bibnamefont {Neubauer}}, \bibinfo
  {author} {\bibfnamefont {M.}~\bibnamefont {Brando}}, \bibinfo {author}
  {\bibfnamefont {C.}~\bibnamefont {Pfleiderer}}, \ and\ \bibinfo {author}
  {\bibfnamefont {F.~M.}\ \bibnamefont {Grosche}},\ }\bibfield  {title}
  {\enquote {\bibinfo {title} {Quantum tricritical points in {{NbFe$_2$}}},}\
  }\href {\doibase 10.1038/nphys4242} {\bibfield  {journal} {\bibinfo
  {journal} {Nature Physics}\ }\textbf {\bibinfo {volume} {14}},\ \bibinfo
  {pages} {62} (\bibinfo {year} {2018})}\BibitemShut {NoStop}%
\bibitem [{\citenamefont {Aoki}\ \emph {et~al.}(2001)\citenamefont {Aoki},
  \citenamefont {Huxley}, \citenamefont {Ressouche}, \citenamefont
  {Braithwaite}, \citenamefont {Flouquet}, \citenamefont {Brison},
  \citenamefont {Lhotel},\ and\ \citenamefont {Paulsen}}]{aoki2001}%
  \BibitemOpen
  \bibfield  {author} {\bibinfo {author} {\bibfnamefont {Dai}\ \bibnamefont
  {Aoki}}, \bibinfo {author} {\bibfnamefont {Andrew}\ \bibnamefont {Huxley}},
  \bibinfo {author} {\bibfnamefont {Eric}\ \bibnamefont {Ressouche}}, \bibinfo
  {author} {\bibfnamefont {Daniel}\ \bibnamefont {Braithwaite}}, \bibinfo
  {author} {\bibfnamefont {Jacques}\ \bibnamefont {Flouquet}}, \bibinfo
  {author} {\bibfnamefont {Jean-Pascal}\ \bibnamefont {Brison}}, \bibinfo
  {author} {\bibfnamefont {Elsa}\ \bibnamefont {Lhotel}}, \ and\ \bibinfo
  {author} {\bibfnamefont {Carley}\ \bibnamefont {Paulsen}},\ }\bibfield
  {title} {{\selectlanguage {english}\enquote {\bibinfo {title} {Coexistence of
  superconductivity and ferromagnetism in {{URhGe}}},}\ }}\href {\doibase
  10.1038/35098048} {\bibfield  {journal} {\bibinfo  {journal} {Nature}\
  }\textbf {\bibinfo {volume} {413}},\ \bibinfo {pages} {613--616} (\bibinfo
  {year} {2001})}\BibitemShut {NoStop}%
\bibitem [{\citenamefont {Huy}\ \emph {et~al.}(2007)\citenamefont {Huy},
  \citenamefont {Gasparini}, \citenamefont {{de Nijs}}, \citenamefont {Huang},
  \citenamefont {Klaasse}, \citenamefont {Gortenmulder}, \citenamefont {{de
  Visser}}, \citenamefont {Hamann}, \citenamefont {G{\"o}rlach},\ and\
  \citenamefont {v.~L{\"o}hneysen}}]{huy2007}%
  \BibitemOpen
  \bibfield  {author} {\bibinfo {author} {\bibfnamefont {N.~T.}\ \bibnamefont
  {Huy}}, \bibinfo {author} {\bibfnamefont {A.}~\bibnamefont {Gasparini}},
  \bibinfo {author} {\bibfnamefont {D.~E.}\ \bibnamefont {{de Nijs}}}, \bibinfo
  {author} {\bibfnamefont {Y.}~\bibnamefont {Huang}}, \bibinfo {author}
  {\bibfnamefont {J.~C.~P.}\ \bibnamefont {Klaasse}}, \bibinfo {author}
  {\bibfnamefont {T.}~\bibnamefont {Gortenmulder}}, \bibinfo {author}
  {\bibfnamefont {A.}~\bibnamefont {{de Visser}}}, \bibinfo {author}
  {\bibfnamefont {A.}~\bibnamefont {Hamann}}, \bibinfo {author} {\bibfnamefont
  {T.}~\bibnamefont {G{\"o}rlach}}, \ and\ \bibinfo {author} {\bibfnamefont
  {H.}~\bibnamefont {v.~L{\"o}hneysen}},\ }\bibfield  {title} {{\selectlanguage
  {english}\enquote {\bibinfo {title} {Superconductivity on the {{Border}} of
  {{Weak Itinerant Ferromagnetism}} in {{UCoGe}}},}\ }}\href {\doibase
  10.1103/PhysRevLett.99.067006} {\bibfield  {journal} {\bibinfo  {journal}
  {Physical Review Letters}\ }\textbf {\bibinfo {volume} {99}},\ \bibinfo
  {pages} {067006} (\bibinfo {year} {2007})}\BibitemShut {NoStop}%
\bibitem [{\citenamefont {Aoki}\ \emph {et~al.}(2019)\citenamefont {Aoki},
  \citenamefont {Nakamura}, \citenamefont {Honda}, \citenamefont {Li},
  \citenamefont {Homma}, \citenamefont {Shimizu}, \citenamefont {Sato},
  \citenamefont {Knebel}, \citenamefont {Brison}, \citenamefont {Pourret},
  \citenamefont {Braithwaite}, \citenamefont {Lapertot}, \citenamefont {Niu},
  \citenamefont {Vali{\v s}ka}, \citenamefont {Harima},\ and\ \citenamefont
  {Flouquet}}]{aoki2019}%
  \BibitemOpen
  \bibfield  {author} {\bibinfo {author} {\bibfnamefont {Dai}\ \bibnamefont
  {Aoki}}, \bibinfo {author} {\bibfnamefont {Ai}~\bibnamefont {Nakamura}},
  \bibinfo {author} {\bibfnamefont {Fuminori}\ \bibnamefont {Honda}}, \bibinfo
  {author} {\bibfnamefont {DeXin}\ \bibnamefont {Li}}, \bibinfo {author}
  {\bibfnamefont {Yoshiya}\ \bibnamefont {Homma}}, \bibinfo {author}
  {\bibfnamefont {Yusei}\ \bibnamefont {Shimizu}}, \bibinfo {author}
  {\bibfnamefont {Yoshiki~J.}\ \bibnamefont {Sato}}, \bibinfo {author}
  {\bibfnamefont {Georg}\ \bibnamefont {Knebel}}, \bibinfo {author}
  {\bibfnamefont {Jean-Pascal}\ \bibnamefont {Brison}}, \bibinfo {author}
  {\bibfnamefont {Alexandre}\ \bibnamefont {Pourret}}, \bibinfo {author}
  {\bibfnamefont {Daniel}\ \bibnamefont {Braithwaite}}, \bibinfo {author}
  {\bibfnamefont {Gerard}\ \bibnamefont {Lapertot}}, \bibinfo {author}
  {\bibfnamefont {Qun}\ \bibnamefont {Niu}}, \bibinfo {author} {\bibfnamefont
  {Michal}\ \bibnamefont {Vali{\v s}ka}}, \bibinfo {author} {\bibfnamefont
  {Hisatomo}\ \bibnamefont {Harima}}, \ and\ \bibinfo {author} {\bibfnamefont
  {Jacques}\ \bibnamefont {Flouquet}},\ }\bibfield  {title} {{\selectlanguage
  {english}\enquote {\bibinfo {title} {Unconventional {{Superconductivity}} in
  {{Heavy Fermion UTe$_2$}}},}\ }}\href {\doibase 10.7566/JPSJ.88.043702}
  {\bibfield  {journal} {\bibinfo  {journal} {Journal of the Physical Society
  of Japan}\ }\textbf {\bibinfo {volume} {88}},\ \bibinfo {pages} {043702}
  (\bibinfo {year} {2019})}\BibitemShut {NoStop}%
\bibitem [{\citenamefont {Pfleiderer}(2009)}]{2009_Pfleiderer_RevModPhys}%
  \BibitemOpen
  \bibfield  {author} {\bibinfo {author} {\bibfnamefont {C.}~\bibnamefont
  {Pfleiderer}},\ }\bibfield  {title} {\enquote {\bibinfo {title}
  {Superconducting phases of f -electron compounds},}\ }\href {\doibase
  10.1103/RevModPhys.81.1551} {\bibfield  {journal} {\bibinfo  {journal} {Rev.
  Mod. Phys.}\ }\textbf {\bibinfo {volume} {81}},\ \bibinfo {pages} {1551}
  (\bibinfo {year} {2009})}\BibitemShut {NoStop}%
\bibitem [{\citenamefont {Haslbeck}\ \emph {et~al.}(2019)\citenamefont
  {Haslbeck}, \citenamefont {S{\"a}ubert}, \citenamefont {Seifert},
  \citenamefont {Franz}, \citenamefont {Schulz}, \citenamefont {Heinemann},
  \citenamefont {Keller}, \citenamefont {Das}, \citenamefont {Thompson},
  \citenamefont {Bauer}, \citenamefont {Pfleiderer},\ and\ \citenamefont
  {Janoschek}}]{2019_Haslbeck_PhysRevB}%
  \BibitemOpen
  \bibfield  {author} {\bibinfo {author} {\bibfnamefont {F.}~\bibnamefont
  {Haslbeck}}, \bibinfo {author} {\bibfnamefont {S.}~\bibnamefont
  {S{\"a}ubert}}, \bibinfo {author} {\bibfnamefont {M.}~\bibnamefont
  {Seifert}}, \bibinfo {author} {\bibfnamefont {C.}~\bibnamefont {Franz}},
  \bibinfo {author} {\bibfnamefont {M.}~\bibnamefont {Schulz}}, \bibinfo
  {author} {\bibfnamefont {A.}~\bibnamefont {Heinemann}}, \bibinfo {author}
  {\bibfnamefont {T.}~\bibnamefont {Keller}}, \bibinfo {author} {\bibfnamefont
  {P.}~\bibnamefont {Das}}, \bibinfo {author} {\bibfnamefont {J.~D.}\
  \bibnamefont {Thompson}}, \bibinfo {author} {\bibfnamefont {E.~D.}\
  \bibnamefont {Bauer}}, \bibinfo {author} {\bibfnamefont {C.}~\bibnamefont
  {Pfleiderer}}, \ and\ \bibinfo {author} {\bibfnamefont {M.}~\bibnamefont
  {Janoschek}},\ }\bibfield  {title} {\enquote {\bibinfo {title}
  {Ultrahigh-resolution neutron spectroscopy of low-energy spin dynamics in
  {UGe$_2$}},}\ }\href {\doibase 10.1103/PhysRevB.99.014429} {\bibfield
  {journal} {\bibinfo  {journal} {Phys. Rev. B}\ }\textbf {\bibinfo {volume}
  {99}},\ \bibinfo {pages} {014429} (\bibinfo {year} {2019})}\BibitemShut
  {NoStop}%
\bibitem [{\citenamefont {Miranda}\ and\ \citenamefont
  {Dobrosavljevi{\'c}}(2005)}]{miranda2005}%
  \BibitemOpen
  \bibfield  {author} {\bibinfo {author} {\bibfnamefont {E}~\bibnamefont
  {Miranda}}\ and\ \bibinfo {author} {\bibfnamefont {V}~\bibnamefont
  {Dobrosavljevi{\'c}}},\ }\bibfield  {title} {{\selectlanguage
  {english}\enquote {\bibinfo {title} {Disorder-driven non-{{Fermi}} liquid
  behaviour of correlated electrons},}\ }}\href {\doibase
  10.1088/0034-4885/68/10/R02} {\bibfield  {journal} {\bibinfo  {journal}
  {Reports on Progress in Physics}\ }\textbf {\bibinfo {volume} {68}},\
  \bibinfo {pages} {2337--2408} (\bibinfo {year} {2005})}\BibitemShut {NoStop}%
\bibitem [{\citenamefont {Vojta}(2010)}]{vojta2010}%
  \BibitemOpen
  \bibfield  {author} {\bibinfo {author} {\bibfnamefont {Thomas}\ \bibnamefont
  {Vojta}},\ }\bibfield  {title} {{\selectlanguage {english}\enquote {\bibinfo
  {title} {Quantum {{Griffiths Effects}} and {{Smeared Phase Transitions}} in
  {{Metals}}: {{Theory}} and {{Experiment}}},}\ }}\href {\doibase
  10.1007/s10909-010-0205-4} {\bibfield  {journal} {\bibinfo  {journal}
  {Journal of Low Temperature Physics}\ }\textbf {\bibinfo {volume} {161}},\
  \bibinfo {pages} {299--323} (\bibinfo {year} {2010})}\BibitemShut {NoStop}%
\bibitem [{\citenamefont {Pfleiderer}\ \emph {et~al.}(2010)\citenamefont
  {Pfleiderer}, \citenamefont {B{\"o}ni}, \citenamefont {Franz}, \citenamefont
  {Keller}, \citenamefont {Neubauer}, \citenamefont {Niklowitz}, \citenamefont
  {Schmakat}, \citenamefont {Schulz}, \citenamefont {Huang}, \citenamefont
  {Mydosh}, \citenamefont {Vojta}, \citenamefont {Duncan}, \citenamefont
  {Grosche}, \citenamefont {Brando}, \citenamefont {Deppe}, \citenamefont
  {Geibel}, \citenamefont {Steglich}, \citenamefont {Krimmel},\ and\
  \citenamefont {Loidl}}]{pfleiderer2010}%
  \BibitemOpen
  \bibfield  {author} {\bibinfo {author} {\bibfnamefont {C.}~\bibnamefont
  {Pfleiderer}}, \bibinfo {author} {\bibfnamefont {P.}~\bibnamefont
  {B{\"o}ni}}, \bibinfo {author} {\bibfnamefont {C.}~\bibnamefont {Franz}},
  \bibinfo {author} {\bibfnamefont {T.}~\bibnamefont {Keller}}, \bibinfo
  {author} {\bibfnamefont {A.}~\bibnamefont {Neubauer}}, \bibinfo {author}
  {\bibfnamefont {P.~G.}\ \bibnamefont {Niklowitz}}, \bibinfo {author}
  {\bibfnamefont {P.}~\bibnamefont {Schmakat}}, \bibinfo {author}
  {\bibfnamefont {M.}~\bibnamefont {Schulz}}, \bibinfo {author} {\bibfnamefont
  {Y.-K.}\ \bibnamefont {Huang}}, \bibinfo {author} {\bibfnamefont {J.~A.}\
  \bibnamefont {Mydosh}}, \bibinfo {author} {\bibfnamefont {M.}~\bibnamefont
  {Vojta}}, \bibinfo {author} {\bibfnamefont {W.}~\bibnamefont {Duncan}},
  \bibinfo {author} {\bibfnamefont {F.~M.}\ \bibnamefont {Grosche}}, \bibinfo
  {author} {\bibfnamefont {M.}~\bibnamefont {Brando}}, \bibinfo {author}
  {\bibfnamefont {M.}~\bibnamefont {Deppe}}, \bibinfo {author} {\bibfnamefont
  {C.}~\bibnamefont {Geibel}}, \bibinfo {author} {\bibfnamefont
  {F.}~\bibnamefont {Steglich}}, \bibinfo {author} {\bibfnamefont
  {A.}~\bibnamefont {Krimmel}}, \ and\ \bibinfo {author} {\bibfnamefont
  {A.}~\bibnamefont {Loidl}},\ }\bibfield  {title} {{\selectlanguage
  {english}\enquote {\bibinfo {title} {Search for {{Electronic Phase
  Separation}} at {{Quantum Phase Transitions}}},}\ }}\href {\doibase
  10.1007/s10909-010-0214-3} {\bibfield  {journal} {\bibinfo  {journal}
  {Journal of Low Temperature Physics}\ }\textbf {\bibinfo {volume} {161}},\
  \bibinfo {pages} {167--181} (\bibinfo {year} {2010})}\BibitemShut {NoStop}%
\bibitem [{\citenamefont {Uemura}\ \emph {et~al.}(2007)\citenamefont {Uemura},
  \citenamefont {Goko}, \citenamefont {{Gat-Malureanu}}, \citenamefont {Carlo},
  \citenamefont {Russo}, \citenamefont {Savici}, \citenamefont {Aczel},
  \citenamefont {MacDougall}, \citenamefont {Rodriguez}, \citenamefont {Luke},
  \citenamefont {Dunsiger}, \citenamefont {McCollam}, \citenamefont {Arai},
  \citenamefont {Pfleiderer}, \citenamefont {B{\"o}ni}, \citenamefont
  {Yoshimura}, \citenamefont {{Baggio-Saitovitch}}, \citenamefont {Fontes},
  \citenamefont {Larrea}, \citenamefont {Sushko},\ and\ \citenamefont
  {Sereni}}]{2007_Uemura_NaturePhys}%
  \BibitemOpen
  \bibfield  {author} {\bibinfo {author} {\bibfnamefont {Y.~J.}\ \bibnamefont
  {Uemura}}, \bibinfo {author} {\bibfnamefont {T.}~\bibnamefont {Goko}},
  \bibinfo {author} {\bibfnamefont {I.~M.}\ \bibnamefont {{Gat-Malureanu}}},
  \bibinfo {author} {\bibfnamefont {J.~P.}\ \bibnamefont {Carlo}}, \bibinfo
  {author} {\bibfnamefont {P.~L.}\ \bibnamefont {Russo}}, \bibinfo {author}
  {\bibfnamefont {A.~T.}\ \bibnamefont {Savici}}, \bibinfo {author}
  {\bibfnamefont {A.}~\bibnamefont {Aczel}}, \bibinfo {author} {\bibfnamefont
  {G.~J.}\ \bibnamefont {MacDougall}}, \bibinfo {author} {\bibfnamefont
  {J.~A.}\ \bibnamefont {Rodriguez}}, \bibinfo {author} {\bibfnamefont {G.~M.}\
  \bibnamefont {Luke}}, \bibinfo {author} {\bibfnamefont {S.~R.}\ \bibnamefont
  {Dunsiger}}, \bibinfo {author} {\bibfnamefont {A.}~\bibnamefont {McCollam}},
  \bibinfo {author} {\bibfnamefont {J.}~\bibnamefont {Arai}}, \bibinfo {author}
  {\bibfnamefont {C.}~\bibnamefont {Pfleiderer}}, \bibinfo {author}
  {\bibfnamefont {P.}~\bibnamefont {B{\"o}ni}}, \bibinfo {author}
  {\bibfnamefont {K.}~\bibnamefont {Yoshimura}}, \bibinfo {author}
  {\bibfnamefont {E.}~\bibnamefont {{Baggio-Saitovitch}}}, \bibinfo {author}
  {\bibfnamefont {M.~B.}\ \bibnamefont {Fontes}}, \bibinfo {author}
  {\bibfnamefont {J.}~\bibnamefont {Larrea}}, \bibinfo {author} {\bibfnamefont
  {Y.~V.}\ \bibnamefont {Sushko}}, \ and\ \bibinfo {author} {\bibfnamefont
  {J.}~\bibnamefont {Sereni}},\ }\bibfield  {title} {\enquote {\bibinfo {title}
  {Phase separation and suppression of critical dynamics at quantum phase
  transitions of {{MnSi}} and {Sr$_{1-x}$Ca$_x$RuO$_3$}},}\ }\href {\doibase
  10.1038/nphys488} {\bibfield  {journal} {\bibinfo  {journal} {Nature
  Physics}\ }\textbf {\bibinfo {volume} {3}},\ \bibinfo {pages} {29} (\bibinfo
  {year} {2007})}\BibitemShut {NoStop}%
\bibitem [{\citenamefont {Schmakat}\ \emph {et~al.}(2015)\citenamefont
  {Schmakat}, \citenamefont {Wagner}, \citenamefont {Ritz}, \citenamefont
  {Bauer}, \citenamefont {Brando}, \citenamefont {Deppe}, \citenamefont
  {Duncan}, \citenamefont {Duvinage}, \citenamefont {Franz}, \citenamefont
  {Geibel}, \citenamefont {Grosche}, \citenamefont {Hirschberger},
  \citenamefont {Hradil}, \citenamefont {Meven}, \citenamefont {Neubauer},
  \citenamefont {Schulz}, \citenamefont {Senyshyn}, \citenamefont {S\"ullow},
  \citenamefont {Pedersen}, \citenamefont {B\"oni},\ and\ \citenamefont
  {Pfleiderer}}]{2015_Schmakat_EPJST}%
  \BibitemOpen
  \bibfield  {author} {\bibinfo {author} {\bibfnamefont {P.}~\bibnamefont
  {Schmakat}}, \bibinfo {author} {\bibfnamefont {M.}~\bibnamefont {Wagner}},
  \bibinfo {author} {\bibfnamefont {R.}~\bibnamefont {Ritz}}, \bibinfo {author}
  {\bibfnamefont {A.}~\bibnamefont {Bauer}}, \bibinfo {author} {\bibfnamefont
  {M.}~\bibnamefont {Brando}}, \bibinfo {author} {\bibfnamefont
  {M.}~\bibnamefont {Deppe}}, \bibinfo {author} {\bibfnamefont
  {W.}~\bibnamefont {Duncan}}, \bibinfo {author} {\bibfnamefont
  {C.}~\bibnamefont {Duvinage}}, \bibinfo {author} {\bibfnamefont
  {C.}~\bibnamefont {Franz}}, \bibinfo {author} {\bibfnamefont
  {C.}~\bibnamefont {Geibel}}, \bibinfo {author} {\bibfnamefont {F.M.}\
  \bibnamefont {Grosche}}, \bibinfo {author} {\bibfnamefont {M.}~\bibnamefont
  {Hirschberger}}, \bibinfo {author} {\bibfnamefont {K.}~\bibnamefont
  {Hradil}}, \bibinfo {author} {\bibfnamefont {M.}~\bibnamefont {Meven}},
  \bibinfo {author} {\bibfnamefont {A.}~\bibnamefont {Neubauer}}, \bibinfo
  {author} {\bibfnamefont {M.}~\bibnamefont {Schulz}}, \bibinfo {author}
  {\bibfnamefont {A.}~\bibnamefont {Senyshyn}}, \bibinfo {author}
  {\bibfnamefont {S.}~\bibnamefont {S\"ullow}}, \bibinfo {author}
  {\bibfnamefont {B.}~\bibnamefont {Pedersen}}, \bibinfo {author}
  {\bibfnamefont {P.}~\bibnamefont {B\"oni}}, \ and\ \bibinfo {author}
  {\bibfnamefont {C.}~\bibnamefont {Pfleiderer}},\ }\bibfield  {title}
  {\enquote {\bibinfo {title} {Spin dynamics and spin freezing at ferromagnetic
  quantum phase transitions},}\ }\href {\doibase 10.1140/epjst/e2015-02445-4}
  {\bibfield  {journal} {\bibinfo  {journal} {Eur. Phys. J. Special Topics}\
  }\textbf {\bibinfo {volume} {224}},\ \bibinfo {pages} {1041} (\bibinfo {year}
  {2015})}\BibitemShut {NoStop}%
\bibitem [{\citenamefont {Benka}\ \emph {et~al.}(2022)\citenamefont {Benka},
  \citenamefont {Bauer}, \citenamefont {Schmakat}, \citenamefont {S\"aubert},
  \citenamefont {Seifert}, \citenamefont {Jorba},\ and\ \citenamefont
  {Pfleiderer}}]{2020_Benka_arXiv}%
  \BibitemOpen
  \bibfield  {author} {\bibinfo {author} {\bibfnamefont {Georg}\ \bibnamefont
  {Benka}}, \bibinfo {author} {\bibfnamefont {Andreas}\ \bibnamefont {Bauer}},
  \bibinfo {author} {\bibfnamefont {Philipp}\ \bibnamefont {Schmakat}},
  \bibinfo {author} {\bibfnamefont {Steffen}\ \bibnamefont {S\"aubert}},
  \bibinfo {author} {\bibfnamefont {Marc}\ \bibnamefont {Seifert}}, \bibinfo
  {author} {\bibfnamefont {Pau}\ \bibnamefont {Jorba}}, \ and\ \bibinfo
  {author} {\bibfnamefont {Christian}\ \bibnamefont {Pfleiderer}},\ }\bibfield
  {title} {\enquote {\bibinfo {title} {Interplay of itinerant magnetism and
  spin-glass behavior in {Fe$_{1-x}$Cr$_x$}},}\ }\href {\doibase
  10.1103/PhysRevMaterials.6.044407} {\bibfield  {journal} {\bibinfo  {journal}
  {Phys. Rev. Materials}\ }\textbf {\bibinfo {volume} {6}},\ \bibinfo {pages}
  {044407} (\bibinfo {year} {2022})}\BibitemShut {NoStop}%
\bibitem [{\citenamefont {Niklowitz}\ \emph {et~al.}(2005)\citenamefont
  {Niklowitz}, \citenamefont {Beckers}, \citenamefont {Lonzarich},
  \citenamefont {Knebel}, \citenamefont {Salce}, \citenamefont {Thomasson},
  \citenamefont {Bernhoeft}, \citenamefont {Braithwaite},\ and\ \citenamefont
  {Flouquet}}]{2005_Niklowitz_PhysRevBb}%
  \BibitemOpen
  \bibfield  {author} {\bibinfo {author} {\bibfnamefont {P.~G.}\ \bibnamefont
  {Niklowitz}}, \bibinfo {author} {\bibfnamefont {F.}~\bibnamefont {Beckers}},
  \bibinfo {author} {\bibfnamefont {G.~G.}\ \bibnamefont {Lonzarich}}, \bibinfo
  {author} {\bibfnamefont {G.}~\bibnamefont {Knebel}}, \bibinfo {author}
  {\bibfnamefont {B.}~\bibnamefont {Salce}}, \bibinfo {author} {\bibfnamefont
  {J.}~\bibnamefont {Thomasson}}, \bibinfo {author} {\bibfnamefont
  {N.}~\bibnamefont {Bernhoeft}}, \bibinfo {author} {\bibfnamefont
  {D.}~\bibnamefont {Braithwaite}}, \ and\ \bibinfo {author} {\bibfnamefont
  {J.}~\bibnamefont {Flouquet}},\ }\bibfield  {title} {\enquote {\bibinfo
  {title} {Spin-fluctuation-dominated electrical transport of {Ni$_3$Al} at
  high pressure},}\ }\href {\doibase 10.1103/PhysRevB.72.024424} {\bibfield
  {journal} {\bibinfo  {journal} {Phys. Rev. B}\ }\textbf {\bibinfo {volume}
  {72}},\ \bibinfo {pages} {024424} (\bibinfo {year} {2005})}\BibitemShut
  {NoStop}%
\bibitem [{\citenamefont {Ritz}\ \emph
  {et~al.}(2013{\natexlab{a}})\citenamefont {Ritz}, \citenamefont {Halder},
  \citenamefont {Wagner}, \citenamefont {Franz}, \citenamefont {Bauer},\ and\
  \citenamefont {Pfleiderer}}]{2013_Ritz_Naturea}%
  \BibitemOpen
  \bibfield  {author} {\bibinfo {author} {\bibfnamefont {R.}~\bibnamefont
  {Ritz}}, \bibinfo {author} {\bibfnamefont {M.}~\bibnamefont {Halder}},
  \bibinfo {author} {\bibfnamefont {M.}~\bibnamefont {Wagner}}, \bibinfo
  {author} {\bibfnamefont {C.}~\bibnamefont {Franz}}, \bibinfo {author}
  {\bibfnamefont {A.}~\bibnamefont {Bauer}}, \ and\ \bibinfo {author}
  {\bibfnamefont {C.}~\bibnamefont {Pfleiderer}},\ }\bibfield  {title}
  {\enquote {\bibinfo {title} {Formation of a topological non-{{Fermi}} liquid
  in {{MnSi}}},}\ }\href {\doibase 10.1038/nature12023} {\bibfield  {journal}
  {\bibinfo  {journal} {Nature}\ }\textbf {\bibinfo {volume} {497}},\ \bibinfo
  {pages} {231} (\bibinfo {year} {2013}{\natexlab{a}})}\BibitemShut {NoStop}%
\bibitem [{\citenamefont {Ritz}\ \emph
  {et~al.}(2013{\natexlab{b}})\citenamefont {Ritz}, \citenamefont {Halder},
  \citenamefont {Franz}, \citenamefont {Bauer}, \citenamefont {Wagner},
  \citenamefont {Bamler}, \citenamefont {Rosch},\ and\ \citenamefont
  {Pfleiderer}}]{2013_Ritz_PhysRevBc}%
  \BibitemOpen
  \bibfield  {author} {\bibinfo {author} {\bibfnamefont {R.}~\bibnamefont
  {Ritz}}, \bibinfo {author} {\bibfnamefont {M.}~\bibnamefont {Halder}},
  \bibinfo {author} {\bibfnamefont {C.}~\bibnamefont {Franz}}, \bibinfo
  {author} {\bibfnamefont {A.}~\bibnamefont {Bauer}}, \bibinfo {author}
  {\bibfnamefont {M.}~\bibnamefont {Wagner}}, \bibinfo {author} {\bibfnamefont
  {R.}~\bibnamefont {Bamler}}, \bibinfo {author} {\bibfnamefont
  {A.}~\bibnamefont {Rosch}}, \ and\ \bibinfo {author} {\bibfnamefont
  {C.}~\bibnamefont {Pfleiderer}},\ }\bibfield  {title} {\enquote {\bibinfo
  {title} {Giant generic topological {{Hall}} resistivity of {{MnSi}} under
  pressure},}\ }\href {\doibase 10.1103/PhysRevB.87.134424} {\bibfield
  {journal} {\bibinfo  {journal} {Phys. Rev. B}\ }\textbf {\bibinfo {volume}
  {87}},\ \bibinfo {pages} {134424} (\bibinfo {year}
  {2013}{\natexlab{b}})}\BibitemShut {NoStop}%
\bibitem [{\citenamefont {Gati}\ \emph {et~al.}(2021)\citenamefont {Gati},
  \citenamefont {Wilde}, \citenamefont {Khasanov}, \citenamefont {Xiang},
  \citenamefont {Dissanayake}, \citenamefont {Gupta}, \citenamefont {Matsuda},
  \citenamefont {Ye}, \citenamefont {Haberl}, \citenamefont {Kaluarachchi},
  \citenamefont {McQueeney}, \citenamefont {Kreyssig}, \citenamefont {Bud'ko},\
  and\ \citenamefont {Canfield}}]{2021_Gati_PhysRevB}%
  \BibitemOpen
  \bibfield  {author} {\bibinfo {author} {\bibfnamefont {E.}~\bibnamefont
  {Gati}}, \bibinfo {author} {\bibfnamefont {J.~M.}\ \bibnamefont {Wilde}},
  \bibinfo {author} {\bibfnamefont {R.}~\bibnamefont {Khasanov}}, \bibinfo
  {author} {\bibfnamefont {L.}~\bibnamefont {Xiang}}, \bibinfo {author}
  {\bibfnamefont {S.}~\bibnamefont {Dissanayake}}, \bibinfo {author}
  {\bibfnamefont {R.}~\bibnamefont {Gupta}}, \bibinfo {author} {\bibfnamefont
  {M.}~\bibnamefont {Matsuda}}, \bibinfo {author} {\bibfnamefont
  {F.}~\bibnamefont {Ye}}, \bibinfo {author} {\bibfnamefont {B.}~\bibnamefont
  {Haberl}}, \bibinfo {author} {\bibfnamefont {U.}~\bibnamefont
  {Kaluarachchi}}, \bibinfo {author} {\bibfnamefont {R.~J.}\ \bibnamefont
  {McQueeney}}, \bibinfo {author} {\bibfnamefont {A.}~\bibnamefont {Kreyssig}},
  \bibinfo {author} {\bibfnamefont {S.~L.}\ \bibnamefont {Bud'ko}}, \ and\
  \bibinfo {author} {\bibfnamefont {P.~C.}\ \bibnamefont {Canfield}},\
  }\bibfield  {title} {\enquote {\bibinfo {title} {Formation of short-range
  magnetic order and avoided ferromagnetic quantum criticality in pressurized
  {{LaCrGe$_3$}}},}\ }\href {\doibase 10.1103/PhysRevB.103.075111} {\bibfield
  {journal} {\bibinfo  {journal} {Phys. Rev. B}\ }\textbf {\bibinfo {volume}
  {103}},\ \bibinfo {pages} {075111} (\bibinfo {year} {2021})}\BibitemShut
  {NoStop}%
\bibitem [{\citenamefont {Rana}\ \emph {et~al.}(2021)\citenamefont {Rana},
  \citenamefont {Kotegawa}, \citenamefont {Ullah}, \citenamefont {Gati},
  \citenamefont {Bud'ko}, \citenamefont {Canfield}, \citenamefont {Tou},
  \citenamefont {Taufour},\ and\ \citenamefont
  {Furukawa}}]{2021_Rana_PhysRevB}%
  \BibitemOpen
  \bibfield  {author} {\bibinfo {author} {\bibfnamefont {K.}~\bibnamefont
  {Rana}}, \bibinfo {author} {\bibfnamefont {H.}~\bibnamefont {Kotegawa}},
  \bibinfo {author} {\bibfnamefont {R.~R.}\ \bibnamefont {Ullah}}, \bibinfo
  {author} {\bibfnamefont {E.}~\bibnamefont {Gati}}, \bibinfo {author}
  {\bibfnamefont {S.~L.}\ \bibnamefont {Bud'ko}}, \bibinfo {author}
  {\bibfnamefont {P.~C.}\ \bibnamefont {Canfield}}, \bibinfo {author}
  {\bibfnamefont {H.}~\bibnamefont {Tou}}, \bibinfo {author} {\bibfnamefont
  {V.}~\bibnamefont {Taufour}}, \ and\ \bibinfo {author} {\bibfnamefont
  {Y.}~\bibnamefont {Furukawa}},\ }\bibfield  {title} {\enquote {\bibinfo
  {title} {Magnetic properties of the itinerant ferromagnet {{LaCrGe$_3$}}
  under pressure studied by {{La}}$^{139}$ {{NMR}}},}\ }\href {\doibase
  10.1103/PhysRevB.103.174426} {\bibfield  {journal} {\bibinfo  {journal}
  {Phys. Rev. B}\ }\textbf {\bibinfo {volume} {103}},\ \bibinfo {pages}
  {174426} (\bibinfo {year} {2021})}\BibitemShut {NoStop}%
\bibitem [{\citenamefont {Wendl}\ \emph {et~al.}(2022)\citenamefont {Wendl},
  \citenamefont {Eisenlohr}, \citenamefont {Rucker}, \citenamefont {Duvinage},
  \citenamefont {Kleinhans}, \citenamefont {Vojta},\ and\ \citenamefont
  {Pfleiderer}}]{2022_Wendl_Nature}%
  \BibitemOpen
  \bibfield  {author} {\bibinfo {author} {\bibfnamefont {A.}~\bibnamefont
  {Wendl}}, \bibinfo {author} {\bibfnamefont {H.}~\bibnamefont {Eisenlohr}},
  \bibinfo {author} {\bibfnamefont {F.}~\bibnamefont {Rucker}}, \bibinfo
  {author} {\bibfnamefont {C.}~\bibnamefont {Duvinage}}, \bibinfo {author}
  {\bibfnamefont {M.}~\bibnamefont {Kleinhans}}, \bibinfo {author}
  {\bibfnamefont {M:}~\bibnamefont {Vojta}}, \ and\ \bibinfo {author}
  {\bibfnamefont {C.}~\bibnamefont {Pfleiderer}},\ }\bibfield  {title}
  {\enquote {\bibinfo {title} {Emergence of mesoscale quantum phase transitions
  in a ferromagnet},}\ }\href {\doibase 10.1038/s41586-022-04995-5} {\bibfield
  {journal} {\bibinfo  {journal} {Nature}\ }\textbf {\bibinfo {volume} {609}},\
  \bibinfo {pages} {65} (\bibinfo {year} {2022})}\BibitemShut {NoStop}%
\bibitem [{\citenamefont {Halpern}\ and\ \citenamefont
  {Holstein}(1941)}]{halpern1941}%
  \BibitemOpen
  \bibfield  {author} {\bibinfo {author} {\bibfnamefont {O.}~\bibnamefont
  {Halpern}}\ and\ \bibinfo {author} {\bibfnamefont {T.}~\bibnamefont
  {Holstein}},\ }\bibfield  {title} {{\selectlanguage {english}\enquote
  {\bibinfo {title} {On the {{Passage}} of {{Neutrons Through
  Ferromagnets}}},}\ }}\href {\doibase 10.1103/PhysRev.59.960} {\bibfield
  {journal} {\bibinfo  {journal} {Physical Review}\ }\textbf {\bibinfo {volume}
  {59}},\ \bibinfo {pages} {960--981} (\bibinfo {year} {1941})}\BibitemShut
  {NoStop}%
\bibitem [{\citenamefont {Rauch}\ and\ \citenamefont
  {Weber}(1968)}]{rauch1968a}%
  \BibitemOpen
  \bibfield  {author} {\bibinfo {author} {\bibfnamefont {H.}~\bibnamefont
  {Rauch}}\ and\ \bibinfo {author} {\bibfnamefont {H.W.}\ \bibnamefont
  {Weber}},\ }\bibfield  {title} {{\selectlanguage {english}\enquote {\bibinfo
  {title} {Passage of polarized neutrons through type {{II}}
  superconductors},}\ }}\href {\doibase 10.1016/0375-9601(68)90788-3}
  {\bibfield  {journal} {\bibinfo  {journal} {Physics Letters A}\ }\textbf
  {\bibinfo {volume} {26}},\ \bibinfo {pages} {460--461} (\bibinfo {year}
  {1968})}\BibitemShut {NoStop}%
\bibitem [{\citenamefont {Drabkin}\ \emph {et~al.}(1969)\citenamefont
  {Drabkin}, \citenamefont {Zabidarov}, \citenamefont {Kasman},\ and\
  \citenamefont {Okorokov}}]{drabkin1969}%
  \BibitemOpen
  \bibfield  {author} {\bibinfo {author} {\bibfnamefont {G.~M.}\ \bibnamefont
  {Drabkin}}, \bibinfo {author} {\bibfnamefont {E.~I.}\ \bibnamefont
  {Zabidarov}}, \bibinfo {author} {\bibfnamefont {Y.~A.}\ \bibnamefont
  {Kasman}}, \ and\ \bibinfo {author} {\bibfnamefont {A.~I.}\ \bibnamefont
  {Okorokov}},\ }\bibfield  {title} {{\selectlanguage {english}\enquote
  {\bibinfo {title} {Investigation of a phase transition in nickel with
  polarized neutrons},}\ }}\href@noop {} {\bibfield  {journal} {\bibinfo
  {journal} {Soviet Physics JETP}\ }\textbf {\bibinfo {volume} {29}},\ \bibinfo
  {pages} {261--266} (\bibinfo {year} {1969})}\BibitemShut {NoStop}%
\bibitem [{\citenamefont {Maleev}\ and\ \citenamefont
  {Ruban}(1970)}]{maleev1970}%
  \BibitemOpen
  \bibfield  {author} {\bibinfo {author} {\bibfnamefont {S.~V.}\ \bibnamefont
  {Maleev}}\ and\ \bibinfo {author} {\bibfnamefont {V.~A.}\ \bibnamefont
  {Ruban}},\ }\bibfield  {title} {\enquote {\bibinfo {title} {Depolarization of
  neutrons passing through a ferromagnet},}\ }\href@noop {} {\bibfield
  {journal} {\bibinfo  {journal} {Soviet Physics JETP}\ }\textbf {\bibinfo
  {volume} {31}},\ \bibinfo {pages} {111--116} (\bibinfo {year}
  {1970})}\BibitemShut {NoStop}%
\bibitem [{\citenamefont {Weber}(1974)}]{weber1974}%
  \BibitemOpen
  \bibfield  {author} {\bibinfo {author} {\bibfnamefont {H.~W.}\ \bibnamefont
  {Weber}},\ }\bibfield  {title} {{\selectlanguage {english}\enquote {\bibinfo
  {title} {Properties of the flux line lattice in hysteretic type {{II}}
  superconductors. {{II}}. {{Neutron}} depolarization experiments},}\ }}\href
  {\doibase 10.1007/BF00654543} {\bibfield  {journal} {\bibinfo  {journal}
  {Journal of Low Temperature Physics}\ }\textbf {\bibinfo {volume} {17}},\
  \bibinfo {pages} {49--63} (\bibinfo {year} {1974})}\BibitemShut {NoStop}%
\bibitem [{\citenamefont {Rekveldt}\ and\ \citenamefont {{van
  Schaik}}(1979)}]{rekveldt1979}%
  \BibitemOpen
  \bibfield  {author} {\bibinfo {author} {\bibfnamefont {M.~Th.}\ \bibnamefont
  {Rekveldt}}\ and\ \bibinfo {author} {\bibfnamefont {F.~J.}\ \bibnamefont
  {{van Schaik}}},\ }\bibfield  {title} {{\selectlanguage {english}\enquote
  {\bibinfo {title} {Static and dynamic neutron depolarization studies of
  ferromagnetic domain structures},}\ }}\href {\doibase 10.1063/1.327079}
  {\bibfield  {journal} {\bibinfo  {journal} {Journal of Applied Physics}\
  }\textbf {\bibinfo {volume} {50}},\ \bibinfo {pages} {2122--2127} (\bibinfo
  {year} {1979})}\BibitemShut {NoStop}%
\bibitem [{\citenamefont {{van der Valk}}\ and\ \citenamefont
  {Rekveldt}(1982)}]{vandervalk1982}%
  \BibitemOpen
  \bibfield  {author} {\bibinfo {author} {\bibfnamefont {H.J.L.}\ \bibnamefont
  {{van der Valk}}}\ and\ \bibinfo {author} {\bibfnamefont {M.Th.}\
  \bibnamefont {Rekveldt}},\ }\bibfield  {title} {{\selectlanguage
  {english}\enquote {\bibinfo {title} {Neutron depolarization in ferromagnets
  in terms of correlation functions},}\ }}\href {\doibase
  10.1016/0304-8853(82)90032-4} {\bibfield  {journal} {\bibinfo  {journal}
  {Journal of Magnetism and Magnetic Materials}\ }\textbf {\bibinfo {volume}
  {28}},\ \bibinfo {pages} {88--96} (\bibinfo {year} {1982})}\BibitemShut
  {NoStop}%
\bibitem [{\citenamefont {Mitsuda}\ and\ \citenamefont
  {Endoh}(1985)}]{mitsuda1985}%
  \BibitemOpen
  \bibfield  {author} {\bibinfo {author} {\bibfnamefont {S.}~\bibnamefont
  {Mitsuda}}\ and\ \bibinfo {author} {\bibfnamefont {Y.}~\bibnamefont
  {Endoh}},\ }\bibfield  {title} {{\selectlanguage {english}\enquote {\bibinfo
  {title} {Neutron {{Depolarization Studies}} on {{Magnetization Process Using
  Pulsed Polarized Neutrons}}},}\ }}\href@noop {} {\bibfield  {journal}
  {\bibinfo  {journal} {Journal of the Physical Society of Japan}\ }\textbf
  {\bibinfo {volume} {54}},\ \bibinfo {pages} {1570--1580} (\bibinfo {year}
  {1985})}\BibitemShut {NoStop}%
\bibitem [{\citenamefont {Mirebeau}\ \emph {et~al.}(1986)\citenamefont
  {Mirebeau}, \citenamefont {Jehanno}, \citenamefont {Campbell}, \citenamefont
  {Hippert}, \citenamefont {Hennion},\ and\ \citenamefont
  {Hennion}}]{mirebeau1986}%
  \BibitemOpen
  \bibfield  {author} {\bibinfo {author} {\bibfnamefont {I.}~\bibnamefont
  {Mirebeau}}, \bibinfo {author} {\bibfnamefont {G.}~\bibnamefont {Jehanno}},
  \bibinfo {author} {\bibfnamefont {I.~A.}\ \bibnamefont {Campbell}}, \bibinfo
  {author} {\bibfnamefont {F.}~\bibnamefont {Hippert}}, \bibinfo {author}
  {\bibfnamefont {B.}~\bibnamefont {Hennion}}, \ and\ \bibinfo {author}
  {\bibfnamefont {M.}~\bibnamefont {Hennion}},\ }\bibfield  {title}
  {{\selectlanguage {english}\enquote {\bibinfo {title} {Magnetic order and
  canting in a reentrant alloy studied by magnetization, {{M{\"o}ssbauer}} and
  neutron scattering},}\ }}\href@noop {} {\bibfield  {journal} {\bibinfo
  {journal} {Journal of Magnetism and Magnetic Materials}\ }\textbf {\bibinfo
  {volume} {54}},\ \bibinfo {pages} {99--100} (\bibinfo {year}
  {1986})}\BibitemShut {NoStop}%
\bibitem [{\citenamefont {Rosman}\ and\ \citenamefont
  {Rekveldt}(1990)}]{rosman1990}%
  \BibitemOpen
  \bibfield  {author} {\bibinfo {author} {\bibfnamefont {R.}~\bibnamefont
  {Rosman}}\ and\ \bibinfo {author} {\bibfnamefont {M.~Th.}\ \bibnamefont
  {Rekveldt}},\ }\bibfield  {title} {{\selectlanguage {english}\enquote
  {\bibinfo {title} {Neutron depolarization theory in the {{Larmor}} and the
  scattering approach},}\ }}\href@noop {} {\bibfield  {journal} {\bibinfo
  {journal} {Zeitschrift f{\"u}r Physik B - Condensed Matter}\ }\textbf
  {\bibinfo {volume} {79}},\ \bibinfo {pages} {61--68} (\bibinfo {year}
  {1990})}\BibitemShut {NoStop}%
\bibitem [{\citenamefont {Mirebeau}\ \emph {et~al.}(1992)\citenamefont
  {Mirebeau}, \citenamefont {Bellouard}, \citenamefont {Hennion}, \citenamefont
  {Dormann}, \citenamefont {{Djega-Mariadassou}},\ and\ \citenamefont
  {Tessier}}]{mirebeau1992}%
  \BibitemOpen
  \bibfield  {author} {\bibinfo {author} {\bibfnamefont {I.}~\bibnamefont
  {Mirebeau}}, \bibinfo {author} {\bibfnamefont {C.}~\bibnamefont {Bellouard}},
  \bibinfo {author} {\bibfnamefont {M.}~\bibnamefont {Hennion}}, \bibinfo
  {author} {\bibfnamefont {J.L.}\ \bibnamefont {Dormann}}, \bibinfo {author}
  {\bibfnamefont {C.}~\bibnamefont {{Djega-Mariadassou}}}, \ and\ \bibinfo
  {author} {\bibfnamefont {M.}~\bibnamefont {Tessier}},\ }\bibfield  {title}
  {{\selectlanguage {english}\enquote {\bibinfo {title} {Small angle neutron
  scattering in a superparamagnet},}\ }}\href {\doibase
  10.1016/0304-8853(92)91454-2} {\bibfield  {journal} {\bibinfo  {journal}
  {Journal of Magnetism and Magnetic Materials}\ }\textbf {\bibinfo {volume}
  {104-107}},\ \bibinfo {pages} {1560--1562} (\bibinfo {year}
  {1992})}\BibitemShut {NoStop}%
\bibitem [{\citenamefont {Rekveldt}(1999)}]{rekveldt1999}%
  \BibitemOpen
  \bibfield  {author} {\bibinfo {author} {\bibfnamefont {M.~Th.}\ \bibnamefont
  {Rekveldt}},\ }\bibfield  {title} {{\selectlanguage {english}\enquote
  {\bibinfo {title} {Transmission of polarised neutrons in magnetic
  materials},}\ }}\href {\doibase 10.1016/S0921-4526(99)00005-8} {\bibfield
  {journal} {\bibinfo  {journal} {Physica B: Condensed Matter}\ }\textbf
  {\bibinfo {volume} {267-268}},\ \bibinfo {pages} {60--68} (\bibinfo {year}
  {1999})}\BibitemShut {NoStop}%
\bibitem [{\citenamefont {Yusuf}\ \emph {et~al.}(2000)\citenamefont {Yusuf},
  \citenamefont {Sahana}, \citenamefont {Hegde}, \citenamefont {D\"orr},\ and\
  \citenamefont {M\"uller}}]{2000_Yusuf_PRB}%
  \BibitemOpen
  \bibfield  {author} {\bibinfo {author} {\bibfnamefont {S.~M.}\ \bibnamefont
  {Yusuf}}, \bibinfo {author} {\bibfnamefont {M.}~\bibnamefont {Sahana}},
  \bibinfo {author} {\bibfnamefont {M.~S.}\ \bibnamefont {Hegde}}, \bibinfo
  {author} {\bibfnamefont {K.}~\bibnamefont {D\"orr}}, \ and\ \bibinfo {author}
  {\bibfnamefont {K.-H.}\ \bibnamefont {M\"uller}},\ }\bibfield  {title}
  {\enquote {\bibinfo {title} {Evidence of ferromagnetic domains in the
  {${\mathrm{La}}_{0.67}{\mathrm{Ca}}_{0.33}{\mathrm{Mn}}_{0.9}{\mathrm{Fe}}_{0.1}{\mathrm{O}}_{3}$}
  perovskite},}\ }\href {\doibase 10.1103/PhysRevB.62.1118} {\bibfield
  {journal} {\bibinfo  {journal} {Phys. Rev. B}\ }\textbf {\bibinfo {volume}
  {62}},\ \bibinfo {pages} {1118--1123} (\bibinfo {year} {2000})}\BibitemShut
  {NoStop}%
\bibitem [{\citenamefont {Sato}\ \emph {et~al.}(2004)\citenamefont {Sato},
  \citenamefont {Shinohara}, \citenamefont {Ogawa},\ and\ \citenamefont
  {Takeda}}]{sato2004}%
  \BibitemOpen
  \bibfield  {author} {\bibinfo {author} {\bibfnamefont {T.}~\bibnamefont
  {Sato}}, \bibinfo {author} {\bibfnamefont {T.}~\bibnamefont {Shinohara}},
  \bibinfo {author} {\bibfnamefont {T.}~\bibnamefont {Ogawa}}, \ and\ \bibinfo
  {author} {\bibfnamefont {M.}~\bibnamefont {Takeda}},\ }\bibfield  {title}
  {{\selectlanguage {english}\enquote {\bibinfo {title} {Spin freezing process
  in a reentrant ferromagnet studied by neutron depolarization analysis},}\
  }}\href {\doibase 10.1103/PhysRevB.70.134410} {\bibfield  {journal} {\bibinfo
   {journal} {Physical Review B}\ }\textbf {\bibinfo {volume} {70}},\ \bibinfo
  {pages} {134410} (\bibinfo {year} {2004})}\BibitemShut {NoStop}%
\bibitem [{\citenamefont {De~Teresa}\ \emph {et~al.}(2006)\citenamefont
  {De~Teresa}, \citenamefont {Ritter}, \citenamefont {Algarabel}, \citenamefont
  {Yusuf}, \citenamefont {Blasco}, \citenamefont {Kumar}, \citenamefont
  {Marquina},\ and\ \citenamefont {Ibarra}}]{2006_DeTeresa_PRB}%
  \BibitemOpen
  \bibfield  {author} {\bibinfo {author} {\bibfnamefont {J.~M.}\ \bibnamefont
  {De~Teresa}}, \bibinfo {author} {\bibfnamefont {C.}~\bibnamefont {Ritter}},
  \bibinfo {author} {\bibfnamefont {P.~A.}\ \bibnamefont {Algarabel}}, \bibinfo
  {author} {\bibfnamefont {S.~M.}\ \bibnamefont {Yusuf}}, \bibinfo {author}
  {\bibfnamefont {J.}~\bibnamefont {Blasco}}, \bibinfo {author} {\bibfnamefont
  {A.}~\bibnamefont {Kumar}}, \bibinfo {author} {\bibfnamefont
  {C.}~\bibnamefont {Marquina}}, \ and\ \bibinfo {author} {\bibfnamefont
  {M.~R.}\ \bibnamefont {Ibarra}},\ }\bibfield  {title} {\enquote {\bibinfo
  {title} {Detailed neutron study of the crossover from long-range to
  short-range magnetic ordering in
  {${({\mathrm{Nd}}_{1\ensuremath{-}x}{\mathrm{Tb}}_{x})}_{0.55}{\mathrm{Sr}}_{0.45}\mathrm{Mn}{\mathrm{O}}_{3}$}
  manganites},}\ }\href {\doibase 10.1103/PhysRevB.74.224442} {\bibfield
  {journal} {\bibinfo  {journal} {Phys. Rev. B}\ }\textbf {\bibinfo {volume}
  {74}},\ \bibinfo {pages} {224442} (\bibinfo {year} {2006})}\BibitemShut
  {NoStop}%
\bibitem [{\citenamefont {Rekveldt}\ \emph {et~al.}(2006)\citenamefont
  {Rekveldt}, \citenamefont {{van Dijk}}, \citenamefont {Grigoriev},
  \citenamefont {Kraan},\ and\ \citenamefont {Bouwman}}]{rekveldt2006}%
  \BibitemOpen
  \bibfield  {author} {\bibinfo {author} {\bibfnamefont {M.~Theo}\ \bibnamefont
  {Rekveldt}}, \bibinfo {author} {\bibfnamefont {Niels~H.}\ \bibnamefont {{van
  Dijk}}}, \bibinfo {author} {\bibfnamefont {Serguei~V.}\ \bibnamefont
  {Grigoriev}}, \bibinfo {author} {\bibfnamefont {Wicher~H.}\ \bibnamefont
  {Kraan}}, \ and\ \bibinfo {author} {\bibfnamefont {Wim~G.}\ \bibnamefont
  {Bouwman}},\ }\bibfield  {title} {{\selectlanguage {english}\enquote
  {\bibinfo {title} {Three-dimensional magnetic spin-echo small-angle neutron
  scattering and neutron depolarization: {{A}} comparison},}\ }}\href {\doibase
  10.1063/1.2204579} {\bibfield  {journal} {\bibinfo  {journal} {Review of
  Scientific Instruments}\ }\textbf {\bibinfo {volume} {77}},\ \bibinfo {pages}
  {073902} (\bibinfo {year} {2006})}\BibitemShut {NoStop}%
\bibitem [{\citenamefont {Treimer}\ \emph
  {et~al.}(2012{\natexlab{a}})\citenamefont {Treimer}, \citenamefont
  {Ebrahimi},\ and\ \citenamefont {Karakas}}]{treimer2012}%
  \BibitemOpen
  \bibfield  {author} {\bibinfo {author} {\bibfnamefont {W.}~\bibnamefont
  {Treimer}}, \bibinfo {author} {\bibfnamefont {O.}~\bibnamefont {Ebrahimi}}, \
  and\ \bibinfo {author} {\bibfnamefont {N.}~\bibnamefont {Karakas}},\
  }\bibfield  {title} {{\selectlanguage {english}\enquote {\bibinfo {title}
  {Observation of partial {{Meissner}} effect and flux pinning in
  superconducting lead containing non-superconducting parts},}\ }}\href@noop {}
  {\bibfield  {journal} {\bibinfo  {journal} {Applied Physics Letters}\
  }\textbf {\bibinfo {volume} {101}},\ \bibinfo {pages} {162603} (\bibinfo
  {year} {2012}{\natexlab{a}})}\BibitemShut {NoStop}%
\bibitem [{\citenamefont {Treimer}\ \emph
  {et~al.}(2012{\natexlab{b}})\citenamefont {Treimer}, \citenamefont
  {Ebrahimi}, \citenamefont {Karakas},\ and\ \citenamefont
  {Prozorov}}]{treimer2012a}%
  \BibitemOpen
  \bibfield  {author} {\bibinfo {author} {\bibfnamefont {Wolfgang}\
  \bibnamefont {Treimer}}, \bibinfo {author} {\bibfnamefont {Omid}\
  \bibnamefont {Ebrahimi}}, \bibinfo {author} {\bibfnamefont {Nursel}\
  \bibnamefont {Karakas}}, \ and\ \bibinfo {author} {\bibfnamefont {Ruslan}\
  \bibnamefont {Prozorov}},\ }\bibfield  {title} {\enquote {\bibinfo {title}
  {Polarized neutron imaging and three-dimensional calculation of magnetic flux
  trapping in bulk of superconductors},}\ }\href {\doibase
  10.1103/PhysRevB.85.184522} {\bibfield  {journal} {\bibinfo  {journal} {Phys.
  Rev. B}\ }\textbf {\bibinfo {volume} {85}},\ \bibinfo {pages} {184522}
  (\bibinfo {year} {2012}{\natexlab{b}})}\BibitemShut {NoStop}%
\bibitem [{\citenamefont {Treimer}\ \emph {et~al.}(2013)\citenamefont
  {Treimer}, \citenamefont {Ebrahimi},\ and\ \citenamefont
  {Karakas}}]{treimer2013}%
  \BibitemOpen
  \bibfield  {author} {\bibinfo {author} {\bibfnamefont {W.}~\bibnamefont
  {Treimer}}, \bibinfo {author} {\bibfnamefont {O.}~\bibnamefont {Ebrahimi}}, \
  and\ \bibinfo {author} {\bibfnamefont {N.}~\bibnamefont {Karakas}},\
  }\bibfield  {title} {{\selectlanguage {english}\enquote {\bibinfo {title}
  {Imaging of {{Quantum Mechanical Effects}} in {{Superconductors}} by
  {{Means}} of {{Polarized Neutron Radiography}}},}\ }}\href {\doibase
  10.1016/j.phpro.2013.03.028} {\bibfield  {journal} {\bibinfo  {journal}
  {Physics Procedia}\ }\textbf {\bibinfo {volume} {43}},\ \bibinfo {pages}
  {243--253} (\bibinfo {year} {2013})}\BibitemShut {NoStop}%
\bibitem [{\citenamefont {Seifert}\ \emph {et~al.}(2017)\citenamefont
  {Seifert}, \citenamefont {Schulz}, \citenamefont {Benka}, \citenamefont
  {Pfleiderer},\ and\ \citenamefont {Gilder}}]{seifert2017}%
  \BibitemOpen
  \bibfield  {author} {\bibinfo {author} {\bibfnamefont {M}~\bibnamefont
  {Seifert}}, \bibinfo {author} {\bibfnamefont {M}~\bibnamefont {Schulz}},
  \bibinfo {author} {\bibfnamefont {G}~\bibnamefont {Benka}}, \bibinfo {author}
  {\bibfnamefont {C}~\bibnamefont {Pfleiderer}}, \ and\ \bibinfo {author}
  {\bibfnamefont {S}~\bibnamefont {Gilder}},\ }\bibfield  {title}
  {{\selectlanguage {english}\enquote {\bibinfo {title} {Neutron depolarization
  measurements of magnetite in chiton teeth},}\ }}\href {\doibase
  10.1088/1742-6596/862/1/012024} {\bibfield  {journal} {\bibinfo  {journal}
  {Journal of Physics: Conference Series}\ }\textbf {\bibinfo {volume} {862}},\
  \bibinfo {pages} {012024} (\bibinfo {year} {2017})}\BibitemShut {NoStop}%
\bibitem [{\citenamefont {Deepak}\ \emph {et~al.}(2021)\citenamefont {Deepak},
  \citenamefont {Kumar},\ and\ \citenamefont {Yusuf}}]{2021_Kumar_PRM}%
  \BibitemOpen
  \bibfield  {author} {\bibinfo {author} {\bibnamefont {Deepak}}, \bibinfo
  {author} {\bibfnamefont {A.}~\bibnamefont {Kumar}}, \ and\ \bibinfo {author}
  {\bibfnamefont {S.~M.}\ \bibnamefont {Yusuf}},\ }\bibfield  {title} {\enquote
  {\bibinfo {title} {Intertwined magnetization and exchange bias reversals
  across compensation temperature in {$\mathrm{YbCr}{\mathrm{O}}_{3}$}
  compound},}\ }\href {\doibase 10.1103/PhysRevMaterials.5.124402} {\bibfield
  {journal} {\bibinfo  {journal} {Phys. Rev. Materials}\ }\textbf {\bibinfo
  {volume} {5}},\ \bibinfo {pages} {124402} (\bibinfo {year}
  {2021})}\BibitemShut {NoStop}%
\bibitem [{\citenamefont {Bakker}\ \emph {et~al.}(1968)\citenamefont {Bakker},
  \citenamefont {Rekveldt},\ and\ \citenamefont {Van~Loef}}]{bakker1968}%
  \BibitemOpen
  \bibfield  {author} {\bibinfo {author} {\bibfnamefont {H.K.}\ \bibnamefont
  {Bakker}}, \bibinfo {author} {\bibfnamefont {M.Th.}\ \bibnamefont
  {Rekveldt}}, \ and\ \bibinfo {author} {\bibfnamefont {J.J.}\ \bibnamefont
  {Van~Loef}},\ }\bibfield  {title} {{\selectlanguage {english}\enquote
  {\bibinfo {title} {Neutron depolarization measurements in nickel near the
  curie point},}\ }}\href {\doibase 10.1016/0375-9601(68)91123-7} {\bibfield
  {journal} {\bibinfo  {journal} {Physics Letters A}\ }\textbf {\bibinfo
  {volume} {27}},\ \bibinfo {pages} {69--70} (\bibinfo {year}
  {1968})}\BibitemShut {NoStop}%
\bibitem [{\citenamefont {Drabkin}\ \emph {et~al.}(1968)\citenamefont
  {Drabkin}, \citenamefont {Okorokov}, \citenamefont {Zabidarov},\ and\
  \citenamefont {Kasman}}]{drabkin1968}%
  \BibitemOpen
  \bibfield  {author} {\bibinfo {author} {\bibfnamefont {G.~M.}\ \bibnamefont
  {Drabkin}}, \bibinfo {author} {\bibfnamefont {A.~I.}\ \bibnamefont
  {Okorokov}}, \bibinfo {author} {\bibfnamefont {E.~I.}\ \bibnamefont
  {Zabidarov}}, \ and\ \bibinfo {author} {\bibfnamefont {Y.~A.}\ \bibnamefont
  {Kasman}},\ }\bibfield  {title} {{\selectlanguage {english}\enquote {\bibinfo
  {title} {Influence of the magnetic field on the phase transition in
  nickel},}\ }}\href@noop {} {\bibfield  {journal} {\bibinfo  {journal} {ZhETF
  Pisma Redaktsiiu}\ }\textbf {\bibinfo {volume} {8}},\ \bibinfo {pages} {549}
  (\bibinfo {year} {1968})}\BibitemShut {NoStop}%
\bibitem [{\citenamefont {Takahashi}\ \emph {et~al.}(1995)\citenamefont
  {Takahashi}, \citenamefont {Itoh},\ and\ \citenamefont
  {Takeda}}]{takahashi1995}%
  \BibitemOpen
  \bibfield  {author} {\bibinfo {author} {\bibfnamefont {M.}~\bibnamefont
  {Takahashi}}, \bibinfo {author} {\bibfnamefont {S.}~\bibnamefont {Itoh}}, \
  and\ \bibinfo {author} {\bibfnamefont {M.}~\bibnamefont {Takeda}},\
  }\bibfield  {title} {{\selectlanguage {english}\enquote {\bibinfo {title}
  {Neutron depolarization study on the magnetic critical fluctuation in
  {{Rb$_2$CrCl$_4$}}},}\ }}\href@noop {} {\bibfield  {journal} {\bibinfo
  {journal} {Journal of the Physical Society of Japan}\ }\textbf {\bibinfo
  {volume} {64}},\ \bibinfo {pages} {268--274} (\bibinfo {year}
  {1995})}\BibitemShut {NoStop}%
\bibitem [{\citenamefont {Schulz}\ \emph {et~al.}(2010)\citenamefont {Schulz},
  \citenamefont {Neubauer}, \citenamefont {Masalovich}, \citenamefont
  {M{\"u}hlbauer}, \citenamefont {Calzada}, \citenamefont {Schillinger},
  \citenamefont {Pfleiderer},\ and\ \citenamefont {B{\"o}ni}}]{schulz2010}%
  \BibitemOpen
  \bibfield  {author} {\bibinfo {author} {\bibfnamefont {Michael}\ \bibnamefont
  {Schulz}}, \bibinfo {author} {\bibfnamefont {Andreas}\ \bibnamefont
  {Neubauer}}, \bibinfo {author} {\bibfnamefont {Sergey}\ \bibnamefont
  {Masalovich}}, \bibinfo {author} {\bibfnamefont {Martin}\ \bibnamefont
  {M{\"u}hlbauer}}, \bibinfo {author} {\bibfnamefont {Elbio}\ \bibnamefont
  {Calzada}}, \bibinfo {author} {\bibfnamefont {Burkhard}\ \bibnamefont
  {Schillinger}}, \bibinfo {author} {\bibfnamefont {Christian}\ \bibnamefont
  {Pfleiderer}}, \ and\ \bibinfo {author} {\bibfnamefont {Peter}\ \bibnamefont
  {B{\"o}ni}},\ }\bibfield  {title} {{\selectlanguage {english}\enquote
  {\bibinfo {title} {Towards a tomographic reconstruction of neutron
  depolarization data},}\ }}\href {\doibase 10.1088/1742-6596/211/1/012025}
  {\bibfield  {journal} {\bibinfo  {journal} {Journal of Physics: Conference
  Series}\ }\textbf {\bibinfo {volume} {211}},\ \bibinfo {pages} {012025}
  (\bibinfo {year} {2010})}\BibitemShut {NoStop}%
\bibitem [{\citenamefont {Schulz}(2010)}]{schulz_phd}%
  \BibitemOpen
  \bibfield  {author} {\bibinfo {author} {\bibfnamefont {Michael}\ \bibnamefont
  {Schulz}},\ }{\selectlanguage {english}\emph {\bibinfo {title} {Radiography
  with polarized neutrons}}},\ \href@noop {} {\bibinfo {type} {{{PhD}}
  thesis}},\ \bibinfo  {school} {Technische Universit{\"a}t M{\"u}nchen},
  \bibinfo {address} {{Munich}} (\bibinfo {year} {2010})\BibitemShut {NoStop}%
\bibitem [{\citenamefont {Schulz}\ \emph {et~al.}(2016)\citenamefont {Schulz},
  \citenamefont {Neubauer}, \citenamefont {B{\"o}ni},\ and\ \citenamefont
  {Pfleiderer}}]{schulz2016}%
  \BibitemOpen
  \bibfield  {author} {\bibinfo {author} {\bibfnamefont {M.}~\bibnamefont
  {Schulz}}, \bibinfo {author} {\bibfnamefont {A.}~\bibnamefont {Neubauer}},
  \bibinfo {author} {\bibfnamefont {P.}~\bibnamefont {B{\"o}ni}}, \ and\
  \bibinfo {author} {\bibfnamefont {C.}~\bibnamefont {Pfleiderer}},\ }\bibfield
   {title} {{\selectlanguage {english}\enquote {\bibinfo {title} {Neutron
  depolarization imaging of the hydrostatic pressure dependence of
  inhomogeneous ferromagnets},}\ }}\href {\doibase 10.1063/1.4950806}
  {\bibfield  {journal} {\bibinfo  {journal} {Applied Physics Letters}\
  }\textbf {\bibinfo {volume} {108}},\ \bibinfo {pages} {202402} (\bibinfo
  {year} {2016})}\BibitemShut {NoStop}%
\bibitem [{\citenamefont {Jorba}\ \emph {et~al.}(2019)\citenamefont {Jorba},
  \citenamefont {Schulz}, \citenamefont {Hussey}, \citenamefont {Abir},
  \citenamefont {Seifert}, \citenamefont {Tsurkan}, \citenamefont {Loidl},
  \citenamefont {Pfleiderer},\ and\ \citenamefont {Khaykovich}}]{jorba2019}%
  \BibitemOpen
  \bibfield  {author} {\bibinfo {author} {\bibfnamefont {Pau}\ \bibnamefont
  {Jorba}}, \bibinfo {author} {\bibfnamefont {Michael}\ \bibnamefont {Schulz}},
  \bibinfo {author} {\bibfnamefont {Daniel~S.}\ \bibnamefont {Hussey}},
  \bibinfo {author} {\bibfnamefont {Muhammad}\ \bibnamefont {Abir}}, \bibinfo
  {author} {\bibfnamefont {Marc}\ \bibnamefont {Seifert}}, \bibinfo {author}
  {\bibfnamefont {Vladimir}\ \bibnamefont {Tsurkan}}, \bibinfo {author}
  {\bibfnamefont {Alois}\ \bibnamefont {Loidl}}, \bibinfo {author}
  {\bibfnamefont {Christian}\ \bibnamefont {Pfleiderer}}, \ and\ \bibinfo
  {author} {\bibfnamefont {Boris}\ \bibnamefont {Khaykovich}},\ }\bibfield
  {title} {{\selectlanguage {english}\enquote {\bibinfo {title}
  {High-resolution neutron depolarization microscopy of the ferromagnetic
  transitions in {{Ni$_3$Al}} and {{HgCr$_2$Se$_4$}} under pressure},}\ }}\href
  {\doibase 10.1016/j.jmmm.2018.11.086} {\bibfield  {journal} {\bibinfo
  {journal} {Journal of Magnetism and Magnetic Materials}\ }\textbf {\bibinfo
  {volume} {475}},\ \bibinfo {pages} {176--183} (\bibinfo {year}
  {2019})}\BibitemShut {NoStop}%
\bibitem [{\citenamefont {Kardjilov}\ \emph {et~al.}(2008)\citenamefont
  {Kardjilov}, \citenamefont {Manke}, \citenamefont {Strobl}, \citenamefont
  {Hilger}, \citenamefont {Treimer}, \citenamefont {Meissner}, \citenamefont
  {Krist},\ and\ \citenamefont {Banhart}}]{kardjilov2008}%
  \BibitemOpen
  \bibfield  {author} {\bibinfo {author} {\bibfnamefont {Nikolay}\ \bibnamefont
  {Kardjilov}}, \bibinfo {author} {\bibfnamefont {Ingo}\ \bibnamefont {Manke}},
  \bibinfo {author} {\bibfnamefont {Markus}\ \bibnamefont {Strobl}}, \bibinfo
  {author} {\bibfnamefont {Andr{\'e}}\ \bibnamefont {Hilger}}, \bibinfo
  {author} {\bibfnamefont {Wolfgang}\ \bibnamefont {Treimer}}, \bibinfo
  {author} {\bibfnamefont {Michael}\ \bibnamefont {Meissner}}, \bibinfo
  {author} {\bibfnamefont {Thomas}\ \bibnamefont {Krist}}, \ and\ \bibinfo
  {author} {\bibfnamefont {John}\ \bibnamefont {Banhart}},\ }\bibfield  {title}
  {{\selectlanguage {english}\enquote {\bibinfo {title} {Three-dimensional
  imaging of magnetic fields with polarized neutrons},}\ }}\href@noop {}
  {\bibfield  {journal} {\bibinfo  {journal} {Nature Physics}\ }\textbf
  {\bibinfo {volume} {4}},\ \bibinfo {pages} {399--403} (\bibinfo {year}
  {2008})}\BibitemShut {NoStop}%
\bibitem [{\citenamefont {Kappler}\ \emph {et~al.}(1988)\citenamefont
  {Kappler}, \citenamefont {Beaurepaire}, \citenamefont {Krill}, \citenamefont
  {Godart}, \citenamefont {Nieva},\ and\ \citenamefont
  {Sereni}}]{1988_Kappler_JPC}%
  \BibitemOpen
  \bibfield  {author} {\bibinfo {author} {\bibfnamefont {J.~P.}\ \bibnamefont
  {Kappler}}, \bibinfo {author} {\bibfnamefont {E.}~\bibnamefont
  {Beaurepaire}}, \bibinfo {author} {\bibfnamefont {G.}~\bibnamefont {Krill}},
  \bibinfo {author} {\bibfnamefont {C.}~\bibnamefont {Godart}}, \bibinfo
  {author} {\bibfnamefont {G.~L.}\ \bibnamefont {Nieva}}, \ and\ \bibinfo
  {author} {\bibfnamefont {J.G.}\ \bibnamefont {Sereni}},\ }\bibfield  {title}
  {\enquote {\bibinfo {title} {Magnetic to non-magnetic transition of {Ce}
  induced by volume in {Ce(Pd, Ni)} and electron concentration in
  {Ce(Pd,Rh)}},}\ }\href@noop {} {\bibfield  {journal} {\bibinfo  {journal} {J.
  Phys. Colloques}\ }\textbf {\bibinfo {volume} {49}},\ \bibinfo {pages}
  {C8--723} (\bibinfo {year} {1988})}\BibitemShut {NoStop}%
\bibitem [{\citenamefont {Kappler}(1991)}]{kappler1991}%
  \BibitemOpen
  \bibfield  {author} {\bibinfo {author} {\bibfnamefont {J~P}\ \bibnamefont
  {Kappler}},\ }\bibfield  {title} {{\selectlanguage {english}\enquote
  {\bibinfo {title} {Crossover between intermediate valence and magnetic order
  in {{CeRh1}}-{{xPdx}}},}\ }}\href {\doibase 10.1016/0921-4526(91)90547-R}
  {\bibfield  {journal} {\bibinfo  {journal} {Physica B: Condensed Matter}\
  }\textbf {\bibinfo {volume} {171}},\ \bibinfo {pages} {346--349} (\bibinfo
  {year} {1991})}\BibitemShut {NoStop}%
\bibitem [{\citenamefont {Sereni}\ \emph {et~al.}(1993)\citenamefont {Sereni},
  \citenamefont {Beaurepaire},\ and\ \citenamefont {Kappler}}]{sereni1993}%
  \BibitemOpen
  \bibfield  {author} {\bibinfo {author} {\bibfnamefont {J.~G.}\ \bibnamefont
  {Sereni}}, \bibinfo {author} {\bibfnamefont {E.}~\bibnamefont {Beaurepaire}},
  \ and\ \bibinfo {author} {\bibfnamefont {J.~P.}\ \bibnamefont {Kappler}},\
  }\bibfield  {title} {{\selectlanguage {english}\enquote {\bibinfo {title}
  {Effects of volume and electronic concentration on the {CePd$_{1-x}$M$_x$})
  compounds ({{M}}={{Ni}}, {{Rh}}, and {{Ag}})},}\ }}\href {\doibase
  10.1103/PhysRevB.48.3747} {\bibfield  {journal} {\bibinfo  {journal}
  {Physical Review B}\ }\textbf {\bibinfo {volume} {48}},\ \bibinfo {pages}
  {3747--3754} (\bibinfo {year} {1993})}\BibitemShut {NoStop}%
\bibitem [{\citenamefont {Sereni}\ \emph {et~al.}(2005)\citenamefont {Sereni},
  \citenamefont {K{\"u}chler},\ and\ \citenamefont {Geibel}}]{sereni2005}%
  \BibitemOpen
  \bibfield  {author} {\bibinfo {author} {\bibfnamefont {J.G.}\ \bibnamefont
  {Sereni}}, \bibinfo {author} {\bibfnamefont {R.}~\bibnamefont {K{\"u}chler}},
  \ and\ \bibinfo {author} {\bibfnamefont {C.}~\bibnamefont {Geibel}},\
  }\bibfield  {title} {{\selectlanguage {english}\enquote {\bibinfo {title}
  {Evidence for a ferromagnetic quantum critical point in
  {{CePd1}}-{{xRhx}}},}\ }}\href {\doibase 10.1016/j.physb.2004.12.050}
  {\bibfield  {journal} {\bibinfo  {journal} {Physica B: Condensed Matter}\
  }\textbf {\bibinfo {volume} {359-361}},\ \bibinfo {pages} {41--43} (\bibinfo
  {year} {2005})}\BibitemShut {NoStop}%
\bibitem [{\citenamefont {Deppe}\ \emph {et~al.}(2006)\citenamefont {Deppe},
  \citenamefont {Pedrazzini}, \citenamefont {{Caroca-Canales}}, \citenamefont
  {Geibel},\ and\ \citenamefont {Sereni}}]{deppe2006}%
  \BibitemOpen
  \bibfield  {author} {\bibinfo {author} {\bibfnamefont {M.}~\bibnamefont
  {Deppe}}, \bibinfo {author} {\bibfnamefont {P.}~\bibnamefont {Pedrazzini}},
  \bibinfo {author} {\bibfnamefont {N.}~\bibnamefont {{Caroca-Canales}}},
  \bibinfo {author} {\bibfnamefont {C.}~\bibnamefont {Geibel}}, \ and\ \bibinfo
  {author} {\bibfnamefont {J.G.}\ \bibnamefont {Sereni}},\ }\bibfield  {title}
  {\enquote {\bibinfo {title} {Investigations of {CePd$_{1-x}$Rh$_x$} single
  crystals located near a ferromagnetic quantum critical point},}\ }\href
  {\doibase 10.1016/j.physb.2006.01.038} {\bibfield  {journal} {\bibinfo
  {journal} {Physica B: Condensed Matter}\ }\textbf {\bibinfo {volume}
  {378-380}},\ \bibinfo {pages} {96--97} (\bibinfo {year} {2006})}\BibitemShut
  {NoStop}%
\bibitem [{\citenamefont {Pikul}\ \emph {et~al.}(2006)\citenamefont {Pikul},
  \citenamefont {{Caroca-Canales}}, \citenamefont {Deppe}, \citenamefont
  {Gegenwart}, \citenamefont {Sereni}, \citenamefont {Geibel},\ and\
  \citenamefont {Steglich}}]{pikul2006}%
  \BibitemOpen
  \bibfield  {author} {\bibinfo {author} {\bibfnamefont {A~P}\ \bibnamefont
  {Pikul}}, \bibinfo {author} {\bibfnamefont {N}~\bibnamefont
  {{Caroca-Canales}}}, \bibinfo {author} {\bibfnamefont {M}~\bibnamefont
  {Deppe}}, \bibinfo {author} {\bibfnamefont {P}~\bibnamefont {Gegenwart}},
  \bibinfo {author} {\bibfnamefont {J~G}\ \bibnamefont {Sereni}}, \bibinfo
  {author} {\bibfnamefont {C}~\bibnamefont {Geibel}}, \ and\ \bibinfo {author}
  {\bibfnamefont {F}~\bibnamefont {Steglich}},\ }\bibfield  {title}
  {{\selectlanguage {english}\enquote {\bibinfo {title} {Non-{{Fermi}}-liquid
  behaviour close to the disappearance of ferromagnetism in
  {CePd$_{1-x}$Rh$_x$}},}\ }}\href {\doibase 10.1088/0953-8984/18/42/L01}
  {\bibfield  {journal} {\bibinfo  {journal} {Journal of Physics: Condensed
  Matter}\ }\textbf {\bibinfo {volume} {18}},\ \bibinfo {pages} {L535--L542}
  (\bibinfo {year} {2006})}\BibitemShut {NoStop}%
\bibitem [{\citenamefont {Sereni}\ \emph {et~al.}(2006)\citenamefont {Sereni},
  \citenamefont {K{\"u}chler},\ and\ \citenamefont {Geibel}}]{sereni2006}%
  \BibitemOpen
  \bibfield  {author} {\bibinfo {author} {\bibfnamefont {J.G.}\ \bibnamefont
  {Sereni}}, \bibinfo {author} {\bibfnamefont {R.}~\bibnamefont {K{\"u}chler}},
  \ and\ \bibinfo {author} {\bibfnamefont {C.}~\bibnamefont {Geibel}},\
  }\bibfield  {title} {{\selectlanguage {english}\enquote {\bibinfo {title}
  {Peculiar quantum criticality in ferromagnetic {{CePd1}}-{{xRhx}}},}\ }}\href
  {\doibase 10.1016/j.physb.2006.01.185} {\bibfield  {journal} {\bibinfo
  {journal} {Physica B: Condensed Matter}\ }\textbf {\bibinfo {volume}
  {378-380}},\ \bibinfo {pages} {648--649} (\bibinfo {year}
  {2006})}\BibitemShut {NoStop}%
\bibitem [{\citenamefont {Sereni}\ \emph {et~al.}(2007)\citenamefont {Sereni},
  \citenamefont {Westerkamp}, \citenamefont {K{\"u}chler}, \citenamefont
  {{Caroca-Canales}}, \citenamefont {Gegenwart},\ and\ \citenamefont
  {Geibel}}]{sereni2007}%
  \BibitemOpen
  \bibfield  {author} {\bibinfo {author} {\bibfnamefont {J.~G.}\ \bibnamefont
  {Sereni}}, \bibinfo {author} {\bibfnamefont {T.}~\bibnamefont {Westerkamp}},
  \bibinfo {author} {\bibfnamefont {R.}~\bibnamefont {K{\"u}chler}}, \bibinfo
  {author} {\bibfnamefont {N.}~\bibnamefont {{Caroca-Canales}}}, \bibinfo
  {author} {\bibfnamefont {P.}~\bibnamefont {Gegenwart}}, \ and\ \bibinfo
  {author} {\bibfnamefont {C.}~\bibnamefont {Geibel}},\ }\bibfield  {title}
  {{\selectlanguage {english}\enquote {\bibinfo {title} {Ferromagnetic quantum
  criticality in the alloy {CePd$_{1-x}$Rh$_x$}},}\ }}\href {\doibase
  10.1103/PhysRevB.75.024432} {\bibfield  {journal} {\bibinfo  {journal}
  {Physical Review B}\ }\textbf {\bibinfo {volume} {75}},\ \bibinfo {pages}
  {024432} (\bibinfo {year} {2007})}\BibitemShut {NoStop}%
\bibitem [{\citenamefont {Westerkamp}\ \emph {et~al.}(2009)\citenamefont
  {Westerkamp}, \citenamefont {Deppe}, \citenamefont {K{\"u}chler},
  \citenamefont {Brando}, \citenamefont {Geibel}, \citenamefont {Gegenwart},
  \citenamefont {Pikul},\ and\ \citenamefont {Steglich}}]{westerkamp2009}%
  \BibitemOpen
  \bibfield  {author} {\bibinfo {author} {\bibfnamefont {T.}~\bibnamefont
  {Westerkamp}}, \bibinfo {author} {\bibfnamefont {M.}~\bibnamefont {Deppe}},
  \bibinfo {author} {\bibfnamefont {R.}~\bibnamefont {K{\"u}chler}}, \bibinfo
  {author} {\bibfnamefont {M.}~\bibnamefont {Brando}}, \bibinfo {author}
  {\bibfnamefont {C.}~\bibnamefont {Geibel}}, \bibinfo {author} {\bibfnamefont
  {P.}~\bibnamefont {Gegenwart}}, \bibinfo {author} {\bibfnamefont {A.~P.}\
  \bibnamefont {Pikul}}, \ and\ \bibinfo {author} {\bibfnamefont
  {F.}~\bibnamefont {Steglich}},\ }\bibfield  {title} {{\selectlanguage
  {english}\enquote {\bibinfo {title} {Kondo-{{Cluster}}-{{Glass State}} near a
  {{Ferromagnetic Quantum Phase Transition}}},}\ }}\href {\doibase
  10.1103/PhysRevLett.102.206404} {\bibfield  {journal} {\bibinfo  {journal}
  {Physical Review Letters}\ }\textbf {\bibinfo {volume} {102}},\ \bibinfo
  {pages} {206404} (\bibinfo {year} {2009})}\BibitemShut {NoStop}%
\bibitem [{\citenamefont {Brando}\ \emph {et~al.}(2010)\citenamefont {Brando},
  \citenamefont {Westerkamp}, \citenamefont {Deppe}, \citenamefont {Gegenwart},
  \citenamefont {Geibel},\ and\ \citenamefont {Steglich}}]{brando2010}%
  \BibitemOpen
  \bibfield  {author} {\bibinfo {author} {\bibfnamefont {M}~\bibnamefont
  {Brando}}, \bibinfo {author} {\bibfnamefont {T}~\bibnamefont {Westerkamp}},
  \bibinfo {author} {\bibfnamefont {M}~\bibnamefont {Deppe}}, \bibinfo {author}
  {\bibfnamefont {P}~\bibnamefont {Gegenwart}}, \bibinfo {author}
  {\bibfnamefont {C}~\bibnamefont {Geibel}}, \ and\ \bibinfo {author}
  {\bibfnamefont {F}~\bibnamefont {Steglich}},\ }\bibfield  {title}
  {{\selectlanguage {english}\enquote {\bibinfo {title} {Quantum {{Griffiths}}
  phase in {CePd$_{1-x}$Rh$_x$} with x {$\approx$} 0.8},}\ }}\href {\doibase
  10.1088/1742-6596/200/1/012016} {\bibfield  {journal} {\bibinfo  {journal}
  {Journal of Physics: Conference Series}\ }\textbf {\bibinfo {volume} {200}},\
  \bibinfo {pages} {012016} (\bibinfo {year} {2010})}\BibitemShut {NoStop}%
\bibitem [{\citenamefont {Schmakat}(2015)}]{schmakat2015a}%
  \BibitemOpen
  \bibfield  {author} {\bibinfo {author} {\bibfnamefont {Philipp}\ \bibnamefont
  {Schmakat}},\ }{\selectlanguage {english}\emph {\bibinfo {title} {Neutron
  {{Depolarisation Measurements}} of {{Ferromagnetic Quantum Phase
  Transitions}} \& {{Wavelength}}-{{Frame Multiplication Chopper System}} for
  the {{Imaging Instrument ODIN}} at the {{ESS}}}}},\ \href@noop {} {\bibinfo
  {type} {{{PhD}} thesis}},\ \bibinfo  {school} {Technische Universit{\"a}t
  M{\"u}nchen}, \bibinfo {address} {{Munich}} (\bibinfo {year}
  {2015})\BibitemShut {NoStop}%
\bibitem [{\citenamefont {Westerkamp}(2008)}]{westerkamp2008}%
  \BibitemOpen
  \bibfield  {author} {\bibinfo {author} {\bibfnamefont {Tanja}\ \bibnamefont
  {Westerkamp}},\ }\emph {\bibinfo {title} {{Quantenphasen{\"u}berg{\"a}nge in
  den Schwere-Fermionen-Systemen Yb(Rh$_{1-x}$M$_x$)$_2$Si$_2$ und
  CePd$_{1-x}$Rh$_x$}}},\ \href@noop {} {\bibinfo {type} {{PhD thesis}}},\
  \bibinfo  {school} {Technische Universit{\"a}t Dresden}, \bibinfo {address}
  {{Dresden}} (\bibinfo {year} {2008})\BibitemShut {NoStop}%
\bibitem [{\citenamefont {Thornton}\ \emph {et~al.}(1998)\citenamefont
  {Thornton}, \citenamefont {Armitage}, \citenamefont {Tomka}, \citenamefont
  {Riedi}, \citenamefont {Mitchell}, \citenamefont {Houshiar}, \citenamefont
  {Adroja}, \citenamefont {Rainford},\ and\ \citenamefont
  {Fort}}]{thornton1998}%
  \BibitemOpen
  \bibfield  {author} {\bibinfo {author} {\bibfnamefont {M~J}\ \bibnamefont
  {Thornton}}, \bibinfo {author} {\bibfnamefont {J~G~M}\ \bibnamefont
  {Armitage}}, \bibinfo {author} {\bibfnamefont {G~J}\ \bibnamefont {Tomka}},
  \bibinfo {author} {\bibfnamefont {P~C}\ \bibnamefont {Riedi}}, \bibinfo
  {author} {\bibfnamefont {R~H}\ \bibnamefont {Mitchell}}, \bibinfo {author}
  {\bibfnamefont {M}~\bibnamefont {Houshiar}}, \bibinfo {author} {\bibfnamefont
  {D~T}\ \bibnamefont {Adroja}}, \bibinfo {author} {\bibfnamefont {B~D}\
  \bibnamefont {Rainford}}, \ and\ \bibinfo {author} {\bibfnamefont
  {D}~\bibnamefont {Fort}},\ }\bibfield  {title} {{\selectlanguage
  {english}\enquote {\bibinfo {title} {Low-temperature thermal and
  high-pressure studies of {{CePd}} and {{CeAgSb$_2$}}},}\ }}\href {\doibase
  10.1088/0953-8984/10/42/014} {\bibfield  {journal} {\bibinfo  {journal}
  {Journal of Physics: Condensed Matter}\ }\textbf {\bibinfo {volume} {10}},\
  \bibinfo {pages} {9485--9493} (\bibinfo {year} {1998})}\BibitemShut {NoStop}%
\bibitem [{\citenamefont {Nieva}\ \emph {et~al.}(1988)\citenamefont {Nieva},
  \citenamefont {Tm}, \citenamefont {Afyouni}, \citenamefont {Schmerber},\ and\
  \citenamefont {Kappler}}]{1988_Nieva_ZPhysB}%
  \BibitemOpen
  \bibfield  {author} {\bibinfo {author} {\bibfnamefont {G.~L.}\ \bibnamefont
  {Nieva}}, \bibinfo {author} {\bibfnamefont {J.~G.~S.}\ \bibnamefont {Tm}},
  \bibinfo {author} {\bibfnamefont {M.}~\bibnamefont {Afyouni}}, \bibinfo
  {author} {\bibfnamefont {G.}~\bibnamefont {Schmerber}}, \ and\ \bibinfo
  {author} {\bibfnamefont {J.~P.}\ \bibnamefont {Kappler}},\ }\bibfield
  {title} {\enquote {\bibinfo {title} {From {{Ferromagnetic}} to {{Non-Magnetic
  Singlet Ground State}} in {CePd$_{1-x}$Ni$_x$}},}\ }\href@noop {} {\bibfield
  {journal} {\bibinfo  {journal} {Z. Phys. B}\ }\textbf {\bibinfo {volume}
  {70}},\ \bibinfo {pages} {181} (\bibinfo {year} {1988})}\BibitemShut
  {NoStop}%
\bibitem [{\citenamefont {Koelling}\ \emph {et~al.}(1985)\citenamefont
  {Koelling}, \citenamefont {Dunlap},\ and\ \citenamefont
  {Crabtree}}]{1985_Koelling_PRB}%
  \BibitemOpen
  \bibfield  {author} {\bibinfo {author} {\bibfnamefont {D.~D.}\ \bibnamefont
  {Koelling}}, \bibinfo {author} {\bibfnamefont {B.~D.}\ \bibnamefont
  {Dunlap}}, \ and\ \bibinfo {author} {\bibfnamefont {G.~W.}\ \bibnamefont
  {Crabtree}},\ }\bibfield  {title} {\enquote {\bibinfo {title} {f-electron
  hybridization and heavy-fermion compounds},}\ }\href {\doibase
  10.1103/PhysRevB.31.4966} {\bibfield  {journal} {\bibinfo  {journal} {Phys.
  Rev. B}\ }\textbf {\bibinfo {volume} {31}},\ \bibinfo {pages} {4966--4971}
  (\bibinfo {year} {1985})}\BibitemShut {NoStop}%
\bibitem [{\citenamefont {Marcano}\ \emph {et~al.}(2007)\citenamefont
  {Marcano}, \citenamefont {G\'omez~Sal}, \citenamefont {Espeso}, \citenamefont
  {De~Teresa}, \citenamefont {Algarabel}, \citenamefont {Paulsen},\ and\
  \citenamefont {Iglesias}}]{2007_Marcano_PRL}%
  \BibitemOpen
  \bibfield  {author} {\bibinfo {author} {\bibfnamefont {N.}~\bibnamefont
  {Marcano}}, \bibinfo {author} {\bibfnamefont {J.~C.}\ \bibnamefont
  {G\'omez~Sal}}, \bibinfo {author} {\bibfnamefont {J.~I.}\ \bibnamefont
  {Espeso}}, \bibinfo {author} {\bibfnamefont {J.~M.}\ \bibnamefont
  {De~Teresa}}, \bibinfo {author} {\bibfnamefont {P.~A.}\ \bibnamefont
  {Algarabel}}, \bibinfo {author} {\bibfnamefont {C.}~\bibnamefont {Paulsen}},
  \ and\ \bibinfo {author} {\bibfnamefont {J.~R.}\ \bibnamefont {Iglesias}},\
  }\bibfield  {title} {\enquote {\bibinfo {title} {Mesoscopic magnetic states
  in metallic alloys with strong electronic correlations: A percolative
  scenario for {CeNi$_{1-x}$Cu$_x$}},}\ }\href {\doibase
  10.1103/PhysRevLett.98.166406} {\bibfield  {journal} {\bibinfo  {journal}
  {Phys. Rev. Lett.}\ }\textbf {\bibinfo {volume} {98}},\ \bibinfo {pages}
  {166406} (\bibinfo {year} {2007})}\BibitemShut {NoStop}%
\bibitem [{\citenamefont {Vojta}(2009)}]{vojta2009}%
  \BibitemOpen
  \bibfield  {author} {\bibinfo {author} {\bibfnamefont {Thomas}\ \bibnamefont
  {Vojta}},\ }\bibfield  {title} {{\selectlanguage {english}\enquote {\bibinfo
  {title} {Thermal expansion and {{Gr{\"u}neisen}} parameter in quantum
  {{Griffiths}} phases},}\ }}\href {\doibase 10.1103/PhysRevB.80.041101}
  {\bibfield  {journal} {\bibinfo  {journal} {Physical Review B}\ }\textbf
  {\bibinfo {volume} {80}},\ \bibinfo {pages} {041101(R)} (\bibinfo {year}
  {2009})}\BibitemShut {NoStop}%
\bibitem [{\citenamefont {Schlenker}\ and\ \citenamefont
  {Shull}(1973)}]{schlenker1973}%
  \BibitemOpen
  \bibfield  {author} {\bibinfo {author} {\bibfnamefont {Michel}\ \bibnamefont
  {Schlenker}}\ and\ \bibinfo {author} {\bibfnamefont {C.~G.}\ \bibnamefont
  {Shull}},\ }\bibfield  {title} {{\selectlanguage {english}\enquote {\bibinfo
  {title} {Polarized neutron techniques for the observation of ferromagnetic
  domains},}\ }}\href {\doibase 10.1063/1.1662914} {\bibfield  {journal}
  {\bibinfo  {journal} {Journal of Applied Physics}\ }\textbf {\bibinfo
  {volume} {44}},\ \bibinfo {pages} {4181--4184} (\bibinfo {year}
  {1973})}\BibitemShut {NoStop}%
\bibitem [{\citenamefont {Piegsa}\ \emph {et~al.}(2009)\citenamefont {Piegsa},
  \citenamefont {{van den Brandt}}, \citenamefont {Hautle}, \citenamefont
  {Kohlbrecher},\ and\ \citenamefont {Konter}}]{piegsa2009}%
  \BibitemOpen
  \bibfield  {author} {\bibinfo {author} {\bibfnamefont {F.~M.}\ \bibnamefont
  {Piegsa}}, \bibinfo {author} {\bibfnamefont {B.}~\bibnamefont {{van den
  Brandt}}}, \bibinfo {author} {\bibfnamefont {P.}~\bibnamefont {Hautle}},
  \bibinfo {author} {\bibfnamefont {J.}~\bibnamefont {Kohlbrecher}}, \ and\
  \bibinfo {author} {\bibfnamefont {J.~A.}\ \bibnamefont {Konter}},\ }\bibfield
   {title} {{\selectlanguage {english}\enquote {\bibinfo {title} {Quantitative
  {{Radiography}} of {{Magnetic Fields Using Neutron Spin Phase Imaging}}},}\
  }}\href {\doibase 10.1103/PhysRevLett.102.145501} {\bibfield  {journal}
  {\bibinfo  {journal} {Physical Review Letters}\ }\textbf {\bibinfo {volume}
  {102}},\ \bibinfo {pages} {145501} (\bibinfo {year} {2009})}\BibitemShut
  {NoStop}%
\bibitem [{\citenamefont {Sales}\ \emph {et~al.}(2018)\citenamefont {Sales},
  \citenamefont {Strobl}, \citenamefont {Shinohara}, \citenamefont {Tremsin},
  \citenamefont {Kuhn}, \citenamefont {Lionheart}, \citenamefont {Desai},
  \citenamefont {Dahl},\ and\ \citenamefont {Schmidt}}]{sales2018}%
  \BibitemOpen
  \bibfield  {author} {\bibinfo {author} {\bibfnamefont {Morten}\ \bibnamefont
  {Sales}}, \bibinfo {author} {\bibfnamefont {Markus}\ \bibnamefont {Strobl}},
  \bibinfo {author} {\bibfnamefont {Takenao}\ \bibnamefont {Shinohara}},
  \bibinfo {author} {\bibfnamefont {Anton}\ \bibnamefont {Tremsin}}, \bibinfo
  {author} {\bibfnamefont {Luise~Theil}\ \bibnamefont {Kuhn}}, \bibinfo
  {author} {\bibfnamefont {William R.~B.}\ \bibnamefont {Lionheart}}, \bibinfo
  {author} {\bibfnamefont {Naeem~M.}\ \bibnamefont {Desai}}, \bibinfo {author}
  {\bibfnamefont {Anders~Bjorholm}\ \bibnamefont {Dahl}}, \ and\ \bibinfo
  {author} {\bibfnamefont {S{\o}ren}\ \bibnamefont {Schmidt}},\ }\bibfield
  {title} {{\selectlanguage {english}\enquote {\bibinfo {title} {Three
  {{Dimensional Polarimetric Neutron Tomography}} of {{Magnetic Fields}}},}\
  }}\href {\doibase 10.1038/s41598-018-20461-7} {\bibfield  {journal} {\bibinfo
   {journal} {Scientific Reports}\ }\textbf {\bibinfo {volume} {8}} (\bibinfo
  {year} {2018}),\ 10.1038/s41598-018-20461-7}\BibitemShut {NoStop}%
\bibitem [{\citenamefont {Treimer}(2014)}]{treimer2014}%
  \BibitemOpen
  \bibfield  {author} {\bibinfo {author} {\bibfnamefont {Wolfgang}\
  \bibnamefont {Treimer}},\ }\bibfield  {title} {{\selectlanguage
  {english}\enquote {\bibinfo {title} {Radiography and tomography with
  polarized neutrons},}\ }}\href {\doibase 10.1016/j.jmmm.2013.09.032}
  {\bibfield  {journal} {\bibinfo  {journal} {Journal of Magnetism and Magnetic
  Materials}\ }\textbf {\bibinfo {volume} {350}},\ \bibinfo {pages} {188--198}
  (\bibinfo {year} {2014})}\BibitemShut {NoStop}%
\bibitem [{\citenamefont {Mitsuda}\ \emph {et~al.}(1992)\citenamefont
  {Mitsuda}, \citenamefont {Yoshizawa},\ and\ \citenamefont
  {Endoh}}]{mitsuda1992}%
  \BibitemOpen
  \bibfield  {author} {\bibinfo {author} {\bibfnamefont {S.}~\bibnamefont
  {Mitsuda}}, \bibinfo {author} {\bibfnamefont {H.}~\bibnamefont {Yoshizawa}},
  \ and\ \bibinfo {author} {\bibfnamefont {Y.}~\bibnamefont {Endoh}},\
  }\bibfield  {title} {{\selectlanguage {english}\enquote {\bibinfo {title}
  {Neutron-depolarization studies on re-entrant spin glass},}\ }}\href
  {\doibase 10.1103/PhysRevB.45.9788} {\bibfield  {journal} {\bibinfo
  {journal} {Physical Review B}\ }\textbf {\bibinfo {volume} {45}},\ \bibinfo
  {pages} {9788--9797} (\bibinfo {year} {1992})}\BibitemShut {NoStop}%
\bibitem [{\citenamefont {Calzada}\ \emph {et~al.}(2009)\citenamefont
  {Calzada}, \citenamefont {Gruenauer}, \citenamefont {M{\"u}hlbauer},
  \citenamefont {Schillinger},\ and\ \citenamefont {Schulz}}]{calzada2009}%
  \BibitemOpen
  \bibfield  {author} {\bibinfo {author} {\bibfnamefont {Elbio}\ \bibnamefont
  {Calzada}}, \bibinfo {author} {\bibfnamefont {Florian}\ \bibnamefont
  {Gruenauer}}, \bibinfo {author} {\bibfnamefont {Martin}\ \bibnamefont
  {M{\"u}hlbauer}}, \bibinfo {author} {\bibfnamefont {Burkhard}\ \bibnamefont
  {Schillinger}}, \ and\ \bibinfo {author} {\bibfnamefont {Michael}\
  \bibnamefont {Schulz}},\ }\bibfield  {title} {{\selectlanguage
  {english}\enquote {\bibinfo {title} {New design for the {{ANTARES}}-{{II}}
  facility for neutron imaging at {{FRM II}}},}\ }}\href {\doibase
  10.1016/j.nima.2009.01.192} {\bibfield  {journal} {\bibinfo  {journal}
  {Nuclear Instruments and Methods in Physics Research Section A: Accelerators,
  Spectrometers, Detectors and Associated Equipment}\ }\textbf {\bibinfo
  {volume} {605}},\ \bibinfo {pages} {50--53} (\bibinfo {year}
  {2009})}\BibitemShut {NoStop}%
\bibitem [{\citenamefont {Schulz}\ and\ \citenamefont
  {Schillinger}(2015)}]{schulz2015}%
  \BibitemOpen
  \bibfield  {author} {\bibinfo {author} {\bibfnamefont {Michael}\ \bibnamefont
  {Schulz}}\ and\ \bibinfo {author} {\bibfnamefont {Burkhard}\ \bibnamefont
  {Schillinger}},\ }\bibfield  {title} {{\selectlanguage {english}\enquote
  {\bibinfo {title} {{{ANTARES}}: {{Cold}} neutron radiography and tomography
  facility},}\ }}\href {\doibase 10.17815/jlsrf-1-42} {\bibfield  {journal}
  {\bibinfo  {journal} {Journal of large-scale research facilities JLSRF}\
  }\textbf {\bibinfo {volume} {1}} (\bibinfo {year} {2015}),\
  10.17815/jlsrf-1-42}\BibitemShut {NoStop}%
\bibitem [{\citenamefont {Hutanu}\ \emph {et~al.}(2016)\citenamefont {Hutanu},
  \citenamefont {Luberstetter}, \citenamefont {{Bourgeat-Lami}}, \citenamefont
  {Meven}, \citenamefont {Sazonov}, \citenamefont {Steffen}, \citenamefont
  {Heger}, \citenamefont {Roth},\ and\ \citenamefont
  {{Leli{\`e}vre-Berna}}}]{hutanu2016}%
  \BibitemOpen
  \bibfield  {author} {\bibinfo {author} {\bibfnamefont {V.}~\bibnamefont
  {Hutanu}}, \bibinfo {author} {\bibfnamefont {W.}~\bibnamefont
  {Luberstetter}}, \bibinfo {author} {\bibfnamefont {E.}~\bibnamefont
  {{Bourgeat-Lami}}}, \bibinfo {author} {\bibfnamefont {M.}~\bibnamefont
  {Meven}}, \bibinfo {author} {\bibfnamefont {A.}~\bibnamefont {Sazonov}},
  \bibinfo {author} {\bibfnamefont {A.}~\bibnamefont {Steffen}}, \bibinfo
  {author} {\bibfnamefont {G.}~\bibnamefont {Heger}}, \bibinfo {author}
  {\bibfnamefont {G.}~\bibnamefont {Roth}}, \ and\ \bibinfo {author}
  {\bibfnamefont {E.}~\bibnamefont {{Leli{\`e}vre-Berna}}},\ }\bibfield
  {title} {{\selectlanguage {english}\enquote {\bibinfo {title} {Implementation
  of a new {{Cryopad}} on the diffractometer {{POLI}} at {{MLZ}}},}\ }}\href
  {\doibase 10.1063/1.4963697} {\bibfield  {journal} {\bibinfo  {journal}
  {Review of Scientific Instruments}\ }\textbf {\bibinfo {volume} {87}},\
  \bibinfo {pages} {105108} (\bibinfo {year} {2016})}\BibitemShut {NoStop}%
\bibitem [{\citenamefont {Endoh}\ and\ \citenamefont
  {Ishikawa}(1986)}]{endoh1986}%
  \BibitemOpen
  \bibfield  {author} {\bibinfo {author} {\bibfnamefont {Y.}~\bibnamefont
  {Endoh}}\ and\ \bibinfo {author} {\bibfnamefont {Y.}~\bibnamefont
  {Ishikawa}},\ }\bibfield  {title} {{\selectlanguage {english}\enquote
  {\bibinfo {title} {Pulsed polarized neutron studies at {{KENS}}},}\
  }}\href@noop {} {\bibfield  {journal} {\bibinfo  {journal} {Physica B:
  Condensed Matter}\ }\textbf {\bibinfo {volume} {136}},\ \bibinfo {pages}
  {64--68} (\bibinfo {year} {1986})}\BibitemShut {NoStop}%
\bibitem [{\citenamefont {Franz}\ \emph {et~al.}(2019)\citenamefont {Franz},
  \citenamefont {S\"aubert}, \citenamefont {Wendl}, \citenamefont {Haslbeck},
  \citenamefont {Soltwedel}, \citenamefont {Jochum}, \citenamefont {Spitz},
  \citenamefont {Kindervater}, \citenamefont {Bauer}, \citenamefont {B\"oni},\
  and\ \citenamefont {Pfleiderer}}]{2019_Franz_JPSJ}%
  \BibitemOpen
  \bibfield  {author} {\bibinfo {author} {\bibfnamefont {C.}~\bibnamefont
  {Franz}}, \bibinfo {author} {\bibfnamefont {S.}~\bibnamefont {S\"aubert}},
  \bibinfo {author} {\bibfnamefont {A.}~\bibnamefont {Wendl}}, \bibinfo
  {author} {\bibfnamefont {F.}~\bibnamefont {Haslbeck}}, \bibinfo {author}
  {\bibfnamefont {O.}~\bibnamefont {Soltwedel}}, \bibinfo {author}
  {\bibfnamefont {J.K.}\ \bibnamefont {Jochum}}, \bibinfo {author}
  {\bibfnamefont {L.}~\bibnamefont {Spitz}}, \bibinfo {author} {\bibfnamefont
  {J.}~\bibnamefont {Kindervater}}, \bibinfo {author} {\bibfnamefont
  {A.}~\bibnamefont {Bauer}}, \bibinfo {author} {\bibfnamefont
  {P.}~\bibnamefont {B\"oni}}, \ and\ \bibinfo {author} {\bibfnamefont
  {C.}~\bibnamefont {Pfleiderer}},\ }\bibfield  {title} {\enquote {\bibinfo
  {title} {{MIEZE neutron spin-echo spectroscopy of strongly correlated
  electron systems}},}\ }\href {\doibase
  https://doi.org/10.7566/JPSJ.88.081002} {\bibfield  {journal} {\bibinfo
  {journal} {J. Phys. Soc. Jpn.}\ }\textbf {\bibinfo {volume} {88}},\ \bibinfo
  {pages} {081002} (\bibinfo {year} {2019})}\BibitemShut {NoStop}%
\end{thebibliography}
%%%%%%%%%%%%%%%%%%%%%%%%%%%%%%%%
%\end{document}
%%%%%%%%%%%%%%%%%%%%%%%%%%%%%%%%


%%%%%%%%%%%%%%%%%%%%%%%%%%%%%%%%
\end{document}